\RequirePackage{fix-cm}
\documentclass[fleqn,usenatbib]{mnras}

\usepackage{newtxtext,newtxmath}
\usepackage{booktabs,chemformula}

\usepackage[T1]{fontenc}

\DeclareRobustCommand{\VAN}[3]{#2}
\let\VANthebibliography\thebibliography
\def\thebibliography{\DeclareRobustCommand{\VAN}[3]{##3}\VANthebibliography}
\usepackage{xcolor} % add this in your preamble

\newcommand{\brfreq}[3]{#1{\scriptstyle\,^{+#2}_{-#3}}}

\usepackage{graphicx}	% Including figure files
\usepackage{amsmath}	% Advanced maths commands

\usepackage{amssymb}	% Extra maths symbols
\usepackage{tikz}
\usepackage{enumitem}
\usepackage{multicol}
\usepackage{needspace}
\usepackage{placeins}
\usepackage{float}
\usepackage{caption}
\title[New insights on classical radio galaxies]{
A new look at old devils. II: New insights on classical radio galaxies from MeerKAT and uGMRT}

\author[P.~P.~Legodi et al.]{
Portia~P.~Legodi$^{1}$\thanks{E-mail: legodpp@unisa.ac.za},
Tiziana~Venturi$^{2,5}$,
Dharam~V.~Lal$^{3}$,
Bernie~Fanaroff$^{4}$,
Oleg~M.~Smirnov$^{5,4,2}$
\newauthor
and Simona~Giacintucci$^{6}$ \\
$^{1}$UNISA Centre for Astrophysics and Space Sciences (UCASS), College of Science, Engineering and Technology,\\
University of South Africa, Cnr Christian de Wet Rd and Pioneer Avenue, Florida 1709, P.O. Box 392, 0003 UNISA, South Africa\\
$^{2}$INAF—Istituto di Radioastronomia, Via Gobetti 101, I-40129 Bologna, Italy\\
$^{3}$National Centre for Radio Astrophysics, Tata Institute of Fundamental Research, Post Box 3, Ganeshkhind P.O., Pune 411007, India\\
$^{4}$South African Radio Astronomy Observatory, Cape Town, 7700, South Africa\\
$^{5}$Centre for Radio Astronomy Techniques and Technologies (RATT), Department of Physics and Electronics, Rhodes University, Makhanda, 6140, South Africa\\
$^{6}$Naval Research Laboratory, 4555 Overlook Avenue SW, Code 7213, Washington, DC 20375, USA
}

\date{Accepted XXX. Received YYY; in original form ZZZ}

\pubyear{2026}

\begin{document}
\label{firstpage}
\pagerange{\pageref{firstpage}--\pageref{lastpage}}
\maketitle

% Abstract of the paper
\begin{abstract}
This paper presents a detailed morphological and spectral analysis of a sample of ten radio galaxies observed with the MeerKAT in L-band (856 -- 1712 MHz) and the upgraded Giant Metrewave Radio Telescope in Band-4 (550 -- 850 MHz). Our main goals are to revisit the current classification scheme for classical radio galaxies and study the properties of new radio features in jets, lobes and hot spots, which are becoming increasingly numerous due to the sensitivity and imaging capabilities of current radio interferometers, and suggest previous unexplored interaction mechanisms between the radio plasma and the external medium. {The sample from the 4C catalogue includes FR\,I, FR\,II, wide-angle and narrow-angle tailed sources as well as FR\,0 radio galaxies from FR0CAT, with redshifts ranging from 0.04 -- 0.20. The high angular resolution, $\sim$4$^{\prime\prime}$ -- 10$^{\prime\prime}$ total intensity images are presented, revealing complex structures in the jets, lobes and hotspots of these sources. The integrated spectra of the sources, constructed using flux density measurements from our observations and archival data, reveal spectral breaks and slopes indicative of radiative ageing, with ages spanning $\sim$40 – 242 Myr. While the sample broadly confirms the classical FR classification as a useful first-order scheme, the substructures within the radio emission of sources highlight the need for models that incorporate environmental complexity and episodic jet activity to fully describe radio-galaxy evolution.}

\end{abstract}

% Select between one and six entries from the list of approved keywords.
% Don't make up new ones.

%%%%%%%%%%%%%%%%%%%%%%%%%%%%%%%%%%%%%%%%%%%%%%%%%%

%%%%%%%%%%%%%%%%% BODY OF PAPER %%%%%%%%%%%%%%%%%%

\section{Introduction}
Radio galaxies are one of the most spectacular manifestations of accretion and ejection phenomena on supermassive
black holes at the centres of elliptical galaxies \citep{blandford1977electromagnetic, krause2019probing}. They bear information on the astrophysics at play in the
inner regions of active galactic nuclei, where the energy is extracted, and on the mechanisms transporting the
relativistic plasma out to distances which can cover up to Mpc scales \citep{hardcastle2020radio} . Moreover, the interaction of the radio
emission and the surrounding medium (interstellar or intergalactic, depending on the scale under consideration)
provides essential clues on the evolution of the nuclear component's radio properties and the thermal
gas in the groups and clusters in which these objects reside \citep{fabian2012observational, saikia2022jets}.

A revived interest in radio galaxies has blossomed thanks to the current generation of radio interferometers.
Low-Frequency Array (LOFAR, \citealt{lofar}), Karl G. Jansky Very Large Array (JVLA, \citealt{jvla}), MeerKAT \citep{Jonas2016}, Australian Square Kilometre Array Pathfinder (ASKAP, \citealt{askap}), and upgraded Giant Metrewave Radio Telescope (uGMRT, \citealt{ugmrt}), with their superb sensitivity and ($u,v$) coverage,
as well as a broad range of simultaneous angular resolutions, now offer a unique view of radio galaxies on scales from sub-arcsecond
to tens of arcseconds in a seamless range of frequencies, from 144 MHz (LOFAR, Dutch stations and international
baselines) to 8 GHz and beyond (JVLA). {At the same time, the large samples of radio galaxies delivered by
the continuum sky surveys carried out with LOFAR (LoTSS, \citealt{shimwell2026vizier}), ASKAP (EMU, \citealt{emu}), MeerKAT (MIGHTEE \citealt{mightee} and MGCLS, \citealt{mgcls}) and uGMRT (superMIGHTEE, \citealt{Lal2025ApJ})
allow to study the cosmological evolution of such objects through well-defined statistical samples.}

Deep imaging of radio galaxies with the current radio interferometers  shows a complex variety of features in the radio emission:
substructure in the hot spots, with multiple peaks, has been observed in several sources \citep{hardcastle2007chandra, condon2021threads, bruni2021hard, brienza2021snapshot, velovic2023meerkat, 2008MNRAS.390.1105L}; wiggles and thethers along the ridge of radio jets are new features to be understood; thin filaments of radio emission are seen both in the lobes of radio galaxies and emerging from them \citep{rudnick2022intracluster}; thin ``tubes''
of emission have been detected connecting different regions of the radio galaxy at least in projection \citep{ramatsoku2020collimated};
extensive low surface brightness emission has been detected beyond the previously known boundaries of the lobes and
some hot spots \citep{Fanaroff2021, 2021Galax...9...87L}, calling for a revision of the total size of the radio sources and of the statistical occurrence of giant radio galaxies; morphological distortions of the extended low surface brightness emission are
often observed, suggesting substantial interplay with the external medium \citep{hardcastle2019ngc}.

With the purpose of improving our knowledge of the radio galaxy phenomenon and of the interaction between the radio plasma
and the surrounding medium in light of the current exciting imaging capabilities in the radio band, we started an investigation of classical radio galaxies.
%belonging to the  4C catalogue. 
We selected a small sample of 17
FR\,I and FR\,II radio galaxies \citep[][see also Sect.~\ref{sec2}]{fanaroff&riley} to be observed with MeerKAT in L-band (856--1712 MHz) and uGMRT
in Band-4 (550--850 MHz), with the goal to address the following questions:
\\
\newline
\noindent
(a) does the sharp classification in FR\,I and FR\,II types still hold with the improved imaging capabilities available
these days?
\\
(b) Is very low surface brightness emission detectable around the recent class of compact FR\,0 radio galaxies \citep{baldi}?
\\
(c) Do we need more morphological classes or a more refined morphological scheme?
\\
(d) How can we explain the properties of the new features, such as multiple hotspots, filaments, ``tubes'' and several other fine details in the jets and lobes, and which information do they carry on the nuclear engine and on the interplay of the radio plasma with the surrounding medium?

Related to point (c) above, \cite{rudnick2021radio} recently pointed out the need to revise our approach to the classification
of radio galaxies to account for the increasing complexity of the images and the often ambiguous and conflicting
terminology in use by different authors. A new scheme was proposed whereby radio galaxies are not classified following
sharp categories (i.e. FR\,I, FR\,II, compact, wide-angle and narrow-angle tail), but are rather associated
``descriptors'' labelled $\#tags$. \\

Our first results were published in \cite{Fanaroff2021}, hereinafter Paper\,I, where we presented and discussed the
properties of the four radio galaxies 4C\,12.02, 4C\,12.03, CGCG\,044$-$046 (4C\,07.32) and CGCG\,021$-$063 (4C\,00.56)
in the 550$-$1712 MHz and 550--850 MHz frequency ranges. The quality and details of our images (total intensity and spectral imaging) allowed
a considerable advance in our knowledge of all four sources and strengthened the importance of this study. Our observations have revealed a wealth of details, including filamentary emissions in lobes, substructures in hotspots, i.e., multiple peaks, complex morphologies, misalignments, and radio emissions extending beyond the primary hotspot regions. Such abundance of details is a challenge for our understanding of radio galaxies, their evolution and the role of the environment surrounding them.
It is further a warning with respect to the use of automatic classification of radio sources in large-scale surveys.
\newline 
Here, we present total intensity and spectral index images for ten more sources and provide general considerations on the whole sample studies so far. 
The paper is organised as follows. The sample selection criteria is reported in Sect. \ref{sec2}; the observations and data reduction are given in Sect. \ref{sec3}; our new images are presented in Sect. \ref{results}; the results of our investigation are discussed in Sect. \ref{sec5}; conclusions and future outlook are provided in Sect. \ref{sec6}.

{In this paper, we assume the following cosmological parameters for a flat universe: H$_0$ = 69.6 kms$^{-1}$Mpc$^{-1}$, $\Omega_M$ = 0.286 and \noindent$\Omega_\Lambda$ = 0.714 \citep{bennett20141}. We define the spectral index, $\alpha$ in the sense that flux density, S$_\nu$ $\propto \nu^{\alpha}$.}

%______________________________________________________
\section{Target selection} \label{sec2}
{We selected a representative sample of 15 radio galaxies from the 4C catalogue ($-7^{\circ} <\delta < 80^{\circ}$; \citealt{Pilkington}) to investigate their radio morphology and spectral properties using the following criteria (see also \cite{Fanaroff2021}:} \\ [-0.5cm]

\noindent
\begin{enumerate}[leftmargin=*,label=(\roman*)]
\item[(i)]The radio galaxies are hosted by optical galaxies with spectroscopic redshift ranging from 0.04 to 0.20.
This range ensures the detection of extended emission on the megaparsec (Mpc) scale using both MeerKAT and uGMRT Band-4. We point out that the redshift range has been broadened compared to Paper\,I both to increase the sample size and to provide a similar fraction of FR\,I and FR\,II radio galaxies.
\item[(ii)]The radio sources are in the declination range of {[$-7^{\circ}$, +20$^{\circ}$]}.
%\\ [-0.5cm]. 
This ensures sufficiently long visibility while yielding comparable $(u,v)$ coverage for both MeerKAT and uGMRT.
\item[(iii)] The radio sources in the sample exhibit a clear double radio morphology, with evidence of additional surrounding emission visible at the $45^{\prime \prime}$ angular resolution of the NRAO VLA Sky Survey \citep[NVSS,][]{condon}.
\end{enumerate}

{To investigate FR\,0 radio galaxies, question~(b) in our list, we further included two FR\,0 sources from \cite{baldi2018}, selected within the same redshift and declination range, in order to search for faint low surface brightness emission (or lack thereof) potentially associated with previous cycles of activity. This brings the total sample size to 17 sources.}

Table~\ref{log} summarizes the properties of these radio galaxies.
The redshift distribution is sparse, with FR I radio galaxies spanning the range 0.045 $\leq$ z $\leq$ 0.081 and FR II radio galaxies lying in the range 0.0562 $\leq$ z $\leq$ 0.156. Here we report the results of observations carried out for 10 sources, in particular three FR\,Is (4C\,$-$03.43, 3C\,403.1 and 3C\,198), three FR\,IIs (3C\,105, 3C\,227 and 3C\,445), two tailed radio galaxies (CGCG\,047$-$067 and NGC\,7503) and two FR\,0s (SDSS\,J\,0917$+$1331 and SDSS\,J\,1120$+$0407).

\begin{table*}\label{tab1}
  \caption{Our sample.}
  \begin{tabular}{@{}|lcccccc@{}}
    \toprule
 Source Name & 4C Name & RA$_{\rm{J2000}}$  & Dec$_{\rm{J2000}}$  & z & log P$_{\rm{1.4\,GHz}}$ (W Hz$^{-1}$) \\
 \hline
 FR\,0                          &            &            &             &        &       \\
 \textbf{SDSS\, J\,0917$+$1331} &            & 09h 17m 54.3s & $+$13$^\circ$ 31$^{\prime}$ 45$^{\prime\prime}$ & 0.05   & 23.14 \\
 \textbf{SDSS\, J\,1120$+$0407}  &            & 11h 20m 29.2s & $+$04$^\circ$ 07$^\prime$ 42$^{\prime\prime}$ & 0.05   & 21.67 \\
 FR\,I                          &            &            &             &        &       \\
 UGC\,00595*                     &   4C\,$-$01.05         & 00h 57m 34.9s & $-$01$^\circ$ 23$^{\prime}$ 28$^{\prime\prime}$ & 0.045  & 25.39 \\
 \textbf{3C\,198}                &   4C\,$+$06.30         & 08h 22m 31.9s & $+$05$^\circ$ 57$^{\prime}$ 07$^{\prime\prime}$ & 0.081  & 25.61 \\
 \textbf{4C\,$-$03.43}           &   4C\,$-$03.43          & 11h 33m 05.1s & $-$04$^\circ$ 00$^{\prime}$ 48$^{\prime\prime}$ & 0.0519 & 24.49 \\
 CGCG\,021$-$063                 &   4C\,$+$00.56         & 15h 16m 40.2s & $+$00$^\circ$ 15$^{\prime}$ 02$^{\prime\prime}$ & 0.052  & 25.25 \\       
 NGC\,5920*                      &   4C\,$+$07.41         & 15h 21m 51.8s & $+$07$^\circ$ 42$^{\prime}$ 32$^{\prime\prime}$ & 0.045  & 23.71 \\
 \textbf{3C\,403.1}              &   4C\,$-$01.51         & 19h 52m 30.5s & $-$01$^\circ$ 17$^{\prime}$ 21$^{\prime\prime}$ & 0.055  & 25.21 \\
 FR\,I (WAT)                     &            &            &             &        &       \\
 CGCG\,044$-$046                 &   4C\,$+$07.32         & 13h 16m 17.0s & $+$07$^\circ$ 02$^{\prime}$ 47$^{\prime\prime}$ & 0.05   & 25.08 \\
 \textbf{CGCG\,047$-$067}        &   4C\,$+$07.36         & 14h 29m 55.4s & $+$07$^\circ$ 15$^{\prime}$ 13$^{\prime\prime}$ & 0.0548 & 25.04 \\
 FR\,I (NAT)                     &            &            &             &        &       \\
 \textbf{NGC\,7503}              &   4C\,$+$07.61         & 23h 10m 42.3s & $+$07$^\circ$ 34$^{\prime}$ 04$^{\prime\prime}$ & 0.044  & 25.82 \\
 FR\,II                          &            &            &             &        &       \\
 4C\,12.02                       &   4C\,12.02          & 00h 04m 50.2s & $+$12$^\circ$ 48$^{\prime}$ 40$^{\prime\prime}$ & 0.143  & 26.06 \\
 4C\,12.03                       &   4C\,12.03         & 00h 09m 52.6s & $+$02$^\circ$ 44$^{\prime}$ 05$^{\prime\prime}$ & 0.156  & 25.08 \\
 \textbf{3C\,105}                &   4C\,$+$03.08        & 04h 07m 16.5s & $+$03$^\circ$ 42$^{\prime}$ 26$^{\prime\prime}$ & 0.089  & 25.89 \\
 \textbf{3C\,227}                &   4C\,$+$07.29         & 09h 47m 45.1s & $+$07$^\circ$ 25$^{\prime}$ 21$^{\prime\prime}$ & 0.085  & 26.15 \\
 4C\,$+$20.25*                   &   4C\,$+$20.25         & 11h 25m 58.7s & $+$20$^\circ$ 05$^{\prime}$ 54$^{\prime\prime}$ & 0.133  & 25.51 \\
 \textbf{3C\,445}                &   4C\,$-$02.83         & 22h 23m 49.6s & $-$02$^\circ$ 06$^{\prime}$ 12$^{\prime\prime}$ & 0.0562 & 25.64 \\
 \hline
  \end{tabular}\\
 {\footnotesize {Note.} Objects in boldface are those we report in this paper. (*) marks no observation with either arrays. The remaining sources are those presented in Paper I.}
 \label{log}
\end{table*}

%.........................

\section{Observations and data reduction}\label{sec3}

Our observations were conducted using MeerKAT in L-band (856 -- 1712 MHz) and the uGMRT in Band-4 (550 -- 850 MHz). The combination of these two arrays provides a comparable ($u,v$) coverage, with sensitivities of approximately 5 -- 23 $\mu$Jy~beam$^{-1}$ for MeerKAT and 18 -- 116  $\mu$Jy~beam$^{-1}$ for the uGMRT across the frequency range 550 -- 1712 MHz. Owing to the complementary characteristics of these radio interferometers, the resulting data achieve similar angular resolution and broadly comparable sensitivity across the source structures.

Details of the individual observations, including observing dates, durations, and frequency setups, are summarized in Table \ref{tabsum}. Only a subset of sources in our sample was observed with both arrays, due to differences in time allocation between the two facilities.
In the following sections, we describe the observations and data reduction procedures for MeerKAT and the uGMRT.

\subsection{MeerKAT}
The MeerKAT array \citep{jonas2016meerkat, mauch20201} observed the sources 3C\,105, 3C\,445, 4C\,$-$03.43, CGCG\,047$-$067, SDSS J 0917$+$1331 and SDSS J\,1120$+$0407 in May 2019 and August 2021 at frequencies 856 -- 1712 MHz (L band) in full Stokes. The observation used at least 63 of 64 antennas, and each session spanned $\sim$5 hours, including scans of the calibrator sources. The correlator was configured for 8 seconds of integration time. A standard strategy for the observing sequence was used, including 10 minute tracks on the bandpass calibrator (i.e.,  PKS\,0408$-$65 or J\,1939$-$6342) and regular visits to the gain calibrator.

The data are calibrated using the CARacal pipeline \citep{cara}. The pipeline makes a pass of RFI excision using the Tricolour package \citep{hugo2022tricolour} and CASA's tfcrop. Following this, a round of direction-independent calibration using CASA tasks was performed. For all observations employing PKS\,0408$-$65 and J\,1939$-$6342, we used a custom component-based field model provided by the pipeline converted into model visibilities through the Meqtrees package \citep{noordam2010meqtrees}. We iteratively imaged the calibrated data in Stokes I, using multiscale and multifrequency cleaning of WSClean \citep{Offringa2014} and automaking enabled. The gridded visibility data were weighted using the Briggs method with a robust value of 0.0 resulting in an angular resolution of 4$^{\prime \prime}$ -- 10$^{\prime \prime}$. This is followed by a round of phase and delay self calibration using the CubiCal\footnote{https://github.com/ratt-ru/CubiCal} package \citep{kenyon2018cubical}. The 3C\,105 and 3C\,445 images showed strong calibration artefacts, and direction-dependent self-calibration was successfully applied to peel off-axis sources using the QuartiCal\footnote{https://github.com/ratt-ru/QuartiCal} package \citep{kenyon2023quartical}. The 4C\,$-$03.43 and CGCG\,047$-$067  images did not require any direction dependent calibration. The final images were then primary beam corrected by dividing them by the MeerKAT primary beam’s frequency-dependent
model.

\subsection{uGMRT}
We followed up eight sources from the sample, comprising three FR~I (3C\,198, 4C\,$-$03.43 and 3C\,403.1), three FR II (3C\,105, 3C\,227 and 3C\,445), and two tailed (CGCG\,047$-$067 and NGC\,7503) radio galaxies, {using the uGMRT} \citep{gmrt,ugmrt}. 
The observations, carried out in Band-4 (550 -- 850 MHz), were performed in two different observing cycles  (see Table \ref{tabsum}). All the uGMRT data were recorded in spectral line mode, with 8192 channels in the dual-polarisation (RR and LL).

We calibrated the data using the CASA Pipeline‑cum‑Toolkit for uGMRT data reduction \citep[CAPTURE,][]{kale2021capture}. The algorithm uses CASA tasks for this purpose. The absolute flux density scale was established according to \cite{perley2017accurate}. The data corrupted by persistent RFI were flagged along the channels using automated techniques in CASA. 
Following the flagging procedure, after bad data were flagged, we followed the standard calibration procedure in CAPTURE-CASA6 using default parameters. 
The calibrated data were imaged using TCLEAN, with 2 Taylor coefficients (nterms = 2), Briggs weighting (robust = 0.0), multi-term-multifrequency synthesis (MT-MFS) deconvolution algorithm implemented by the multiscale clean CAPTURE default parameters. A region of $\sim$2.3$^\circ$ $\times$ 2.3$^\circ$ was imaged for each source. A round of phase and amplitude self-calibration was applied. Consequently, a round of direction dependent calibration was performed, peeling the off-axis sources in all images with Quartical and later  imaged with WSClean using multiscale and multifrequency cleaning. Primary beam attenuations are corrected using the CASA task wbpbgmrt\footnote{https://github.com/ruta-k/uGMRTprimarybeam} for uGMRT corrections.

\begin{table*}
  \caption{uGMRT and MeerKAT observation overview.}
  \centering
\label{tabsum}
  \begin{tabular}{@{}|lccccccccc@{}}
    \toprule
Object name  & Proposal code	 & 	Obs$\_$Date 	& 	$\nu$   &    $\Delta \nu$   & Ch.-width    &  t$_{\rm int}$   &      FWHM, PA    &       RMS\\

   	                                 &                &                &          (MHz)  &     (MHz)           &      (kHz)      &  ($\sim$hr) &    $^{\prime\prime}\times ^{\prime\prime}, ^{\circ}$&       mJy~beam$^{-1}$\\      

\hline
%-------------------------------------------------------------------------------------------------------------------------------------------------------------------------------------------

uGMRT\\
3C\,105 & 36\_032 & 06 Sep 2020 & 700 & 300 & 48.82 & 3.86 & 7.67$\times$2.84, $-$70 & ~0.069 \\
3C\,198 & 43\_058 & 18 Nov 2022 & 700 & 300 & 48.82 & 5.54 & 4.39$\times$3.46, 61.62 & ~0.026 \\
3C\,227 & 43\_058 & 18 Nov 2022 & 700 & 300 & 48.82 & 5.55 & 3.88$\times$2.57, 64.05 & ~0.116 \\
4C\,$-$03.43 & 36\_032 & 25 May 2019 & 700 & 300 & 48.82 & 3.16 & 4.96$\times$4.12, 71.3 & ~0.034 \\
CGCG\,047$-$067 & 43\_058 & 08 Feb 2023 & 700 & 300 & 48.82 & 4.17 & 3.90$\times$3.23, 45 & ~0.018 \\
3C\,403.1 & 43\_058 & 10 Mar 2023 & 700 & 300 & 48.82 & 3.98 & 3.99$\times$3.16, 50 & ~0.021 \\
3C\,445 & 43\_058 & 13 Dec 2022 & 700 & 300 & 48.82 & 6.21 & 4.42$\times$3.45, 49.34 & 0.081 \\
NGC\,7503 & 43\_058 & 13 Dec 2022 & 700 & 300 & 48.82 & 6.60 & 4.43$\times$3.54, 47 & ~0.021 \\
\\
MeerKAT\\
3C\,105 & MKT-20137 & 4 July 2021 & 1283 & 856 & 208.98 & 5.0 & 7.62$\times$6.03, 0.89 & ~0.110 \\
4C\,$-$03.43 & MKT-20137 & 23 May 2021 & 1283 & 856 & 208.98 & 1.89 & 8.64$\times$7.12, 145 & 0.023 \\
 &  & 21 Aug 2021 & 1283 & 856 & 208.98 & 6.94 & - & - \\
CGCG\,047$-$067 & MKT-20137 & 31 July 2021 & 1283 & 856 & 208.98 & 5.04 & 10.03$\times$6.93, 164.13 & ~0.010 \\
3C\,445 & MKT-20137 & 29 May 2021 & 1283 & 856 & 208.98 & 4.9 & 8.30$\times$6.69, 151.7 & 0.020 \\
SDSS\, J0917$+$1331 & MKT-20137 & 25 June 2021 & 1283 & 856 & 208.98 & 5.20 & 10.75$\times$6.09, 159.05 & 0.006 \\
SDSS\, J\,1120$+$0407 & MKT-20137 & 23 May 2021 & 1283 & 856 & 208.98 & 1.89 & 9.524$\times$5.52, $-$19.55 & 0.005 \\
 &  & 21 Aug 2021 & 1283 & 856 & 208.98 & 6.53 & - & - \\

\hline
  \end{tabular}
 % \caption{Sample table}
\end{table*}
%\end{flushleft}

When calculating the flux density S$_\nu$, we adopted an empirical 3\%  and 5\% calibration error $\xi_{cal}$ for the MeerKAT and the uGMRT  absolute flux density, {respectively \citep{hugo2021reference, chandra2004late}}. For an extended source, the flux density error is given by  
\begin{equation}
\Delta S= \sqrt{\left(\sigma_{\rm{rms}} \sqrt{N_{\rm{beam}}}\right)^2 + \left(\xi_{\rm{cal}} S_\nu \right)^2}
\end{equation}
where  $\sigma$ is the local rms noise extracted from the source free region of the image and $N_{\rm{beam}}$ is the number of beams covering the source extent. 

\section{Results}
\label{results}
\begin{figure*}
    \centering
    \begin{tabular}{cc}

    % Create a tabular layout for side-by-side images
    \begin{tikzpicture}
        % Main image
       \hspace*{-0.5cm} \node[anchor=south west, inner sep=0] (mainimage) at (0,0) {
            \includegraphics[width=0.49\linewidth]{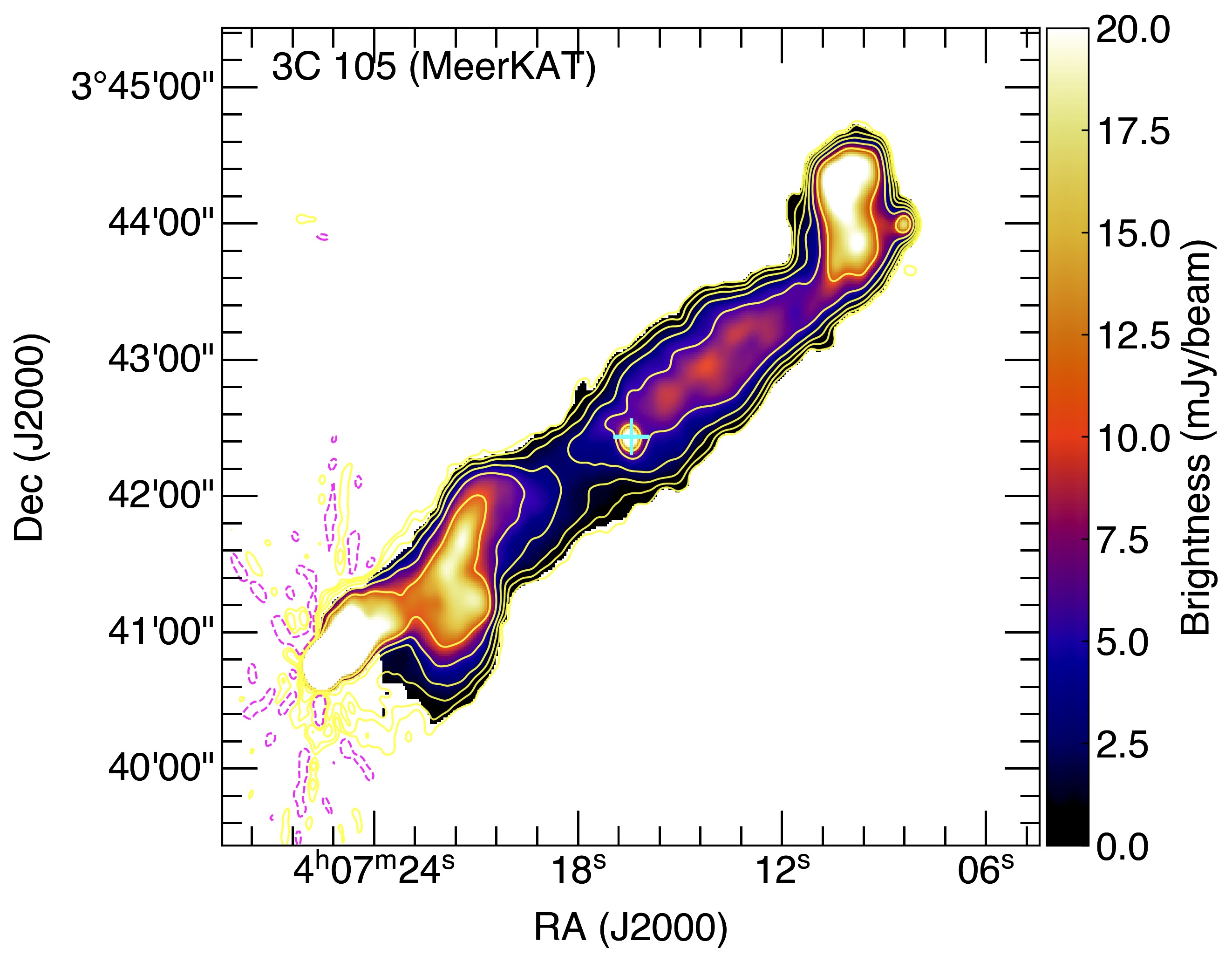}
            
        };
        % Inset at bottom right
        \begin{scope}[x={(mainimage.south east)}, y={(mainimage.north west)}]
            \node[anchor=south west] at (0.57,0.15) { % slightly inset from the corner
                \includegraphics[width=0.1\textwidth]{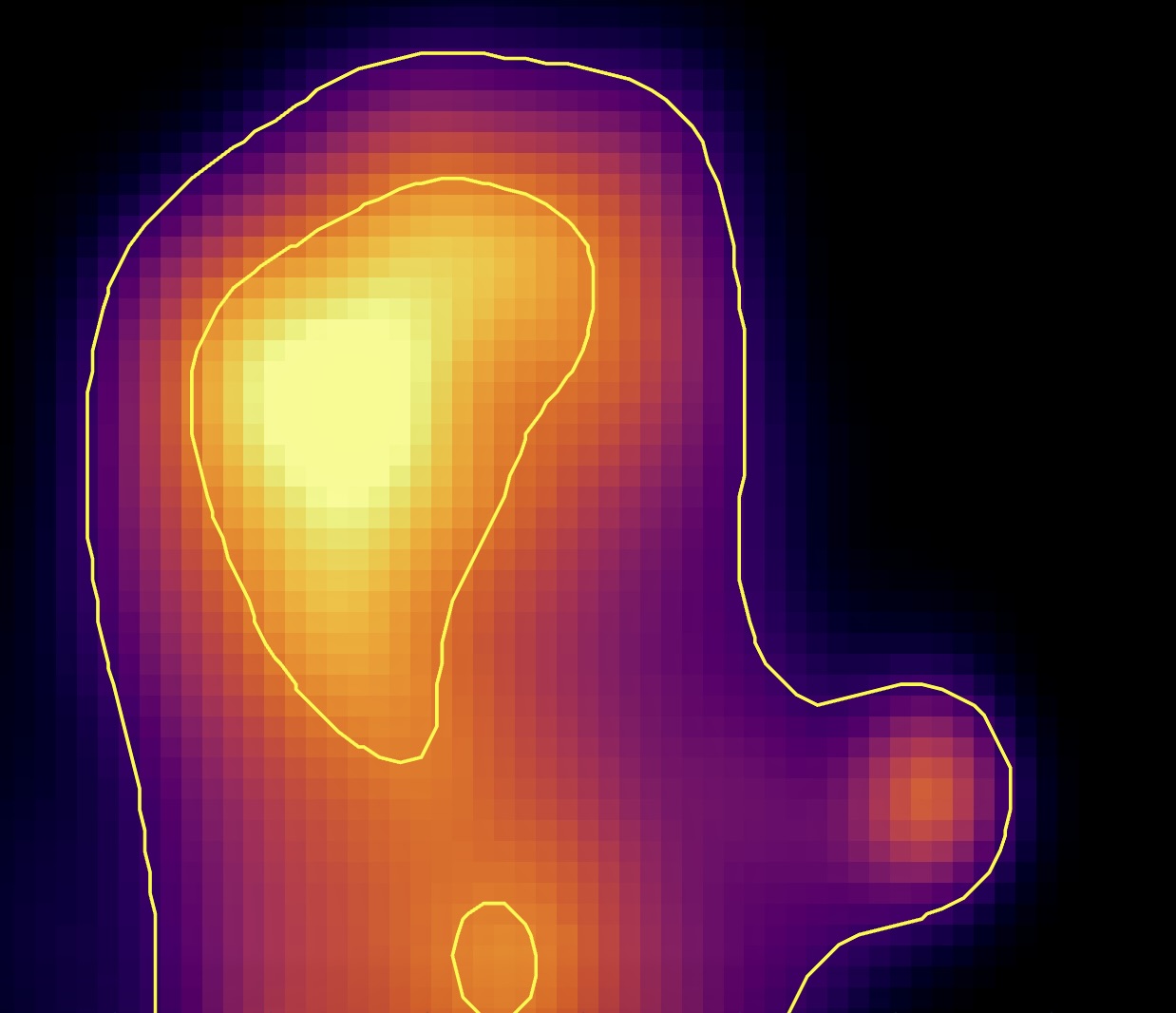}
            };
        \end{scope}

    \end{tikzpicture}&
    %%%%%%%%%%%%%%%%%%%%%%%%%%%
    
 \hspace*{-1cm}\includegraphics[width=0.49\linewidth]{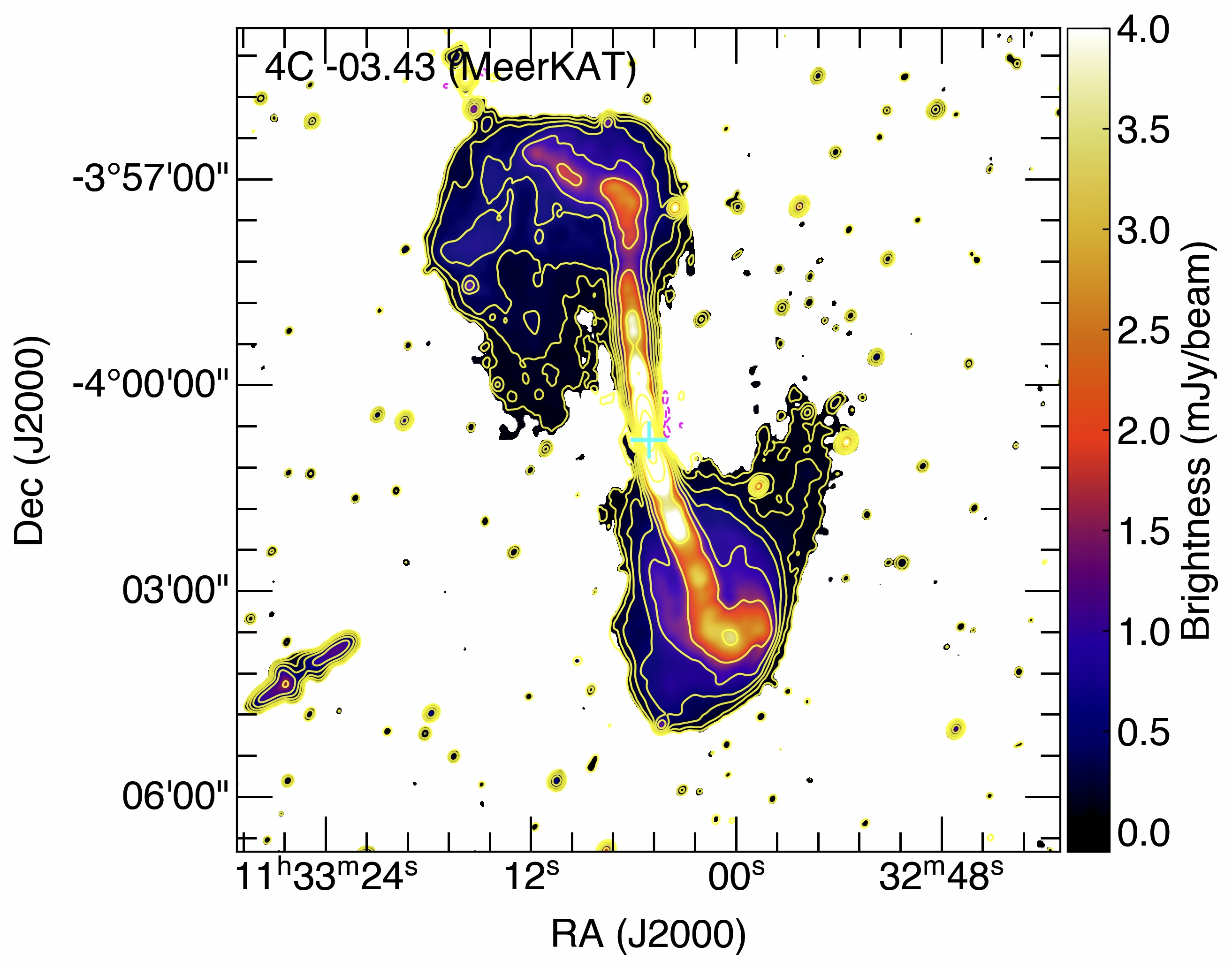}\\%
    \hspace*{-1.4cm}\includegraphics[width=0.49\linewidth]{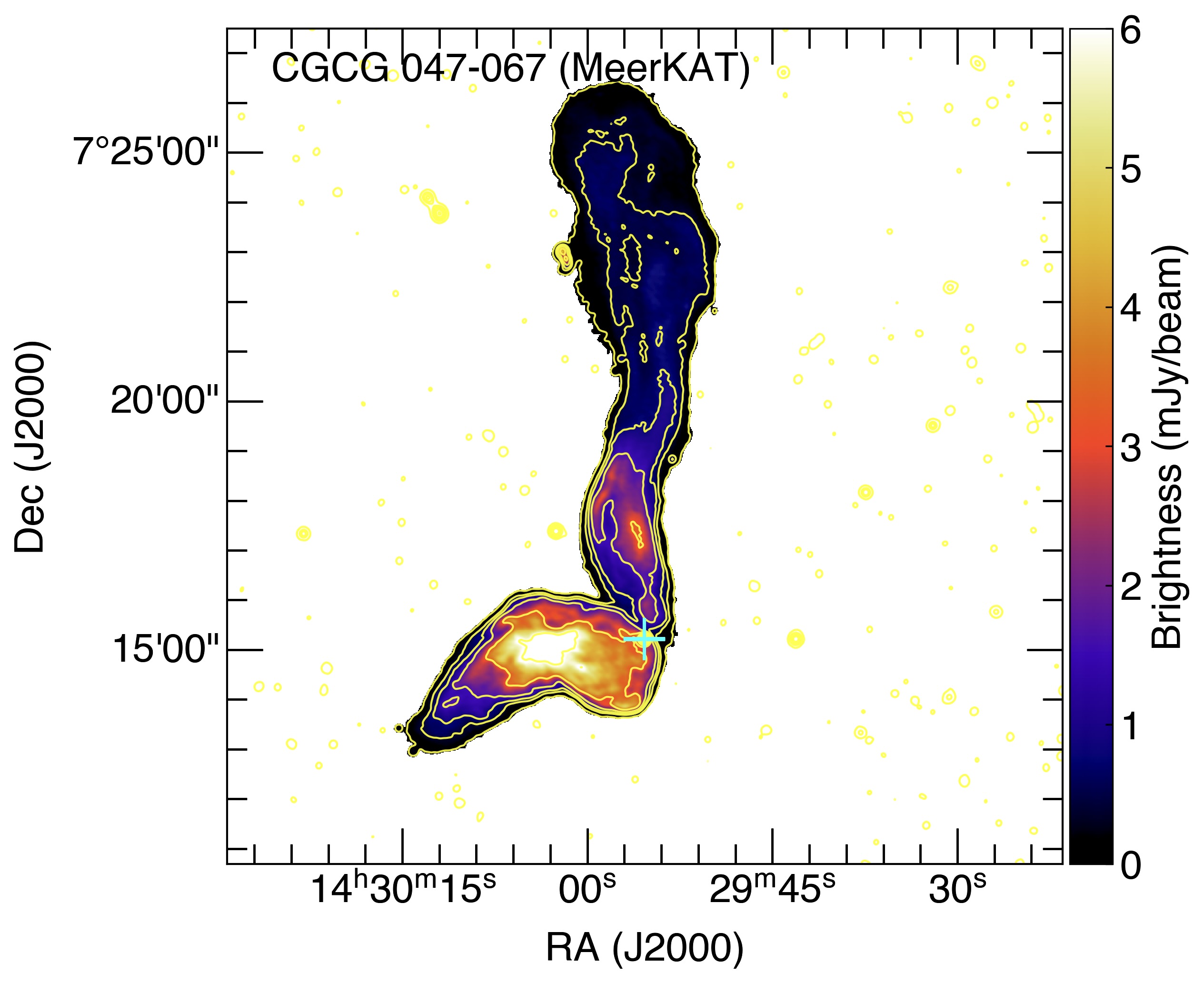}&%
    \begin{tikzpicture}
        % Main image
        \hspace*{-0.5cm} \node[anchor=south west, inner sep=0] (mainimage) at (0,0) {
          \includegraphics[width=0.49\linewidth]{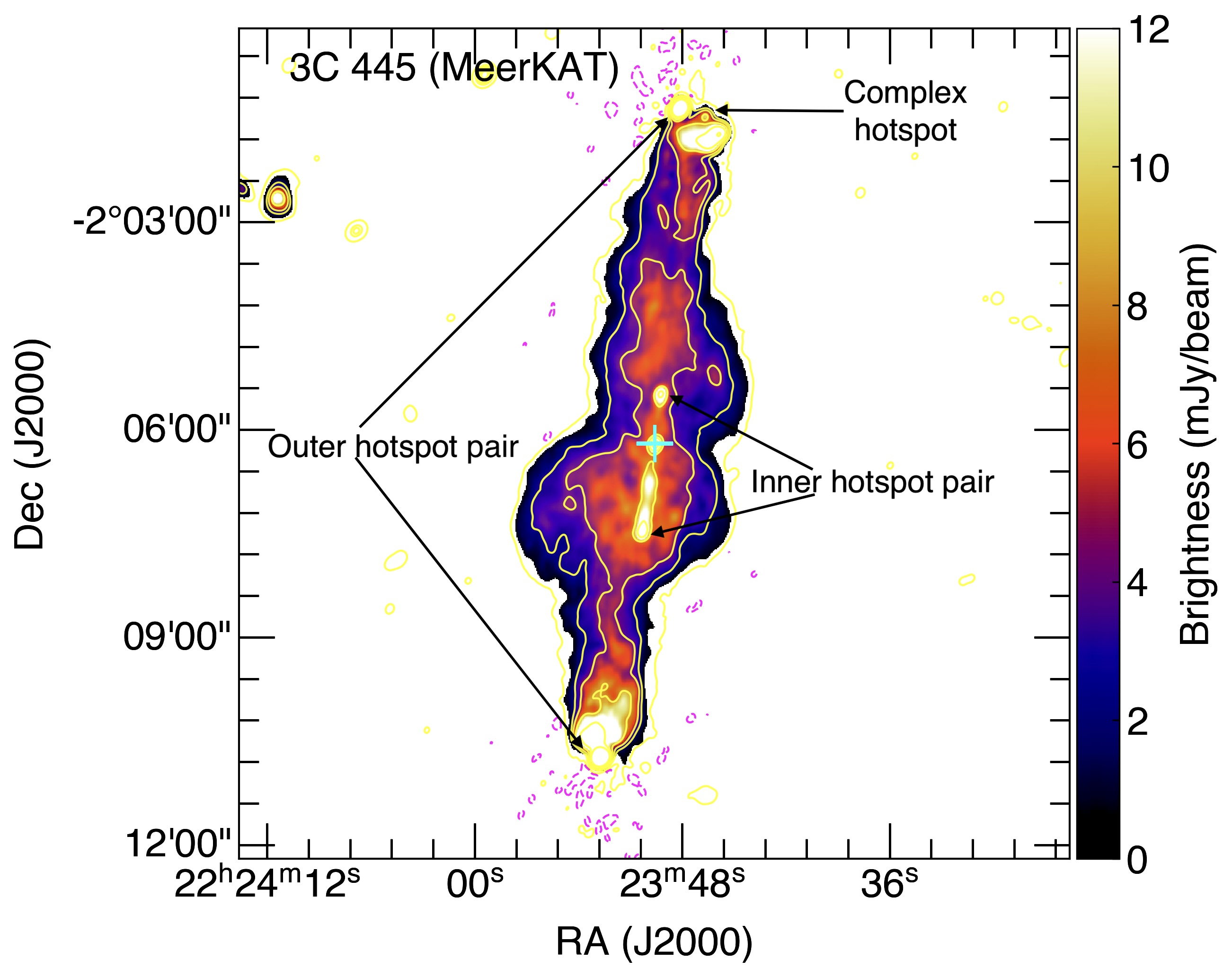}
        };
        % Inset at bottom right
        \begin{scope}[x={(mainimage.south east)}, y={(mainimage.north west)}]
            \node[anchor=south west] at (0.62,0.15) { % slightly inset from the corner
              \includegraphics[width=0.1\textwidth]{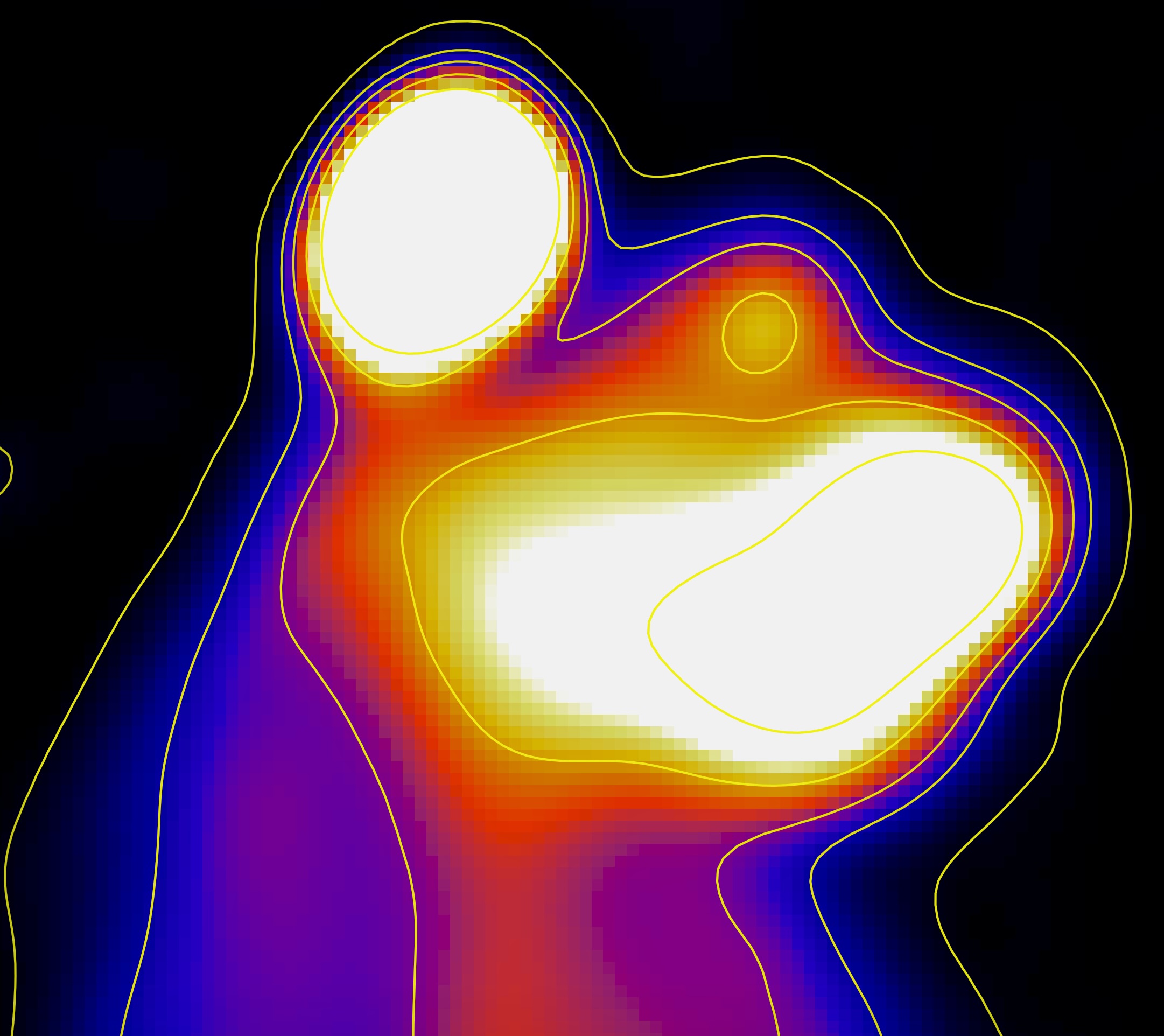}
            };
        \end{scope}

    \end{tikzpicture}\\
\end{tabular}
    \caption{Images of MeerKAT observations of the sources: 3C\,105 (top left panel), inserted north-west hotspot region; 4C\,$-$03.43 (top right panel); CGCG\,046$-$067 (bottom left panel); 3C\,445 (bottom right panel) with an insert of the northern hotspot. The lowest radio contour represents three times the total RMS noise (see Table \ref{tabsum}), with subsequent contours increasing by a factor of two. The negative contours are presented in magenta at $-3\, \times$RMS. The cyan cross represents the position of the optical host. }
    \label{INtesity1}
\end{figure*}

\begin{figure*}
    \includegraphics[width=0.499\textwidth]{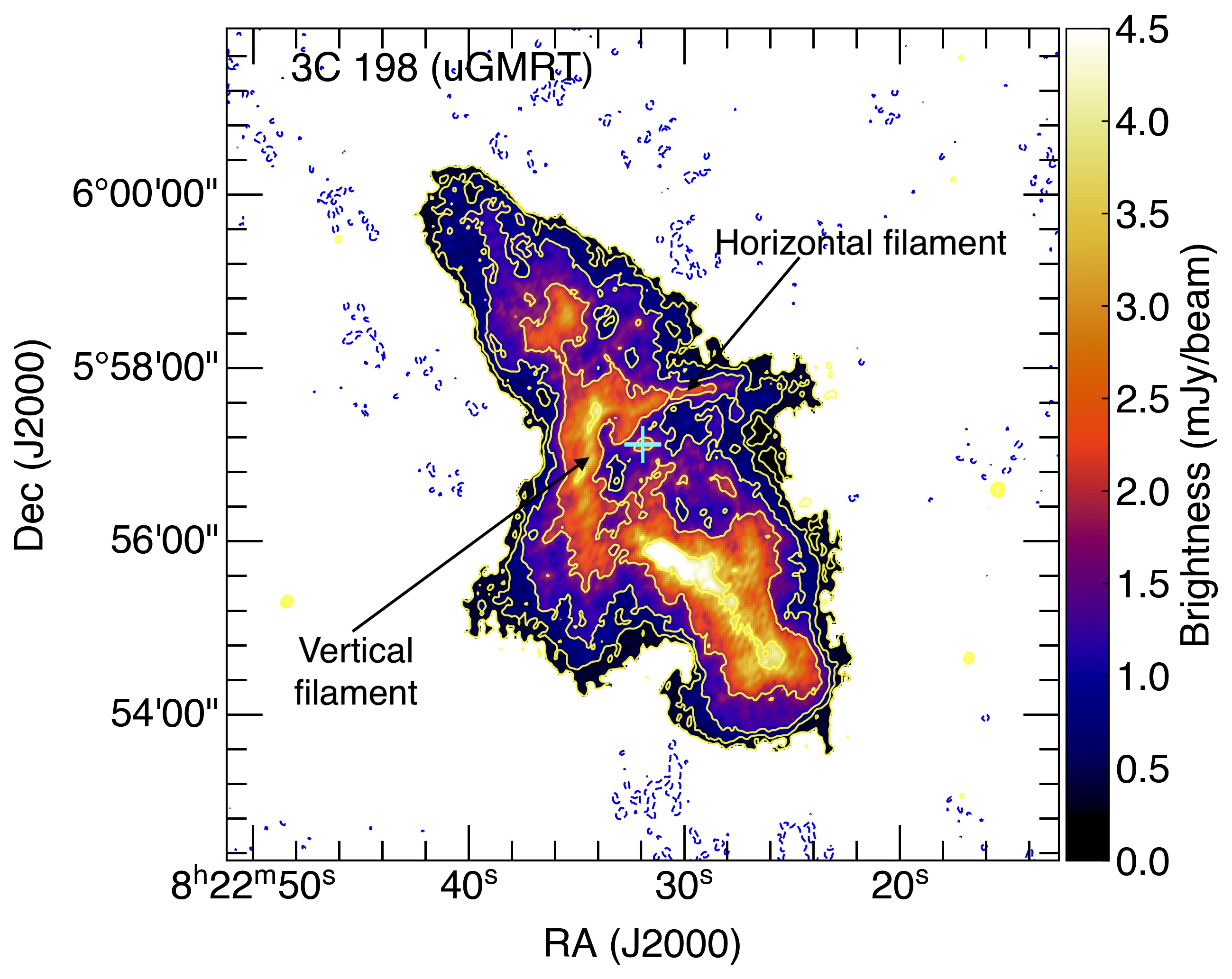}\hfill%
        \begin{tikzpicture}
        % Main image
        \hspace*{-0.2cm}\node[anchor=south west, inner sep=0] (mainimage) at (0,0) {
            \includegraphics[width=0.488\linewidth]{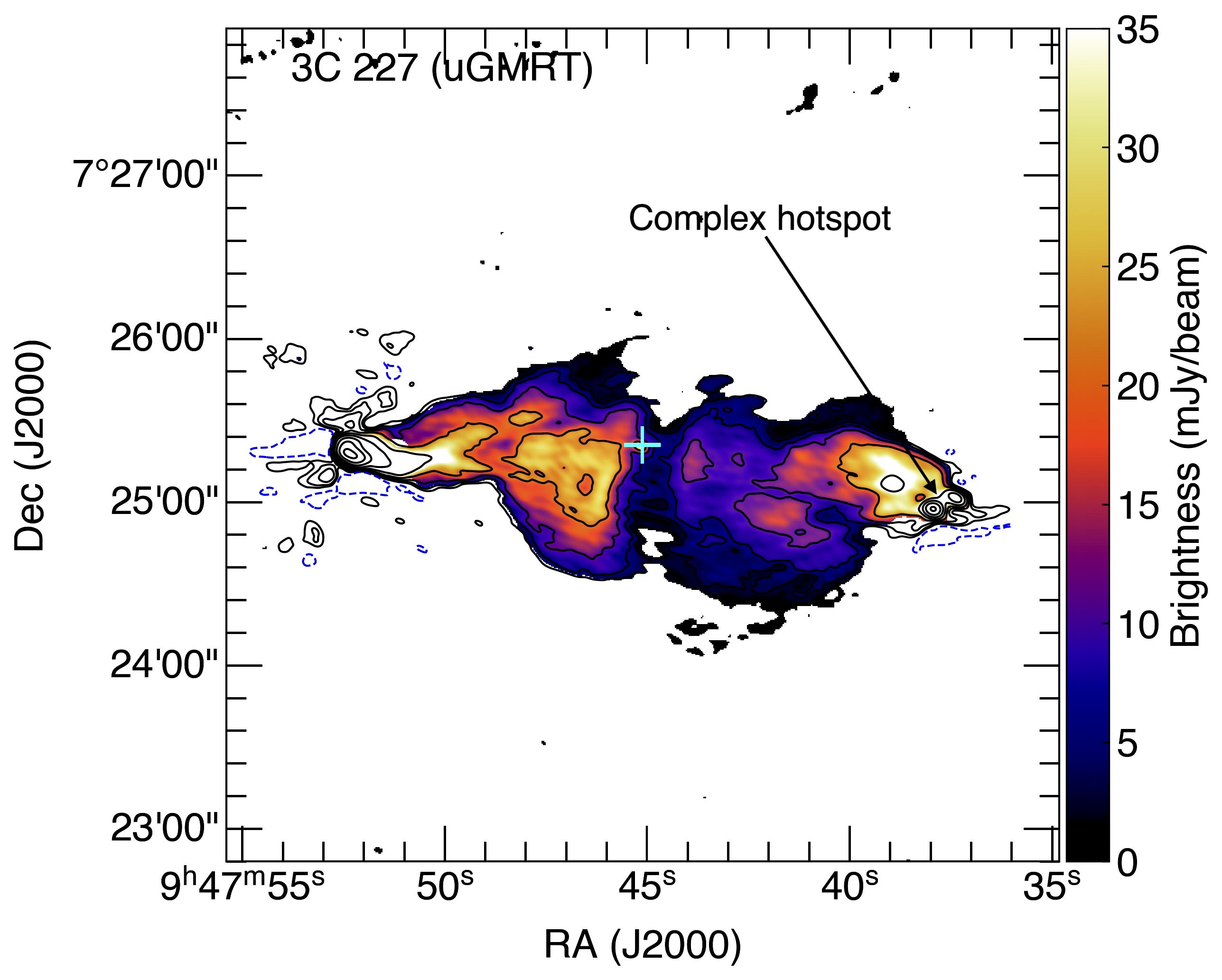}
        };
        % Inset at bottom right
        \begin{scope}[x={(mainimage.south east)}, y={(mainimage.north west)}]
            \node[anchor=south west] at (0.65,0.13) { % slightly inset from the corner
                \includegraphics[width=0.1\textwidth]{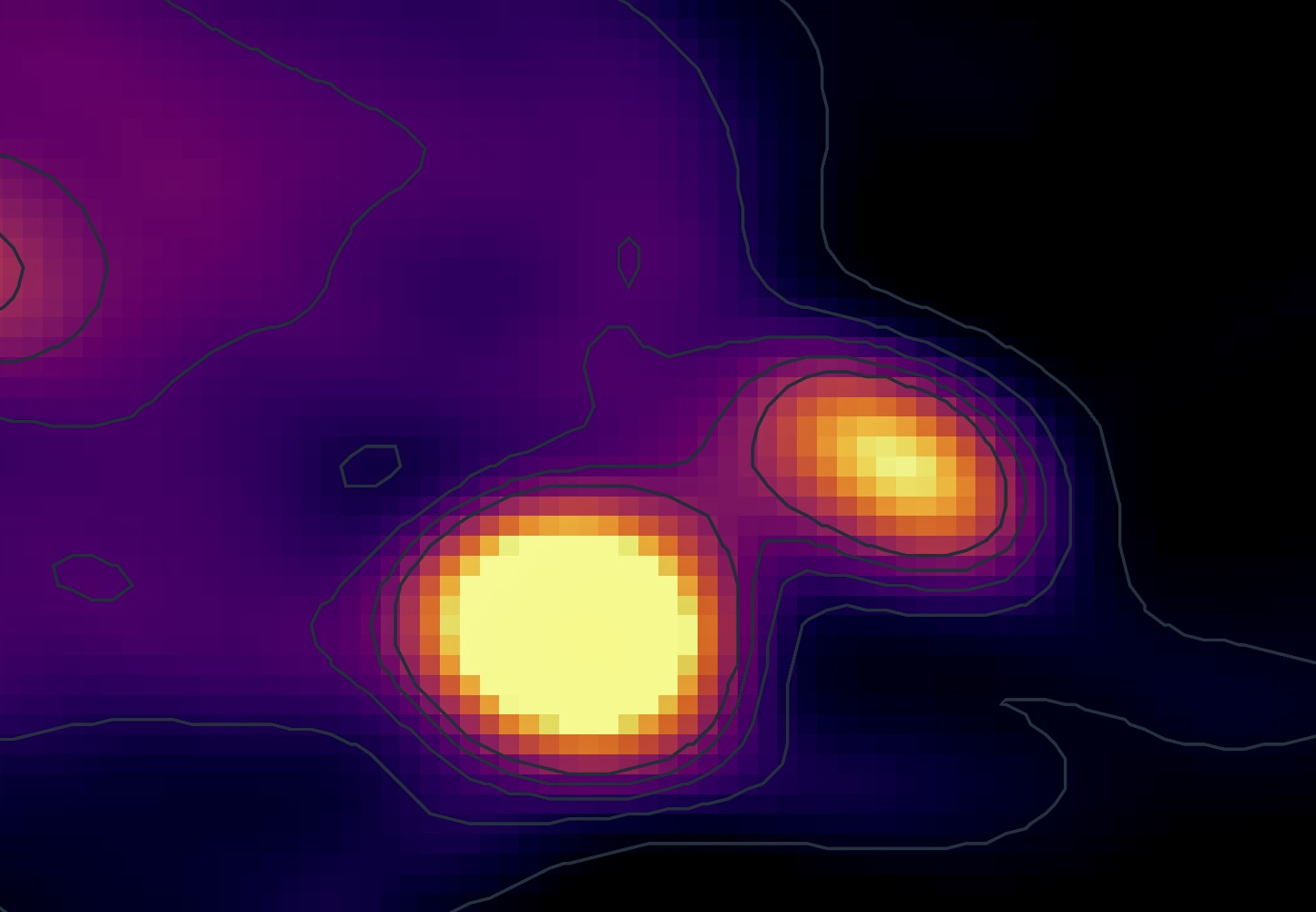}
            };
        \end{scope}\hfill    
    \end{tikzpicture} 
    \smallskip
    \includegraphics[width=0.5\textwidth]{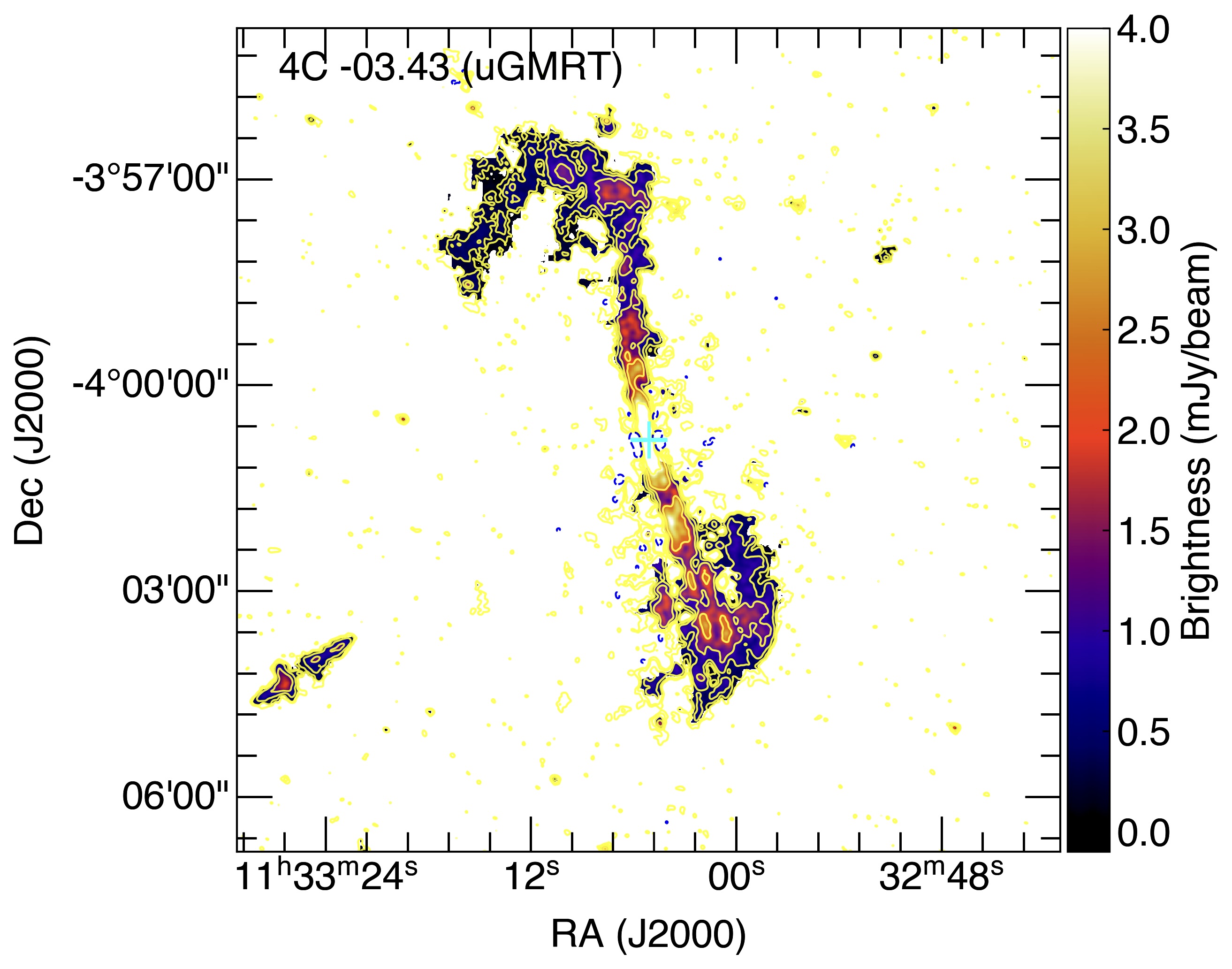}\hfill%
    \includegraphics[width=0.5\linewidth]{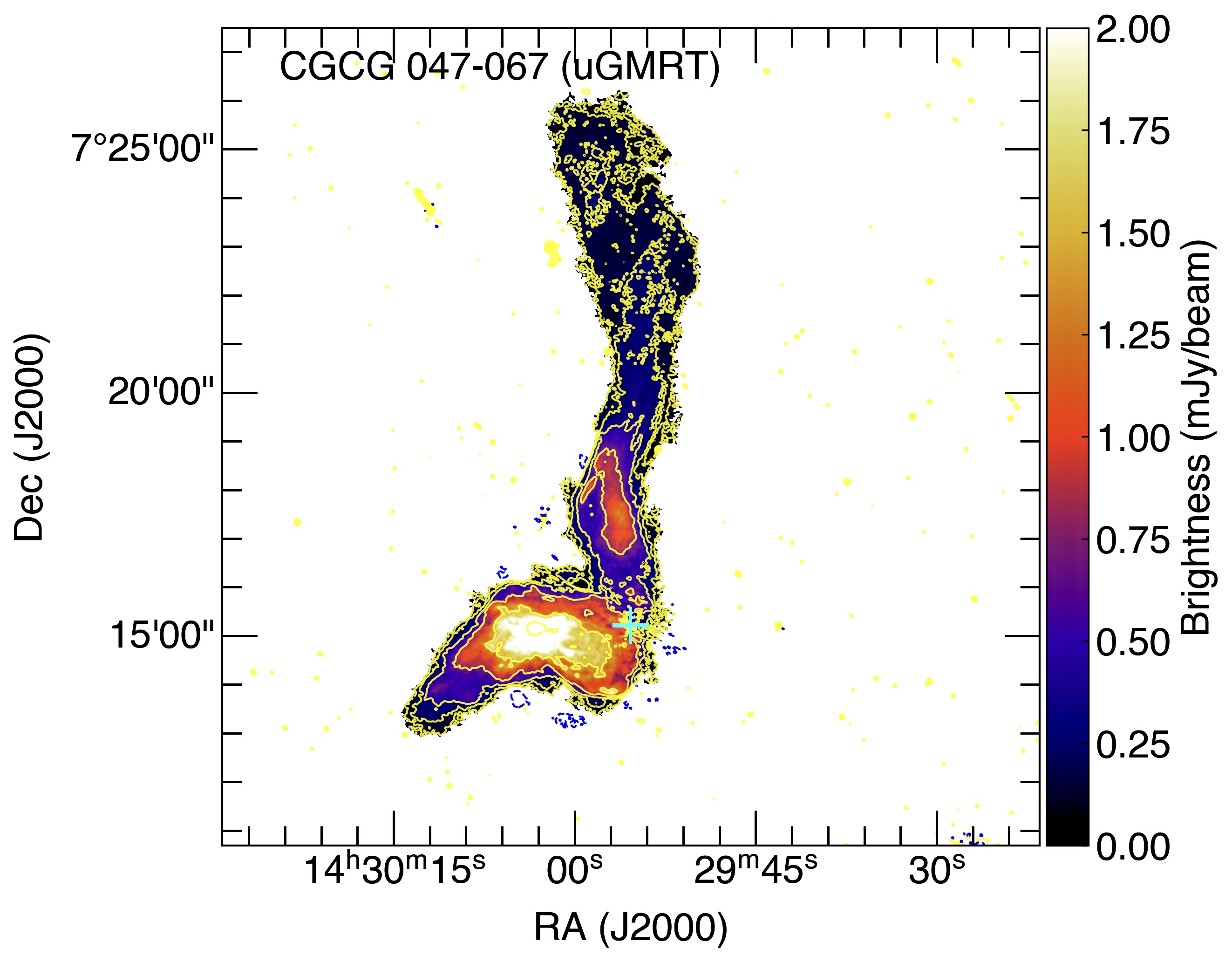}\hfill%
    \smallskip
   \hspace*{-0.5cm} \includegraphics[width=0.489\textwidth]{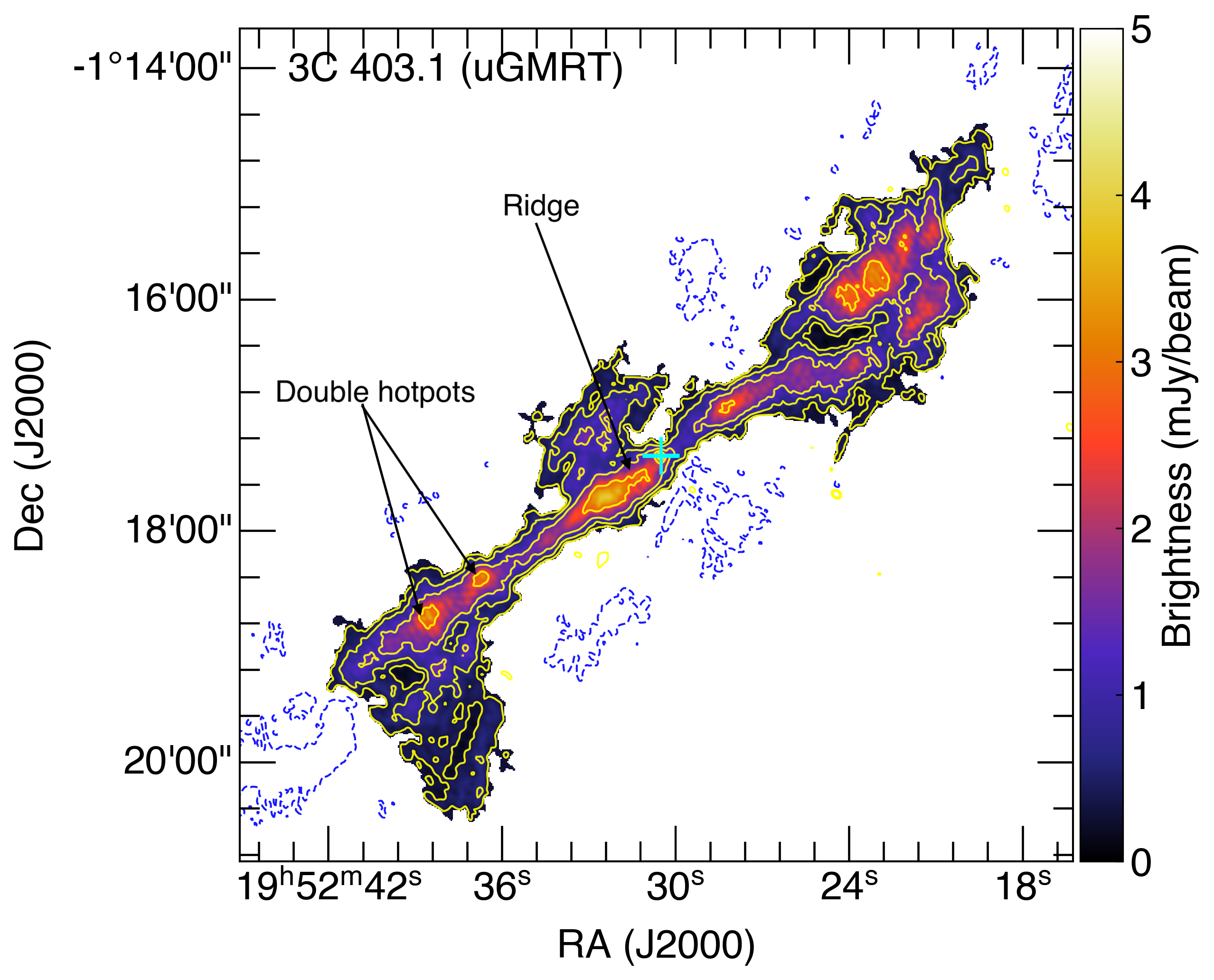}
    \includegraphics[width=0.48\textwidth]{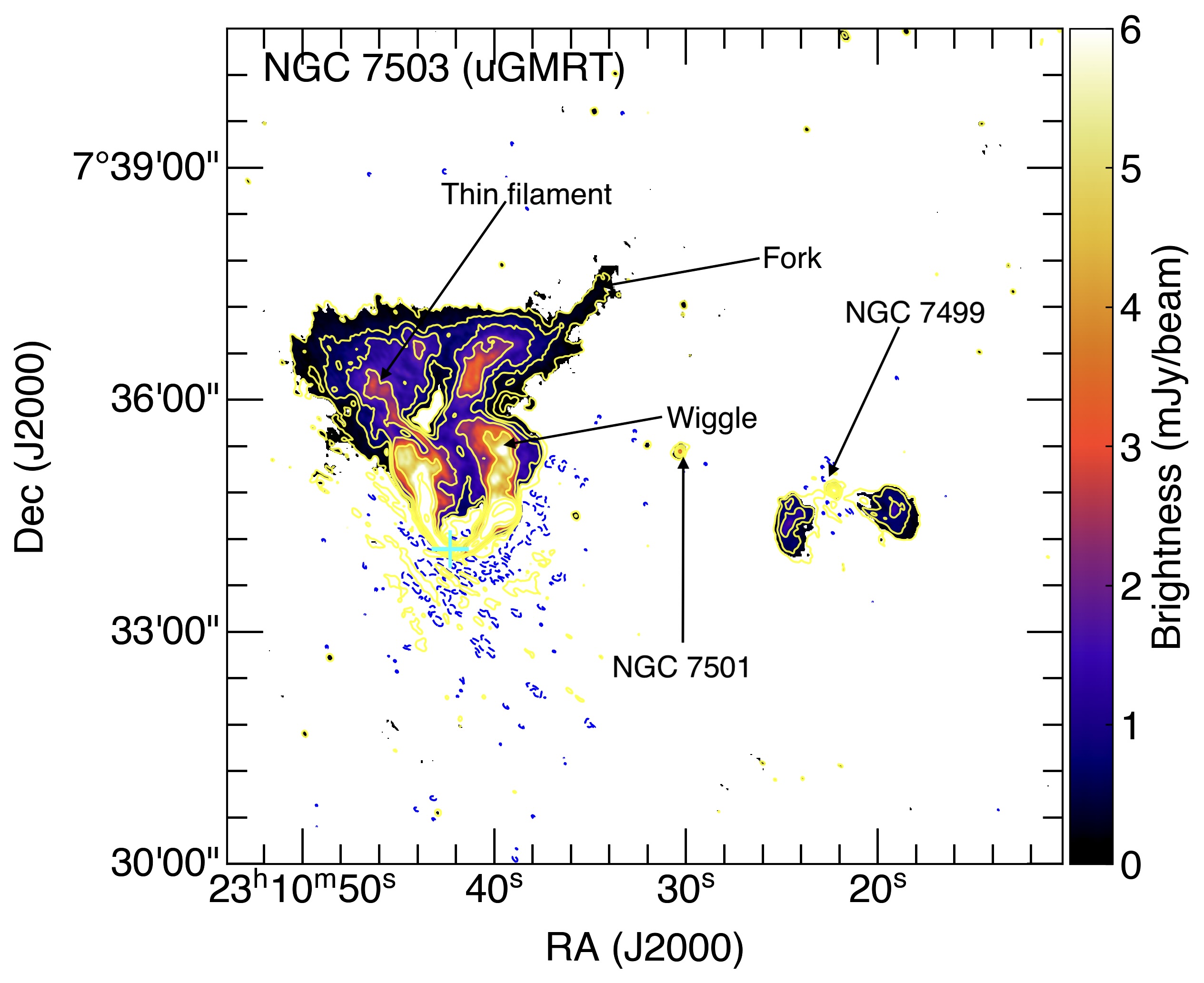}%
    \caption{Images of uGMRT Band-4 observations of the sources: 3C\,198 (top left panel); 3C\,227 (top right panel) with insert of the complex hotspot region; 4C\,$-$03.43 (middle left panel);  CGCG\,046$-$067 (middle right panel); 3C\,403.1 (bottom left panel) and NGC\,7503 (bottom right panel). The lowest radio contour represents three times the total RMS noise (see Table \ref{tabsum}), with subsequent contours increasing by a factor of two. The negative contours are shown in blue at $-3\, \times$RMS. The cyan cross represents the position of the optical host.}
    \label{INtensity2}
\end{figure*}

\begin{figure*}
\begin{center}
\begin{tabular}{ccc}
   \hspace*{-0.5cm}\includegraphics[width=0.95\textwidth]{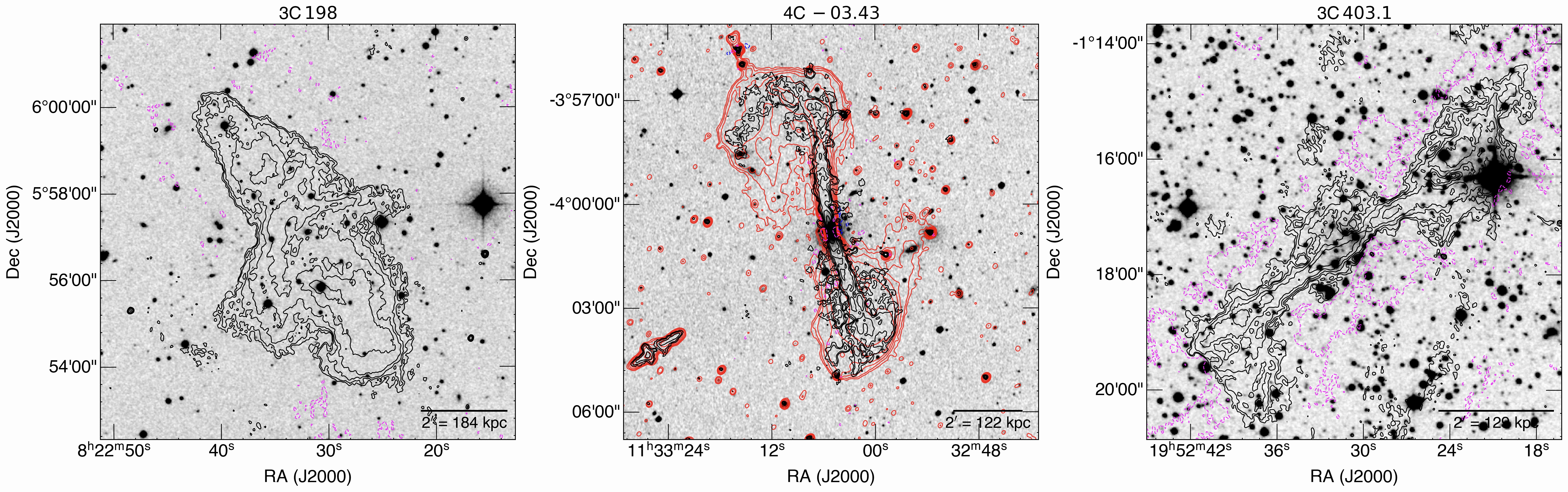}\\ 
    \hspace*{-0.5cm}\includegraphics[width=0.95\textwidth]{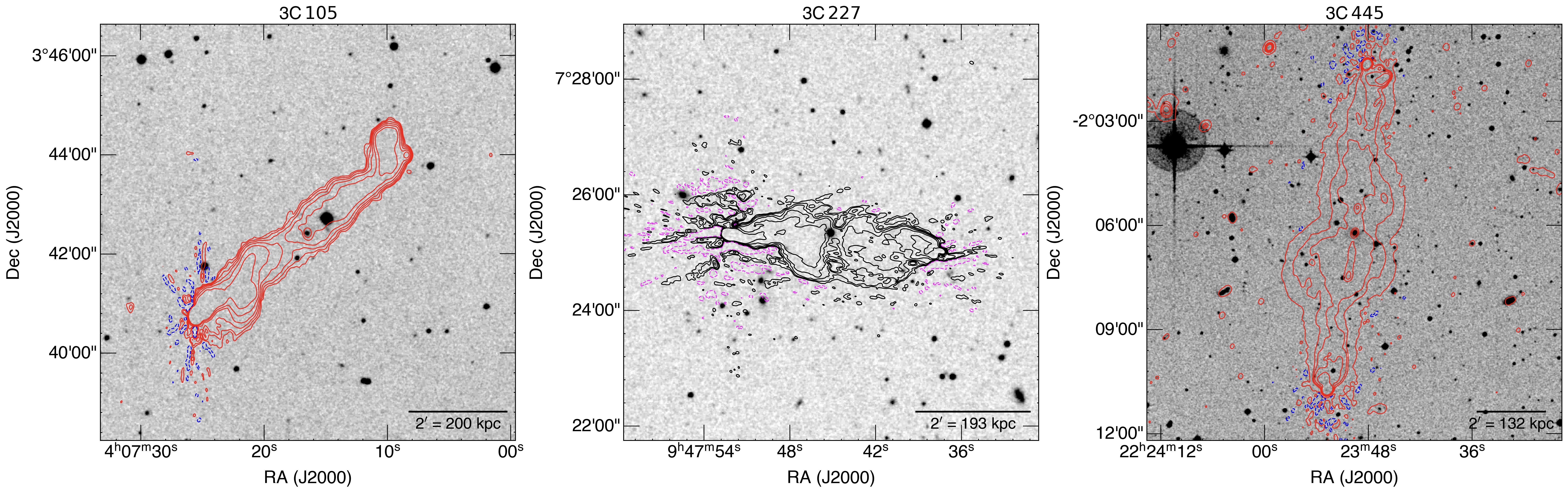}\\
    \hspace*{-0.5cm}\includegraphics[width=0.95\textwidth]{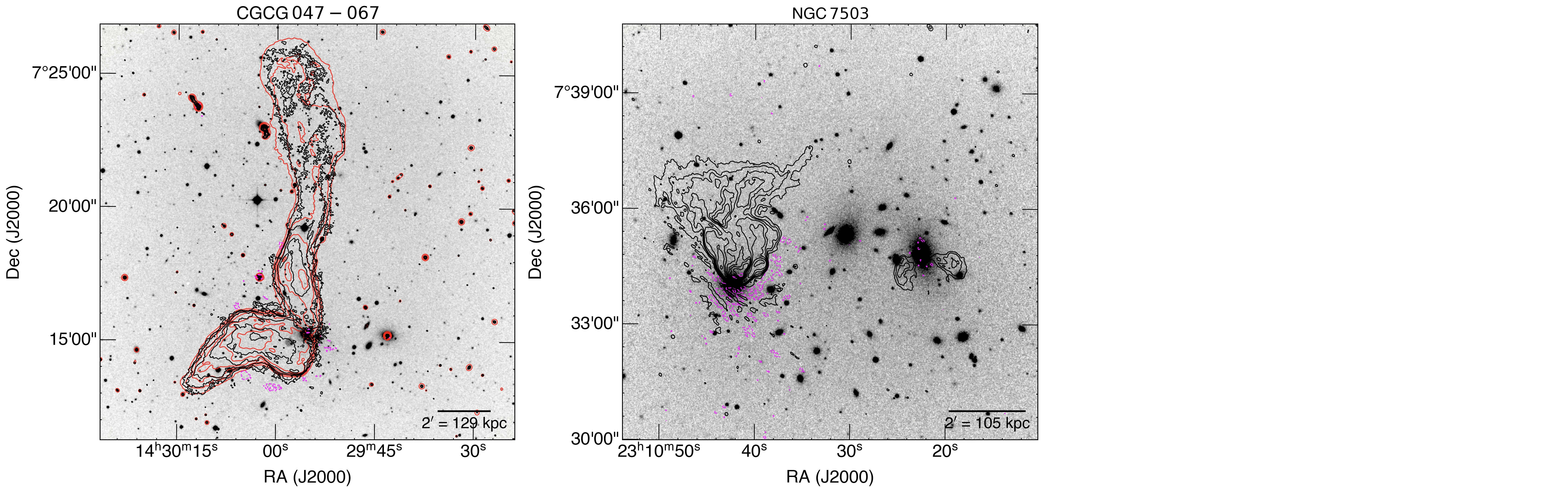}\\
    \hspace*{-0.5cm}\includegraphics[width=0.95\textwidth]{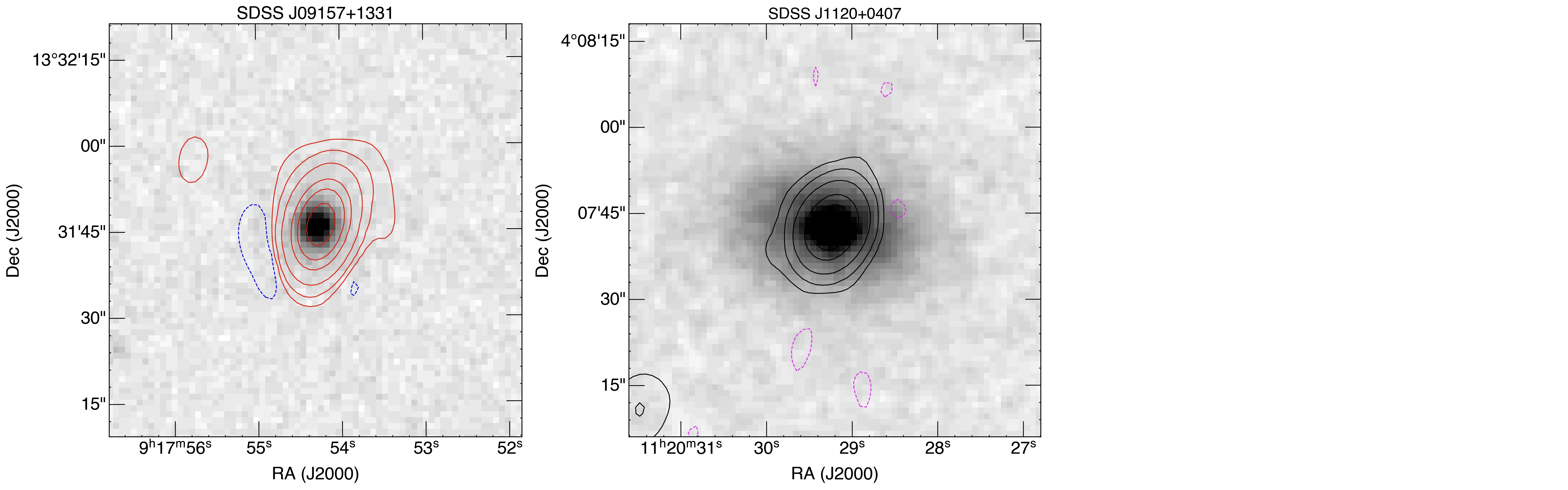}\\

\end{tabular}    
\end{center}
    \caption{The MeerKAT and uGMRT Band-4 radio surface brightness contours of eight sources are shown: 3 FR\,I (first row), 3 FR\,II (second row), 2 FR\,I tailed radio galaxies (third row), and 2 FR\,0 (fourth row). From left to right and top to bottom, the sources are 3C\,198, 4C\,$-$03.43, 3C\,403.1, 3C\,105, 3C\,227, 3C\,445, CGCG\,047$-$067, NGC\,7503, SDSS J0917$+$1331, and SDSS J1120$+$0407. Black (uGMRT) and red (MeerKAT) radio surface brightness contours are overlaid on DSS-II red-band optical images (grayscale). The first negative contours are shown in magenta (uGMRT) and blue (MeerKAT)  at $-3\, \times$RMS. Contour levels drawn from $3\times$RMS and increase by factors of 2.}
    \label{overlay1}
\end{figure*}

{The total intensity images for the sources presented here are reported in Figure \ref{INtesity1} and Figure \ref{INtensity2}, where the radio galaxies observed with MeerKAT and uGMRT, respectively, are shown. The parameters of the images are given in Table \ref{tabsum}. Flux densities and radio power for each radio galaxy are given in Table \ref{tab:radio_properties}. Figure \ref{overlay1} shows the radio emission of each target overlaid on the optical ESO--DSS2 frame.}

{In the following sections, we provide a morphological description of each radio galaxy together with its total source size, defined as the projected largest linear extent of the radio emission. The sizes were measured at the 9$\sigma$ contour level to avoid spurious features.}

\subsection{\texorpdfstring{3C\,105}{3C 105}}
The MeerKAT image is shown in the top left panel of Figure~\ref{INtesity1}. Due to strong artefacts, the uGMRT image could not be satisfactorily corrected and is therefore not used. The total source size is $\sim$485 kpc, consistent with previous measurements \citep{leahy,Fanaroff2021}.
A marked asymmetry is observed in the projected lengths of the two sides of the radio emission. The radio galaxy exhibits a fairly straight north-west jet, which is more extended than the south-eastern jet and shows several emission peaks aligned along the jet ridge, before terminating at the hotspot. The south-eastern hotspot is compact, with two parallel ridges of emission extending from it toward the core, which \citet{leahy} described as a `hammerhead' structure. The radio emission of 3C\,105 is discussed further in Sect. \ref{spec}.

\subsection{\texorpdfstring{3C\,198}{3C 198}}

The total intensity image is shown in Figure~\ref{INtensity2} (top left panel) and Figure~\ref{overlay1} (top right panel). {The source has a largest linear size of $\sim$650 kpc, qualifying it as a large (but not giant) radio galaxy}. Compared to lower-resolution images in the literature, our high-resolution, high-sensitivity data reveal new features, i.e., no bright hotspots are detected, and the southern lobe is brighter and shows some substructure.
In the central region, coincident with the optical counterpart, the radio emission extends perpendicular to the lobe axis and is extremely filamentary in nature.

\subsection{\texorpdfstring{3C\,227}{3C 227}}
Only uGMRT imaging is available for this FR\,II radio galaxy. The radio emission is aligned along the east–west direction, with a total projected size of $\sim$505 kpc. Our image is shown in the upper right panel of Figure~\ref{INtensity2} and the bottom left panel of Figure~\ref{overlay1}.
The angular resolution of the uGMRT image confirms the western hotspot structure previously reported in the literature \citep{miglio}. The two parallel ridges detected at 22~GHz in the eastern hotspot by \citet{2020orienti} are not distinctly resolved in our image. A bifurcation is nevertheless visible at the extremity of the western hotspot region, revealing a trail of emission extending from the hotspot toward the core. Interestingly, this bifurcation is similar to that observed in 3C\,105, and a comparable feature is also seen in the southern lobe of 3C\,198.
The diffuse backflow emission from the lobes broadens as it propagates away from the hotspots. The two lobes are asymmetric in flux density, with the eastern lobe being brighter by a factor of $\sim$1.6 and characterized by filamentary emission.

\subsection{\texorpdfstring{4C\,$-$03.43}{4C --03.43}}

Both MeerKAT and uGMRT images are available for this radio galaxy and are shown in Figure~\ref{INtesity1} (upper right panel) and Figure~\ref{INtensity2} (middle left panel), respectively. The higher sensitivity of the MeerKAT image reveals radio jets extending into low-surface-brightness lobes, whereas the uGMRT image mainly traces the jets and the brightest ridge of the lobes.
The jets are not collinear, but exhibit a mild curvature, bending in opposite directions near their termination points. The lobes follow this bending, extending in opposite directions and giving rise to an overall S-shaped morphology. Such a structure is likely the result of projection effects combined with interactions between the lobes and the intracluster medium. The projected linear size of the source is $\sim$408 kpc.

\subsection{\texorpdfstring{CGCG\,047$-$067}{CGCG 047--067}}
The MeerKAT and uGMRT images are shown in Figure~\ref{INtesity1} (bottom left panel), { Figure~\ref{INtensity2} (middle right panel), and Figure~\ref{overlay1} (bottom left panel)}. The total projected size of CGCG\,047$-$067, defined as the sum of the northern and southern lobe extents, is $\sim$950 kpc, qualifying it as a giant radio galaxy.
The radio morphology is consistent with a wide-angle-tailed radio galaxy. A prominent jet extends northward to a projected length of $\sim$725 kpc, while the opposing jet bends toward the southwest, giving the source an overall L-shaped appearance. The south-eastern lobe contains a very bright emission region with no associated optical counterpart, which we therefore interpret as intrinsic to the radio lobe. A plausible explanation for this enhanced brightness is a line-of-sight effect, where the lobe is viewed edge-on, and the emission is amplified by integration along the line of sight.

\subsection{\texorpdfstring{3C\,403.1}{3C 403.1}}
The uGMRT radio contours and the radio–optical overlay are shown in Figures~\ref{INtensity2} (bottom left panel) and~\ref{overlay1} (top right panel). The source extends over $\sim$390 kpc along the south-east and north-west direction. The north-western jet shows clear substructure, broadening at a projected distance of $\sim$132 kpc from the core and splitting into two roughly parallel ridges.
The south-eastern jet displays a distinctly blobby appearance, characterised by several peaks of emission along its length. Beyond the last peak, the emission diffuses southward, expanding roughly perpendicular to the jet axis and forming a lobe-like structure. This clumpy morphology may arise from recollimation during jet propagation or from episodic activity of the central engine; however, the lack of spectral index imaging prevents further investigation.
In addition, a diffuse ridge of emission is detected north of the inner region of the south-eastern jet (labelled in Figure~\ref{INtensity2}, bottom left panel). While its origin remains uncertain, it may be linked to nuclear activity during a particular phase of the source evolution. The feature is unlikely to be an imaging artefact, as similar structures have been reported in other radio galaxies, such as 3C\,465 \citep{giacintucci2007radio}.

\subsection{\texorpdfstring{3C\,445}{3C 445}}\label{double}

The source has a total size of $\sim$630 kpc. We report only the MeerKAT observations, as strong artefacts in the uGMRT data could not be satisfactorily corrected and are therefore not used. {The total-intensity images shown in Figure~\ref{INtesity1} (bottom right)} and Figure~\ref{overlay1} (second row, left) reveal a north–south alignment, with clear substructure in the northern hotspot. Beyond the most compact component, a more diffuse secondary compact feature is present to the north-west, as also noted by \citet{leahy}, who interpreted this feature in the framework of splatter splitting, in which a secondary shock forms when a supersonic, collimated jet is deflected by interaction with the ambient medium \citep{smith1984jet}. Similar asymmetric features are observed in the north-western hotspot of 3C\,105 and the western hotspot of 3C\,227. In addition, a third, fainter compact component is located between the first/initial and secondary hotspots. At the resolution of our images, the southern hotspot appears as a single compact feature. The backflow emission from both hotspots broadens toward the nuclear region. One of the most intriguing aspects of this radio galaxy is the presence of two inner hotspots aligned along the same major axis; both are resolved along this axis, suggesting restarted activity. If the source is interpreted as a double–double radio galaxy, the size of the restarted emission is $\sim$150 kpc. {Similar characteristics are observed in 3C\,219, where spectral ageing suggests a possible double-double radio structure, interpreted as rapid reorientation of the jet axis, likely due to a minor merger \citep{wolnick}. However, the classification of 3C\,445 as a double-double radio source remains tentative in this work and would require a spectral-ageing analysis of its individual components.}

\subsection{\texorpdfstring{NGC\,7503}{NGC 7503}}

The radio emission associated with NGC\,7503 is shown in the bottom right panel of Figure~\ref{INtensity2} and in the third row (right panel) of Figure~\ref{overlay1}. The source exhibits a narrow-angle-tailed morphology, with fairly collimated jets that break at some distance from the core and form tails extending northward. The eastern and western jets propagate for $\sim$55 and $\sim$78 kpc, respectively. Along the eastern jet, a brightness peak is followed by jet broadening, loss of collimation, and the formation of an extending tail towards north. At the angular resolution of our observations, both tails display fine structure, including wiggles and thin filaments. A particularly thin filament extends from the eastern jet into the corresponding lobe, while another extends from the western lobe into the ICM and shows a `fork' structure, a feature revealed by high-sensitivity, high-resolution imaging. The total flux density of the source is $S_{\rm 700\,MHz}=2.58\pm0.02$ Jy, corresponding to a radio power of P$_{\rm 700\,MHz}=1.17\times10^{25}$ W\,Hz$^{-1}$, consistent with the high optical luminosity of the host galaxy discussed in Appendix~\ref{ngc75}. The radio power of this narrow-angle-tailed source falls within the broad distribution of FR\,I radio galaxies in the radio-optical luminosity correlation \citep{owen1991surface}.

\subsection{\texorpdfstring{SDSS J\,0917$+$1331 and SDSS J\,1120$+$0407}{SDSS J 0917+1331 and SDSS J 1120+0407}}\label{fr0}

The two bottom panels of Figure~\ref{overlay1} show the MeerKAT radio images of SDSS J\,0917$+$1331 (left) and SDSS J\,1120$+$0407 (right), both classified as low-power radio galaxies. The defining morphological feature of FR\,0 radio galaxies, distinguishing them from FR\,I sources, is the lack of extended radio emission on kpc scales \citep{baldi}. SDSS J\,0917$+$1331 exhibits a slight elongation toward the northwest, whereas SDSS J\,1120$+$0407 displays typical FR\,0 properties, i.e. compact and point-like, with no detectable extended emission in the high-resolution, high-sensitivity MeerKAT image. {The restoring beam of the L-band observations is smaller than the angular extents of both sources, each being more than two beam widths, suggesting that they are slightly resolved.}

\begin{table*}
\caption{Radio properties of the sample}
\begin{center}
\begin{tabular}{@{}|l l c c c c c@{}}
\toprule
Source name & Component 
& $S_{700}$ (Jy) 
& $S_{1.28}$ (Jy) 
& LLS (Mpc)
& log P$_{\rm 700~MHz}$ 
& log P$_{\rm 1.28~GHz}$  \\
\midrule
3C\,105 & tot & -- & $6.60\pm0.30$ & 0.48 & -- & 26.12 \\
        & H$_{\mathrm{NW}}$ & -- & 0.02 & -- & -- & -- \\
        & H$_{\mathrm{SE}}$ & -- & 3.20 & -- & -- & -- \\

3C\,198 & tot & $3.91\pm0.07$ & -- & 0.74 & 25.81 & -- \\

3C\,227 & tot & $12.45\pm0.02$ & -- & 0.51 & 26.35 & -- \\
        & H$_\mathrm{W}$ & 0.57 & -- & -- & -- & -- \\
        & H$_\mathrm{E}$ & 3.00 & -- & -- & -- & -- \\

4C$-$03.43 & tot & $1.45\pm0.62$ & $1.08\pm0.05$ & 0.64 & 24.97 & 24.84 \\

CGCG\,047$-$067 & tot & $3.53\pm0.16$ & $2.39\pm0.01$ & 0.88--0.95 & 25.41 & 25.24 \\
3C\,403.1 & tot & $1.64\pm0.01$ & -- & 0.48 & 25.08& -- \\
3C\,445 & tot & -- & $6.46\pm0.32$ & 0.63 & -- & 25.69 \\
       & H$_{\mathrm{S_{outer}}}$ & -- & 0.65 & -- & -- & -- \\
       & H$_{\mathrm{N_{outer}}}$ & -- & 0.46 & -- & -- & -- \\
       & H$_{\mathrm{S_{inner}}}$ & -- & 0.08 & -- & -- & -- \\
       & H$_{\mathrm{N_{inner}}}$ & -- & 0.05 & -- & -- & -- \\

NGC\,7503 & tot & $2.58\pm0.02$ & -- & 0.01 & 25.07 & -- \\

SDSS\,J0917$+$1331 & tot & -- & $0.02\pm0.72$ & -- & -- & 23.08 \\

SDSS\,J1120$+$047 & tot & -- & $0.01\pm0.21$ & -- & -- & 22.78\\
\bottomrule
\end{tabular}

\end{center}
\label{tab:radio_properties}
\begin{flushleft}
\footnotesize {Note.} Column 1 reports the source name.  Column 2, the components for each source were tot denotes the total emission measured. Columns 3 and 4 report\\ on the flux density at the central frequency of uGMRT for Band-4 and MeerKAT, respectively. Column 5 provides the largest linear size. Columns 6 and 7 are the total power of each source at the corresponding central frequencies. 
\end{flushleft}
\end{table*}

\section{Spectral analysis}\label{sec5}
The MeerKAT and uGMRT images in this work cover a frequency range of 550 -- 1712 MHz with very similar angular resolutions and fairly comparable sensitivities, {which in principle make them ideal for the study of the spectral features of the jets and lobes of the radio galaxies in our sample observed with both arrays. Unfortunately, the different quality of the images obtained with the two arrays did not allow such investigation, and MeerKAT in-band spectral index imaging and the study of the ageing of the relativistic plasma was possible only for 3C\,105, 4C\,--03.43 and CGCG\,047$-$067.
On the other hand, for all sources in our sample, a detailed study of the integrated spectral properties was possible with the combination of the MeerKAT and uGMRT  data presented in this paper and the abundant literature information (see Appendix \ref{tab:appendix}).}\\

In the following sections, we will contribute to our knowledge of these sources by means of a global and local spectral study.

\subsection{Integrated spectral properties}
The flux density measurements presented here are supplemented with archival data from the NASA Extragalactic Database. {The values collected from the literature are reported in Appendix B, {together with the appropriate references for the 4C sources in our sample.}  We are aware that these values are heterogeneous; nevertheless, they allow to derive an integrated spectrum for all sources. {Due to the lack of literature information this analysis could not be performed for the two FR\,0 sources.}

We fitted the integrated spectrum with the Synage software package \citep{murgia1999A&A...345..769M}, to test which model for the evolution of the radio emission is best suited to describe our data}.  Synage tries to fit the integrated spectra with three different models. The Kardashev-Pacholczyk (KP) model assumes that the pitch angle of radiating electrons remains constant over time \citep{Kardashev, pacholczyk1970radio}.
On the other hand, the Jaffe-Perola (JP) model considers the single particle to be subject to many scattering events that randomise its pitch angle, assuming a timescale for the isotropization of the electrons much longer than the radiative timescale  \citep{jaffe1973dynamical}. Finally, in the  Continuous Injection (CI) model \citep{Kardashev}, the source is continuously fuelled at a constant rate.  
{All three models were tested, and for all radio galaxies the CI
model provided the best fit. The parameter $\alpha_{\rm inj}$
was initially set free, and for all sources except 3C\,403.1 the
fit converged for values very close to 0.5, which was then fixed.
For 3C\,403.1 the best fit was obtained with $\alpha_{\rm inj}$=0.89.}
One of the key output parameters is the break frequency, $\nu_{\rm br}$, which we use to estimate the time elapsed, since the last re-acceleration of the electron population (see below).

\begin{figure*}
\begin{tabular}{cccc}
 \begin{tikzpicture}
    	\node[anchor=south west,inner sep=0] (image) at (0,0) {\includegraphics[width=0.3\textwidth]{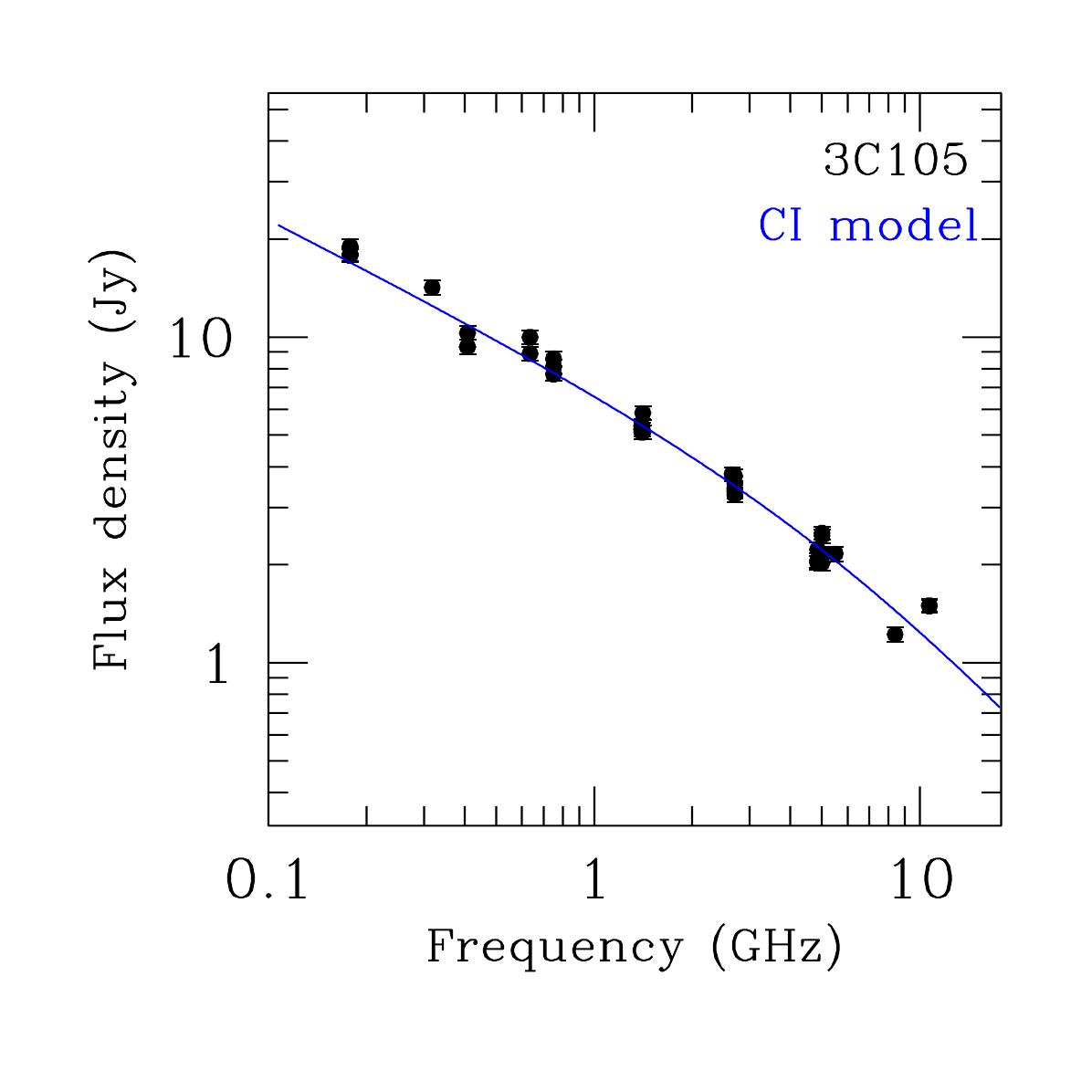}};
    	\begin{scope}[x={(image.south east)},y={(image.north west)}]
        	% Add text at specific coordinates
         	\node[anchor=west]  at (0.33,0.40) {$\nu_{\rm{br}}$ = 6.01 GHz};
           	\end{scope}
	\end{tikzpicture}
&
 \begin{tikzpicture}
    	\node[anchor=south west,inner sep=0] (image) at (0,0) {\includegraphics[width=0.3\textwidth]{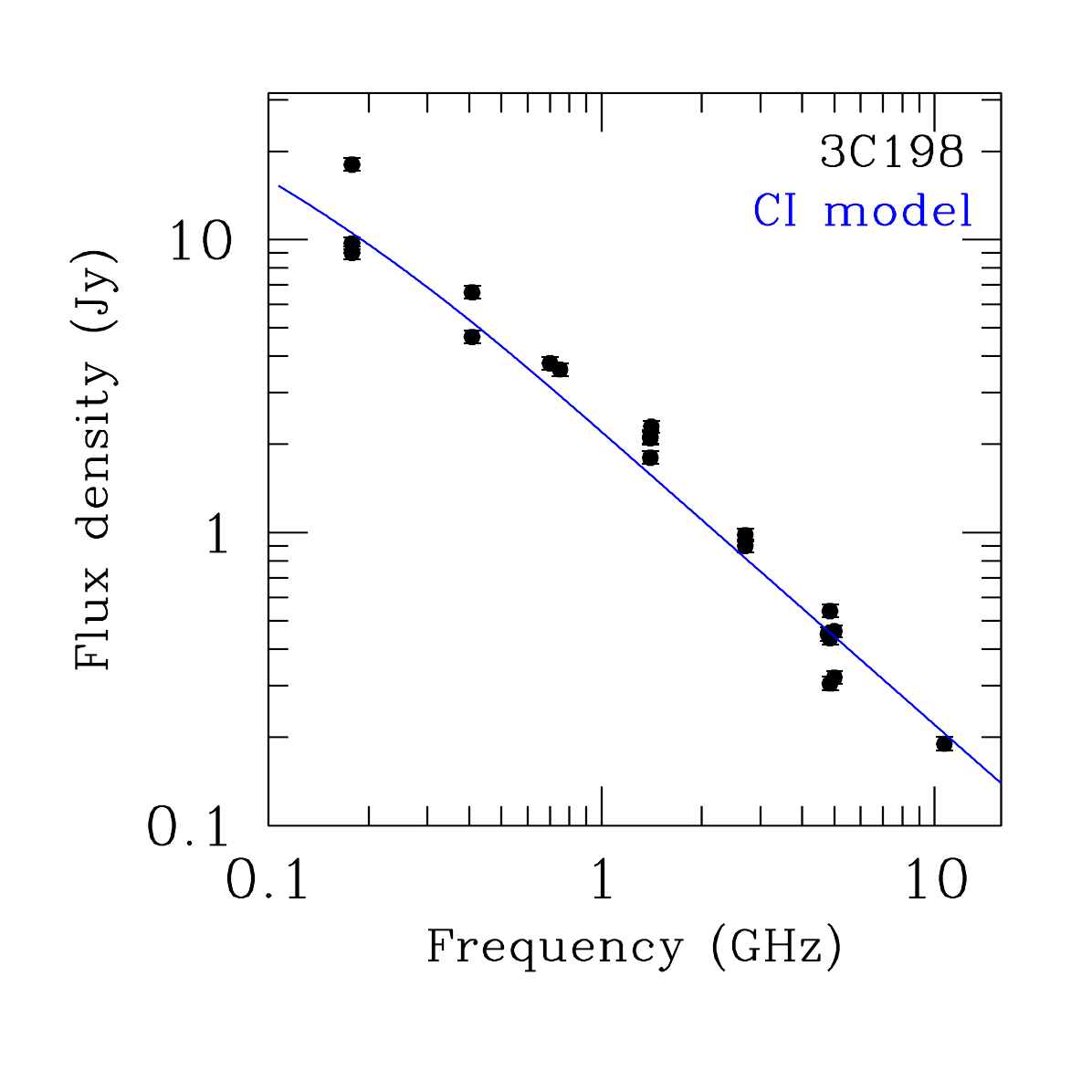}};
    	\begin{scope}[x={(image.south east)},y={(image.north west)}]
        	% Add text at specific coordinates
         	\node[anchor=west]  at (0.33,0.40) {$\nu_{\rm{br}}$ = 0.22 GHz};
           	\end{scope}
	\end{tikzpicture}
&
 \begin{tikzpicture}
    	\node[anchor=south west,inner sep=0] (image) at (0,0) {\includegraphics[width=0.3\textwidth]{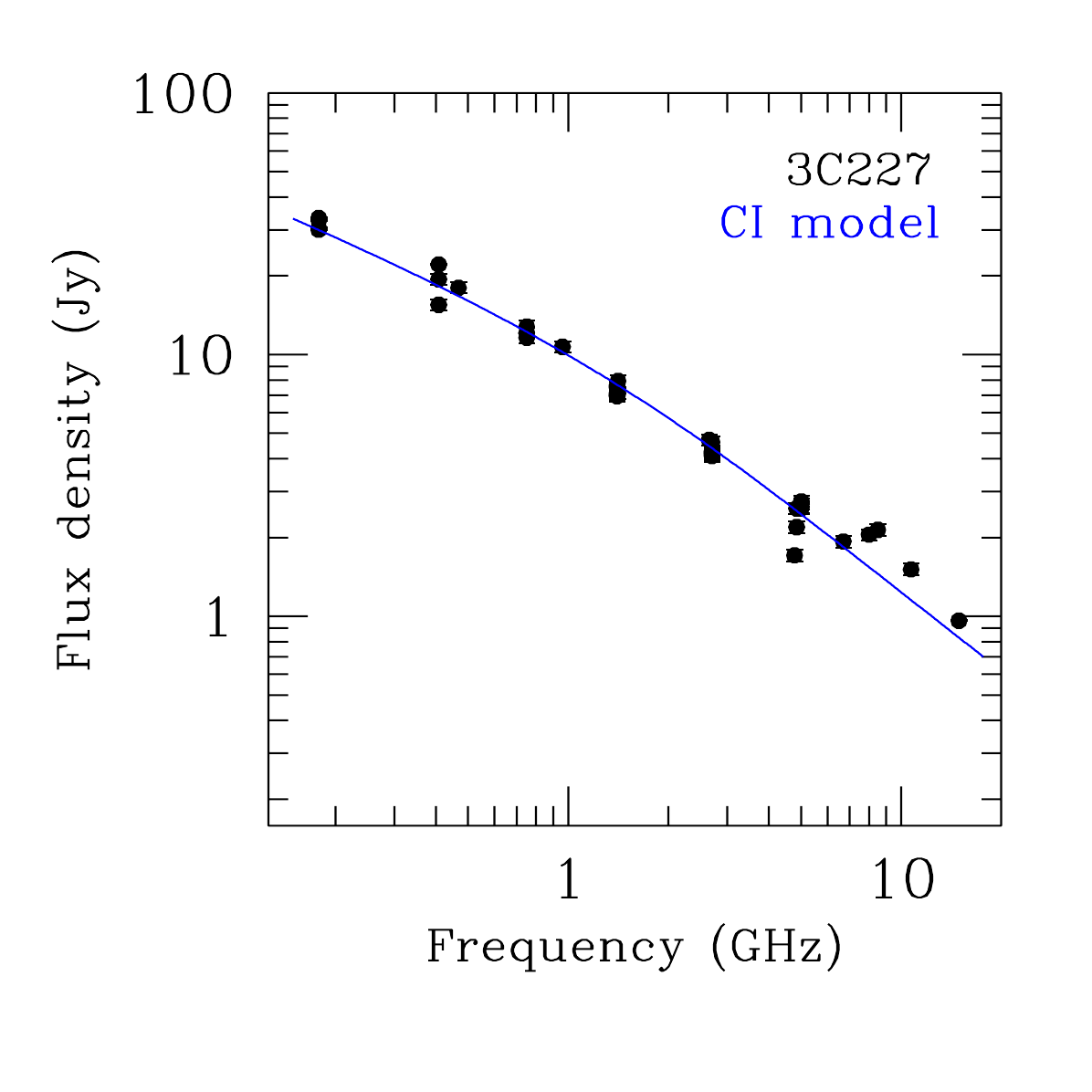}};
    	\begin{scope}[x={(image.south east)},y={(image.north west)}]
        	% Add text at specific coordinates
         	\node[anchor=west]  at (0.33,0.40) {$\nu_{\rm{br}}$ = 1.56 GHz};
           	\end{scope}
	\end{tikzpicture}
\\
    \includegraphics[width=0.3\textwidth]{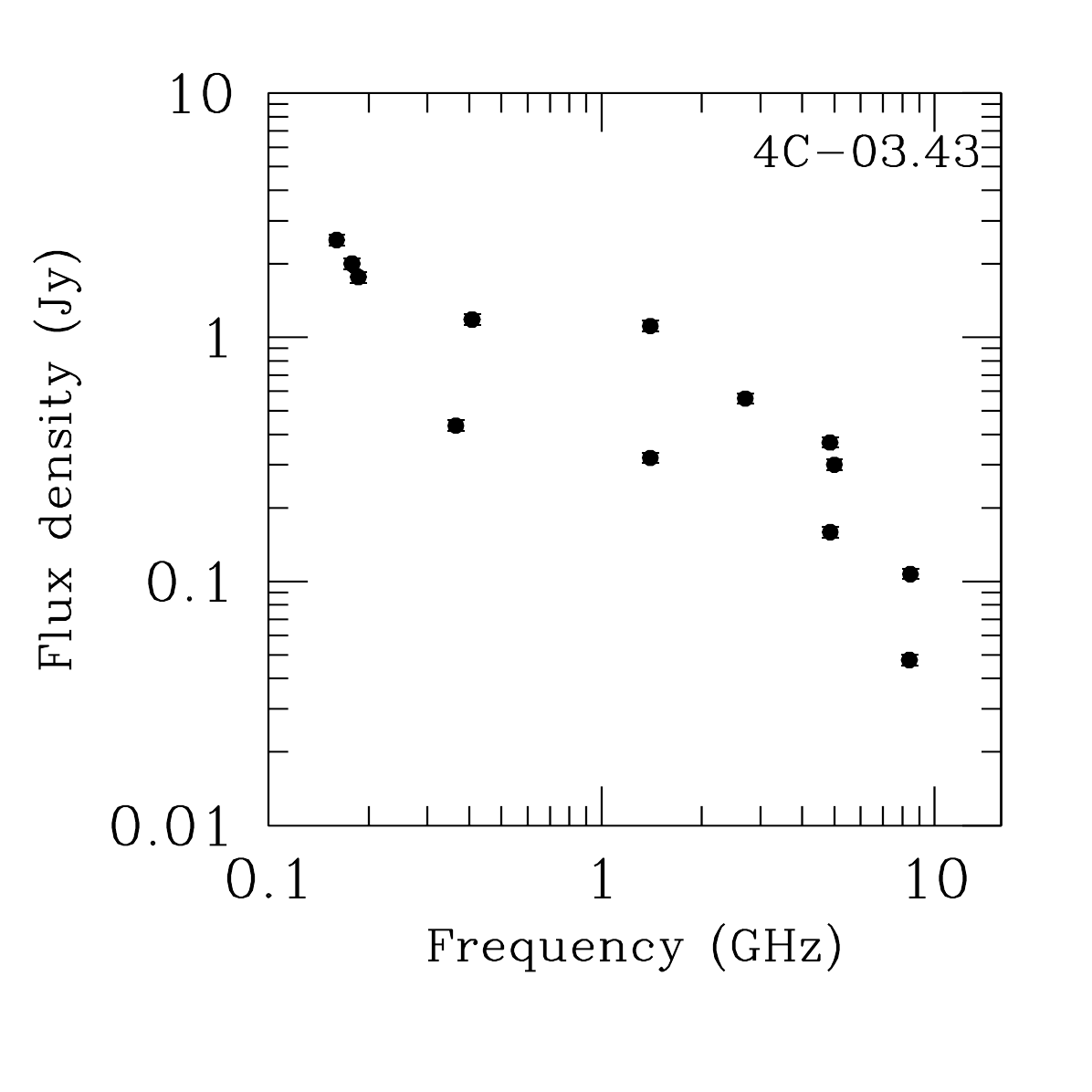}&
     \begin{tikzpicture}
    	\node[anchor=south west,inner sep=0] (image) at (0,0) {\includegraphics[width=0.3\textwidth]{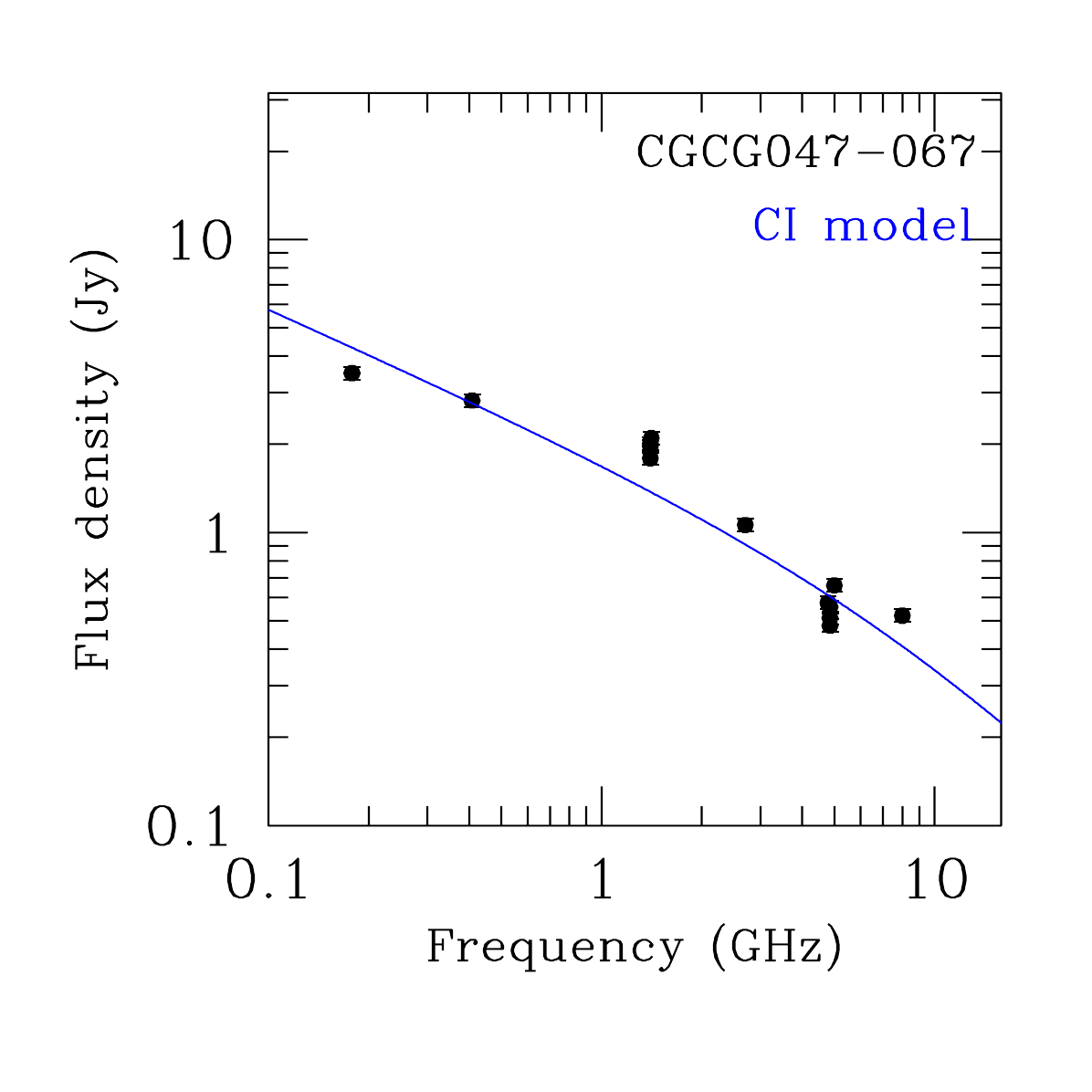}};
    	\begin{scope}[x={(image.south east)},y={(image.north west)}]
        	% Add text at specific coordinates
         	\node[anchor=west]  at (0.33,0.40) {$\nu_{\rm{br}}$ = 7.54 GHz};
           	\end{scope}
	\end{tikzpicture}
&
      
     \begin{tikzpicture}
    	\node[anchor=south west,inner sep=0] (image) at (0,0) { \includegraphics[width=0.3\textwidth]{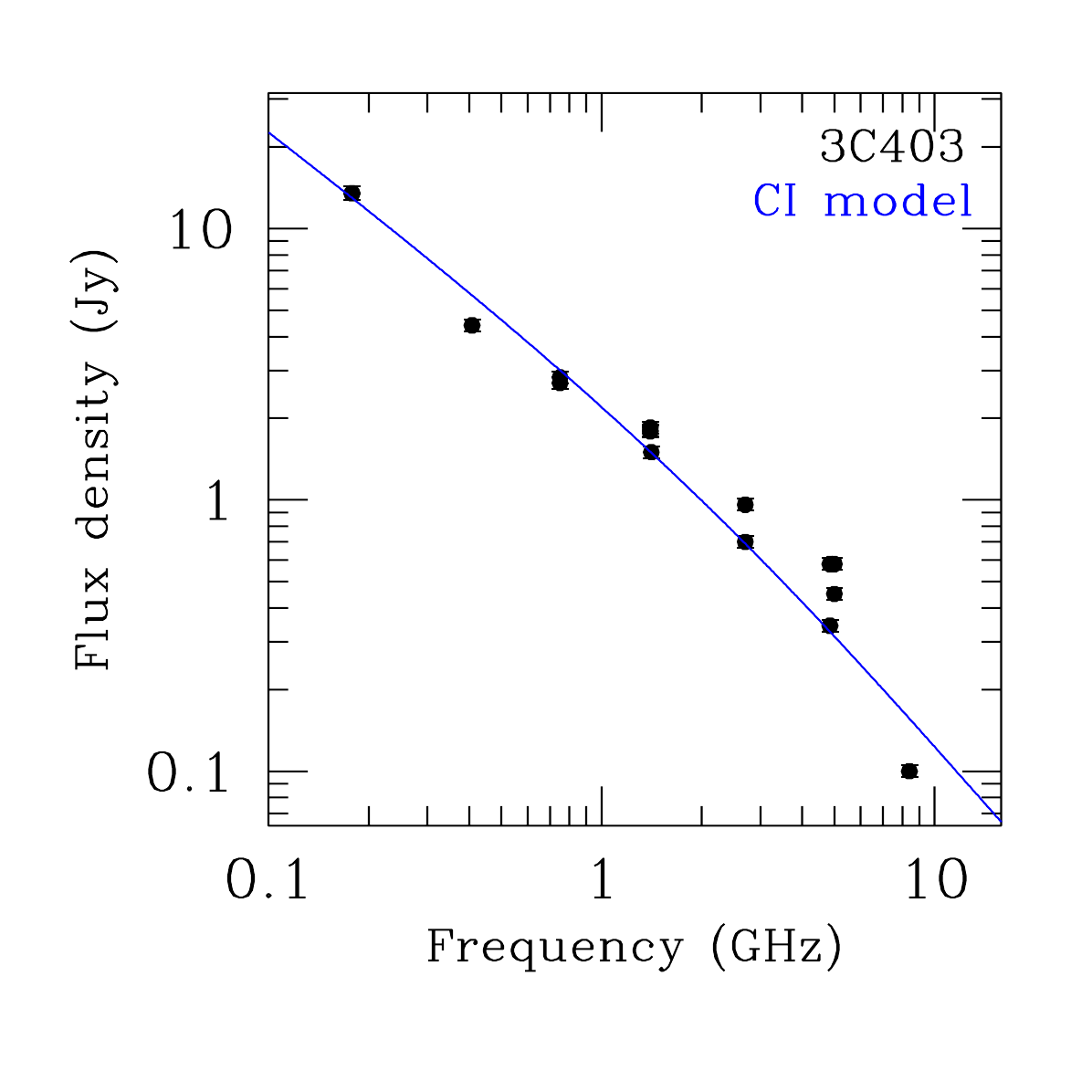}};
    	\begin{scope}[x={(image.south east)},y={(image.north west)}]
        	% Add text at specific coordinates
         	\node[anchor=west]  at (0.33,0.40) {$\nu_{\rm{br}}$ = 3.27 GHz};
           	\end{scope}
	\end{tikzpicture}
\\ 
      
     \begin{tikzpicture}
    	\node[anchor=south west,inner sep=0] (image) at (0,0) {\includegraphics[width=0.3\textwidth]{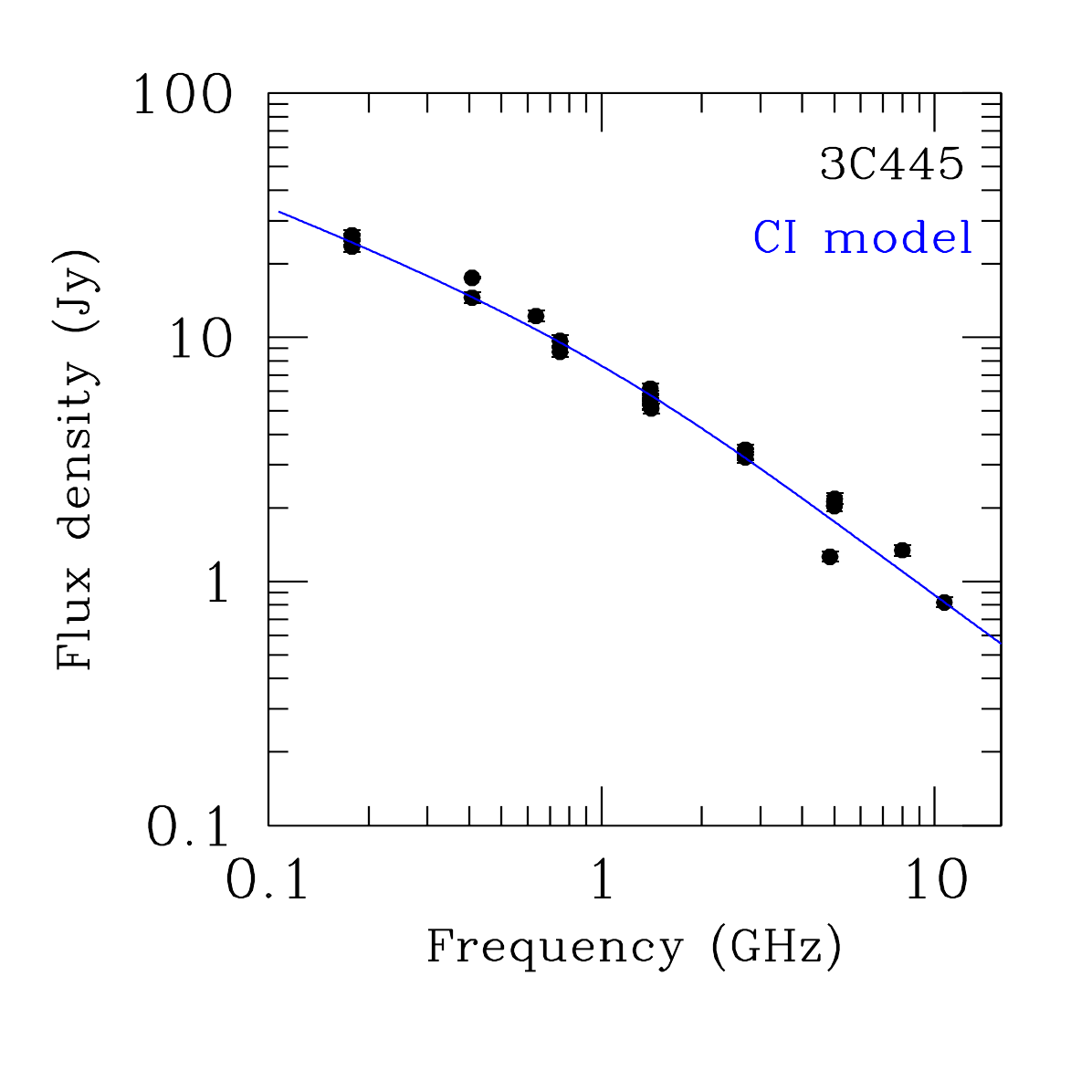}};
    	\begin{scope}[x={(image.south east)},y={(image.north west)}]
        	% Add text at specific coordinates
         	\node[anchor=west]  at (0.33,0.40) {$\nu_{\rm{br}}$ = 1.13 GHz};
           	\end{scope}
	\end{tikzpicture}
    &
        \begin{tikzpicture}
    	\node[anchor=south west,inner sep=0] (image) at (0,0) {\includegraphics[width=0.3\textwidth]{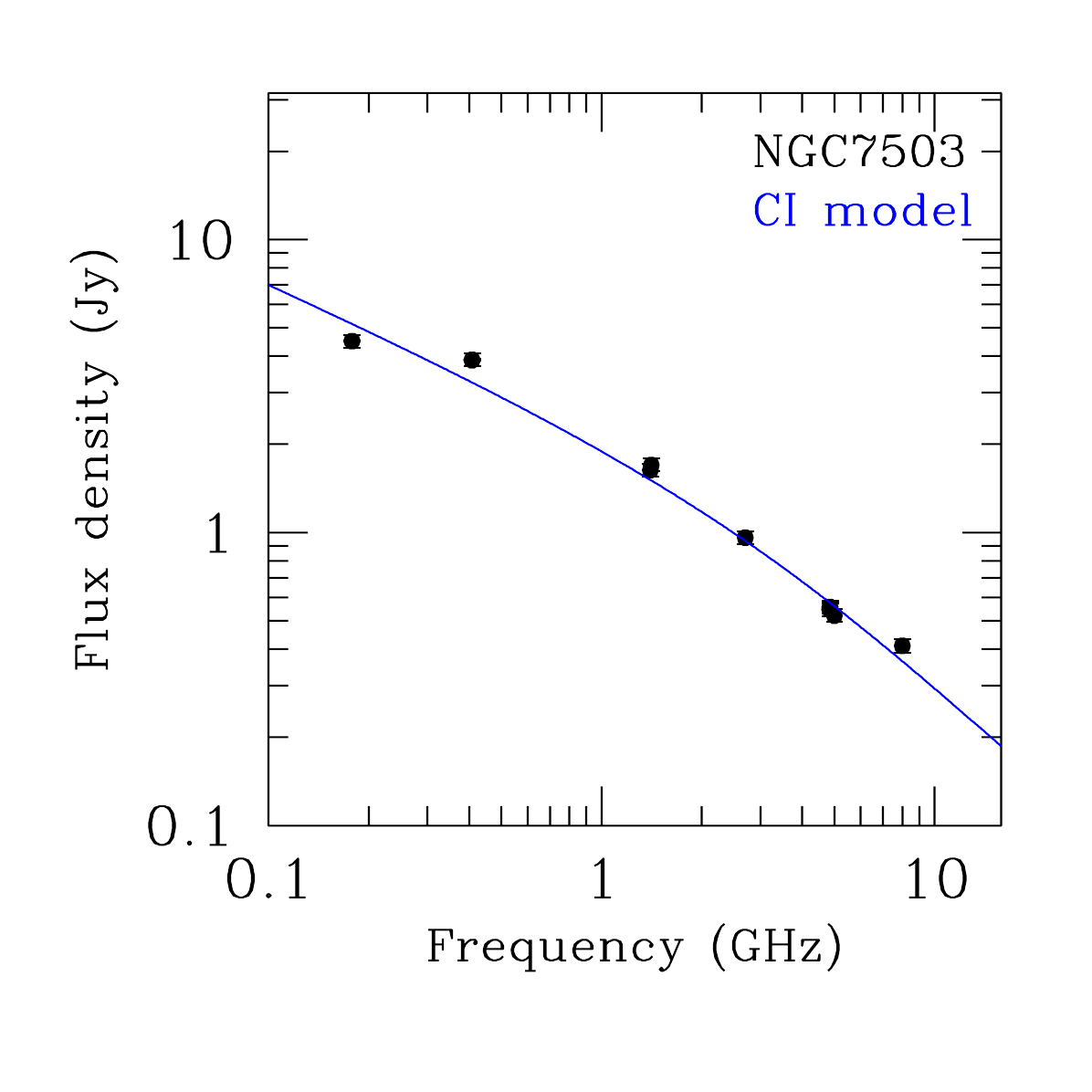}};
    	\begin{scope}[x={(image.south east)},y={(image.north west)}]
        	% Add text at specific coordinates
         	\node[anchor=west]  at (0.33,0.40) {$\nu_{\rm{br}}$ = 3.27 GHz};
           	\end{scope}
	\end{tikzpicture}
\end{tabular}
   \caption{The integrated radio spectra of the radio sources, presented in the order of Table \ref{log}, are shown. The continuous blue curve, representing the CI model fit, indicates the presence of a spectral break at frequency $\nu_{\rm{br}}$.} 
\label{specindex}
\end{figure*}
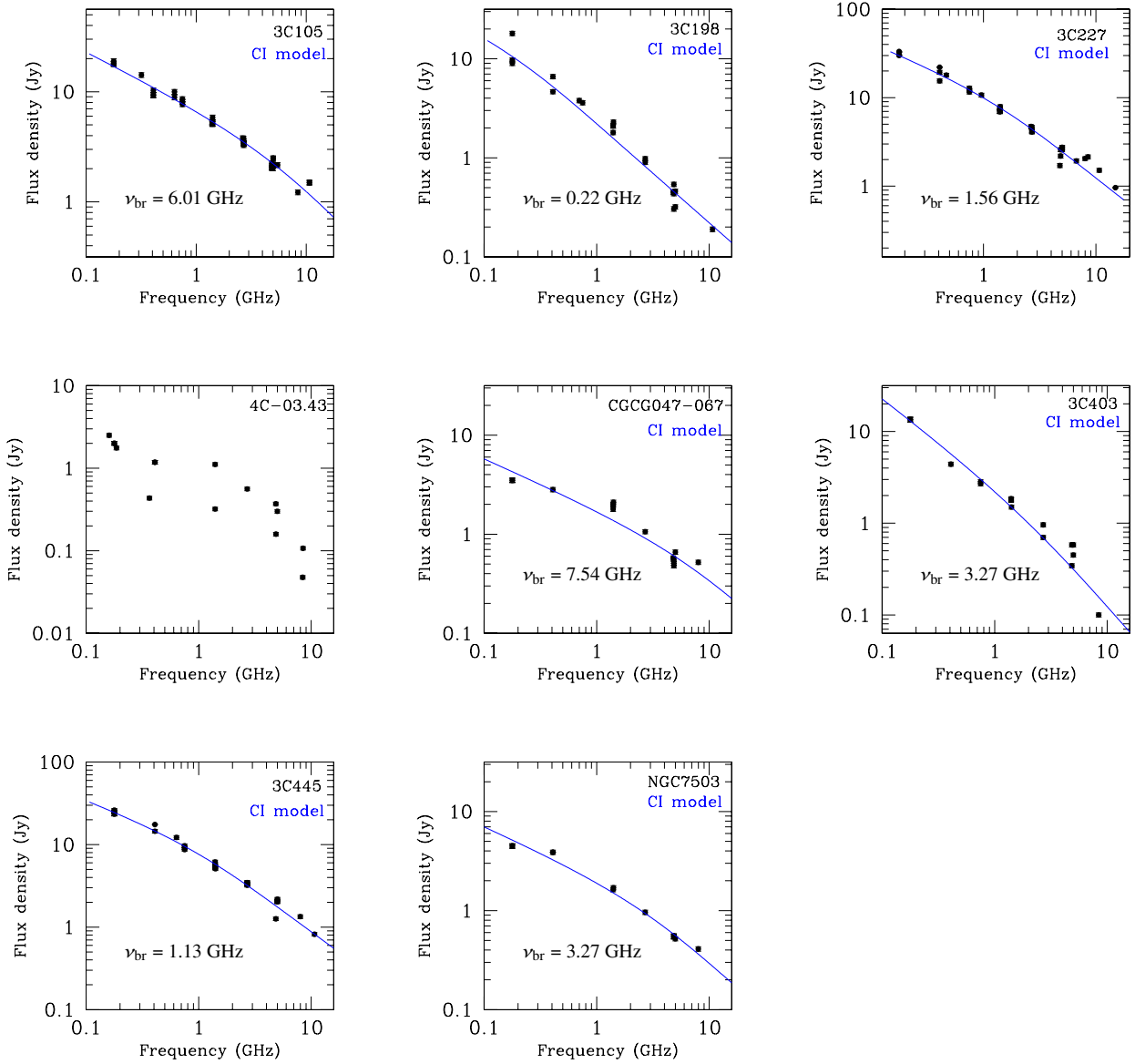

The spectra and the corresponding best fit model are shown in Figure \ref{specindex}. Below, we provide some information on the individual radio galaxies.
The spectral fits and break frequency for all sources are summarised in Table \ref{ages}. 
 
 All sources except 4C\,--03.43 are well described by the CI model with the break frequency varying from case to case, reflecting variation in spectral ageing. { The literature data for 4C\,--03.43 are highly inhomogeneous, as the flux density measurements come from radio interferometers with very different angular resolutions (including very long baseline arrays). As a result, the integrated spectral fit is considered unreliable and unsuitable for further analysis.} 
 %The values obtained from the literature for 4C\,--03.43 are inhomogeneous as they refer to the varying interferometric radio telescope resolutions; hence, the fit of the integrated spectrum is deemed inaccurate and unsuitable for further estimations.
 
 A remarkable case is 3C\,198 whose spectral fit and morphology point to an old radio galaxy. The source has the steepest spectrum among the sources in our sample. The spectral index in the frequency range 0.14 to 10.7 GHz is $\alpha= -1.0 \pm 0.2$ with a very low break frequency ($\nu_ {\rm br}$ = 0.22 GHz).
The overall spectral behaviour of this source is consistent with its morphology, which lacks active/re-energised components, such as core, jets and hotspots.
 
3C\,445 shows signatures of restarted activity  \citep{DDRG, nandi}; however, the overall radio emission is dominated by the extended lobes. The spectral study therefore could be the result of the combination of these two components, and the radiative age of the source derived using the break frequency provided by the fit is most likely a lower limit to the age of the oldest emission.

For each source, we estimated the equipartition magnetic field using the total radio power as measured from NVSS (see Table \ref{log}). {We assumed a filling factor $\Phi$ = 1 and a proton to electron ratio $\kappa$ = 1. We approximated the source volumes with cylindrical components using the MeerKAT images out to the 9$\sigma$ contour level, consistent with the results reported in Sect. 4.} We further used  B$_{\rm eq}$ and the break frequency provided by Synage to estimate the global radiative age of the radio galaxies in our sample, which we derived following Eq. 2 in \citet{murgia1999A&A...345..769M}.

All these quantities are reported in  Table \ref{ages}. {The radiative ages for the radio galaxies in our sample span almost an order of magnitude, from $\sim$4 $\times$ 10$^7$ yr for 3C\,105 to $\sim$2.4 $\times$ 10$^8$ yr for 3C\,198 (see Sect. \ref{dead}), consistent with the radiative lifetimes reported in literature for typical radio sources \citep[e.g,][]{parma2002lives, Lal2007MNRAS, 2008MNRAS.390.1105L, harwood2017fr, mahatma2020, morganti2024learnedlifecycleradio}}.

\subsection{Spectral index distribution imaging}\label{spec}

Spectral index imaging reveals the spatial distribution of spectral indices across a source, providing a more complete understanding of the origin and evolution of the radio emission and its interaction with the external environment.

In principle, our data permit spectral imaging between 600 and 1283 MHz, corresponding to uGMRT Band-4 and MeerKAT L-band, for four sources observed with both arrays. However, due to RFI and dynamic-range limitations, the uGMRT images have lower sensitivity, resulting in 600 -- 1283 MHz spectral index maps that show trends similar to those obtained from the MeerKAT in-band spectral index maps but with substantially larger uncertainties.  Thus, we present spectral analyses based solely on the MeerKAT in-band spectral index maps.
We follow the  standard method of determining the spectral index between  S$_{\rm{\nu_1}}$(x,y)  and S$_{\rm{\nu_2}}$(x,y) at two frequencies  ${\rm{\nu_1}}$ and ${\rm{\nu_2}}$ is given by
\begin{equation}
\alpha_{\nu_1\nu_2}(x,y) = \frac{lnS_{\nu_1}(x,y)-lnS_{\nu_2}(x,y)}{ln\nu_{2}-ln\nu_{1}}
\end{equation}
and the spectral index error estimated as
\begin{equation}
\sigma_\alpha \;=\; 
\frac{1}{\left|\ln\left(\nu_2 / \nu_1\right)\right|} \,
\sqrt{
  \left(\frac{\delta S_1}{S_1}\right)^2
  +
  \left(\frac{\delta S_2}{S_2}\right)^2
};
\end{equation}
where $\delta$S is the absolute flux density scale uncertainty. Our analyses are within a restricted range of 909.39 to 1658.39~MHz, the full band split into eight subbands, each with a width of $\sim$93.6~MHz and a common ($u,v$) coverage in each sub-band.

\begin{figure*}
    \centering
    \begin{tabular}{ccc}
          \includegraphics[height=5.4cm]{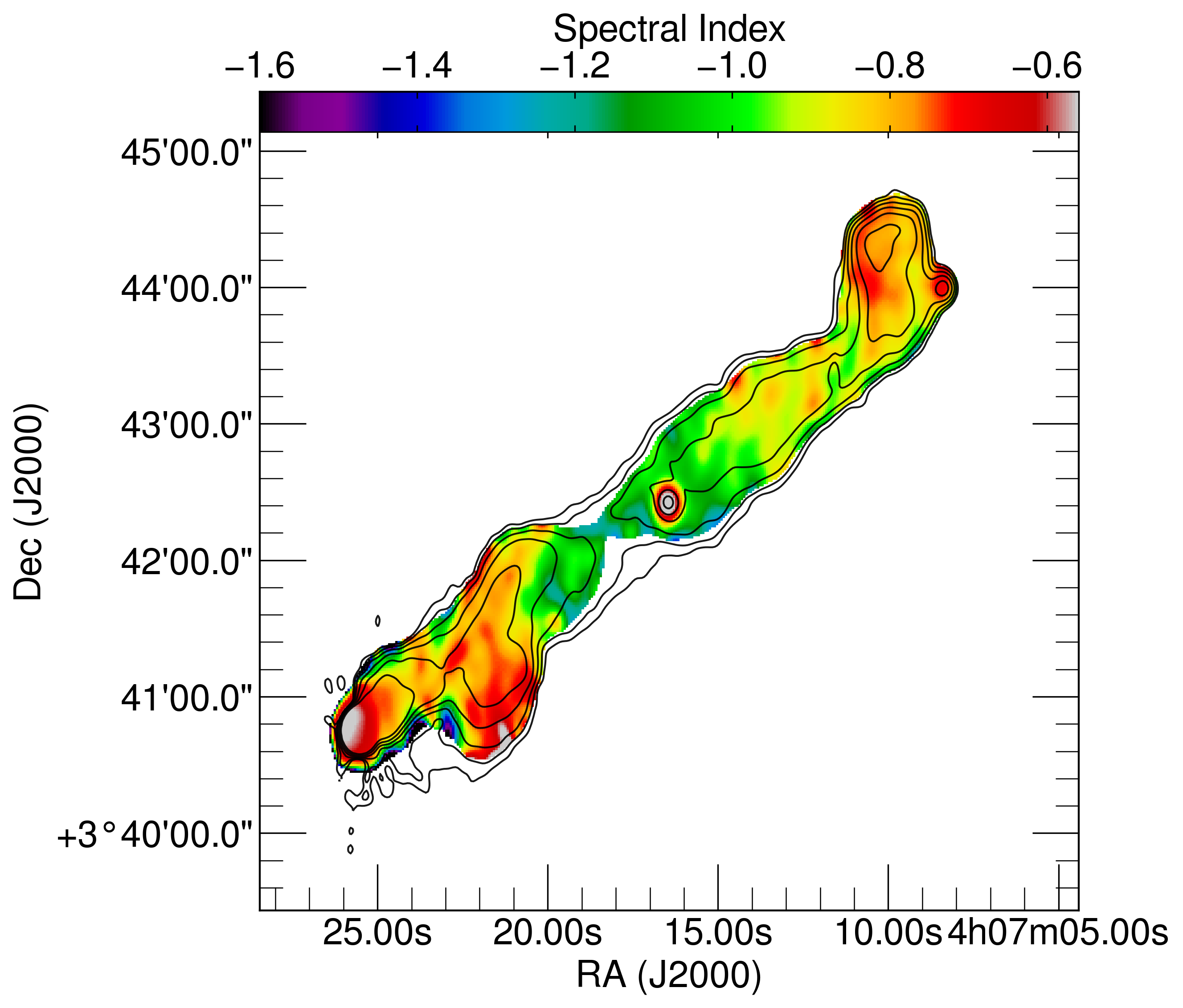} &
          \hspace*{-0.9cm} \includegraphics[height=5.4cm]{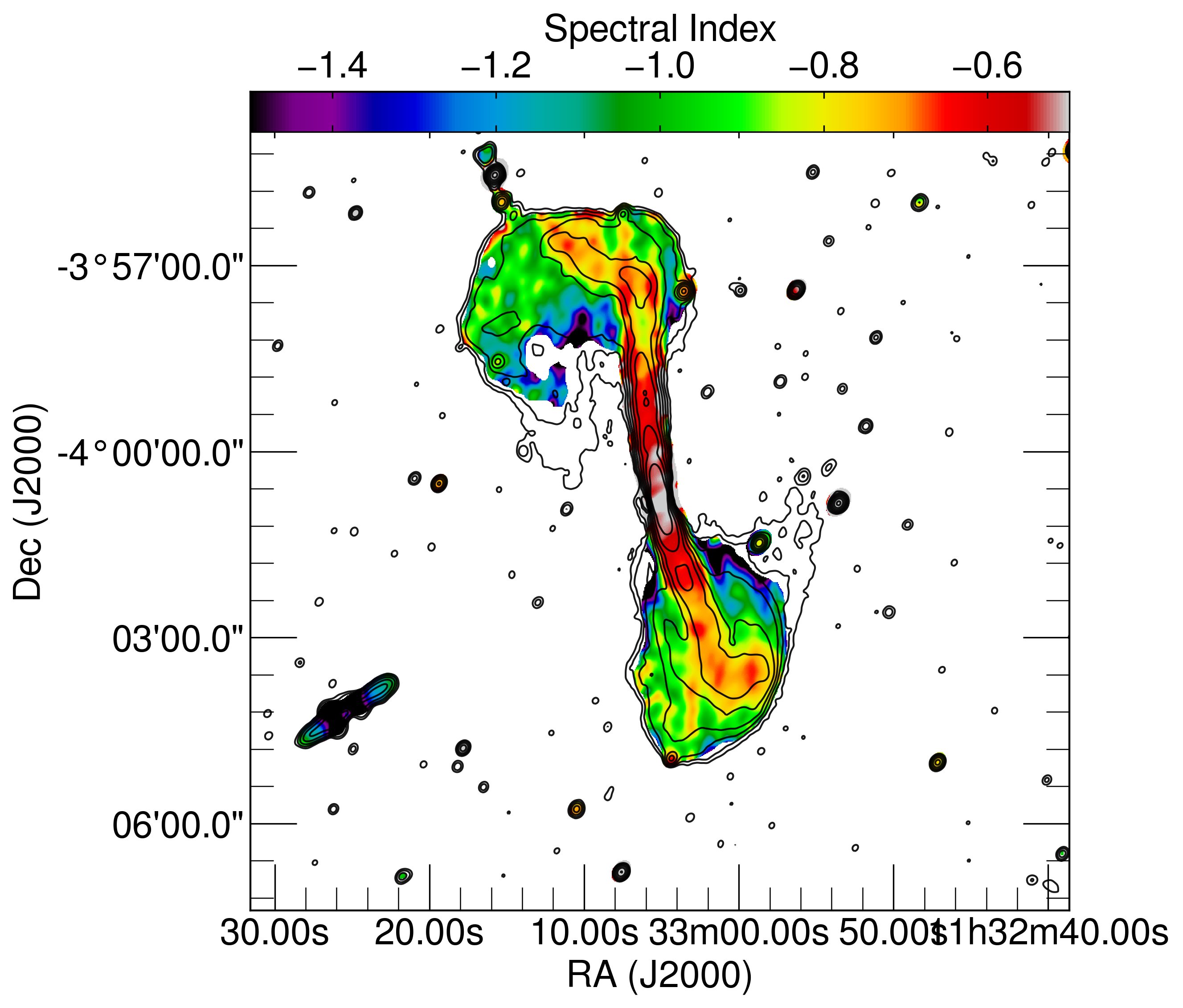} &
          \hspace*{-0.9cm} \includegraphics[height=5.4cm]{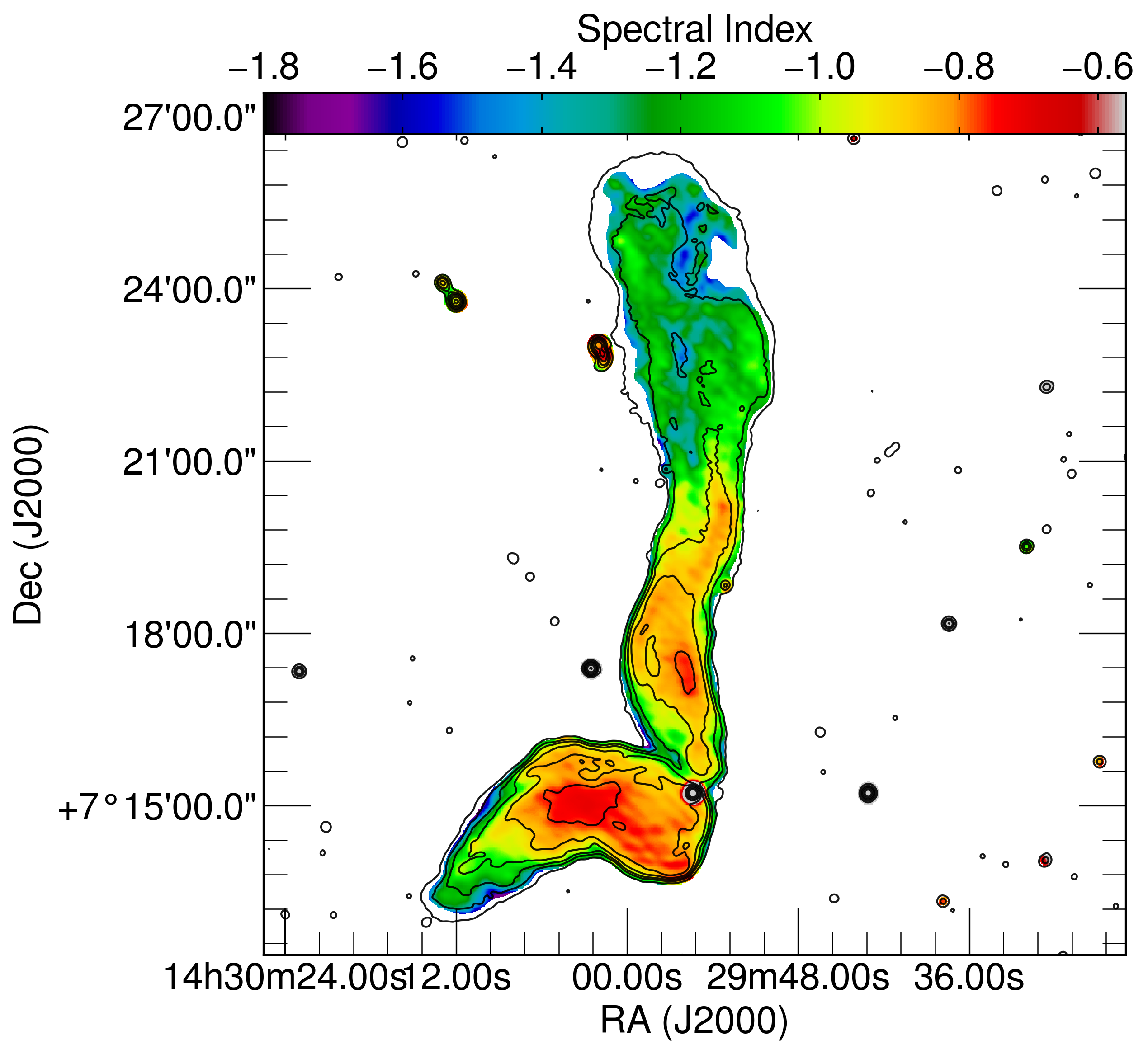}
    \end{tabular}
   \caption{The  MeerKAT spectral index distribution  $\alpha$ (S$\nu \propto  \nu^\alpha$) in range 909.39 to
1658.39 MHz of 3C\,105, 4C\,$-$03.43 and CGCG\,047$-$067. The MeerKAT radio contours at 1.28 GHz are drawn in black at 3$\times$RMS [1, 2, 4, 8].}
\label{specinband}
\end{figure*}

Figure \ref{specinband} illustrates the in-band spectral index distributions of 3C\,105, 4C\,--03.43, and CGCG\,047$-$067. {Unfortunately, dynamic range limitations did not allow us to obtain this information for 3C\,445. 
\newline
The spectral trend of 3C\,105 (left panel) is typical of FR\,II radio galaxies. The core exhibits a flat spectrum ( $\alpha \gtrsim -0.4$), consistent with compact emission}. The hotspots show spectral index in the range --0.6 to --0.8, and the emission of the lobes further steepens gradually inwards towards the core, reaching values in the range [--1.2, --1]; the southeastern hotspot is marginally flatter than its counterpart. 
The southeastern and northwestern lobes display subtly different steepening patterns, indicative of asymmetries between the two sides.

In the in-band spectral index distribution of 4C\,$-$03.43 (middle panel), progressive spectral steepening along the jets from the core to the outer lobes is seen, characteristic of FR\,I radio galaxies \citep{laing2007jet}. In the lobes, the spectrum is moderately steep ($\alpha \approx -0.7$ to $-$1.0) in regions surrounding the jets and steepens further outward, reaching values as low as $\alpha \sim -$1.5. The outer parts of the lobes are too faint for the spectral index to be imaged.

Similarly, in the in-band spectral index map of CGCG\,047$-$067 (right panel), the southeastern lobe exhibits a relatively uniform, moderately steep spectrum ($\alpha \approx -$0.8 to $-$1.0). In contrast, the northern lobe shows a clear and sharp spectral steepening: the inner $\sim$7 arcmin ($\approx$ 440 kpc) has an average index of $\alpha \sim -0.9$, beyond which the spectrum steepens abruptly to values as low as $-$1.5. The spectral change is clearly associated with a morphological change, with a sharp surface brightness drop and loss of collimation. One possibility is that such behaviour reflects the transition of the jet into a less dense intracluster environment. The marked asymmetry in size,  morphology and spectral behaviour between the two lobes suggests that orientation effects might play a relevant role for this source.

\section{General considerations}\label{sec6}
In this section, we present a general discussion of the overall observational properties of our sample of radio galaxies, including the four sources already published in Paper I. We focus on the morphology, size, and radiative ages.
\subsection{Source size distribution}

\begin{figure}
    \centering
    \includegraphics[width=1.0\linewidth]{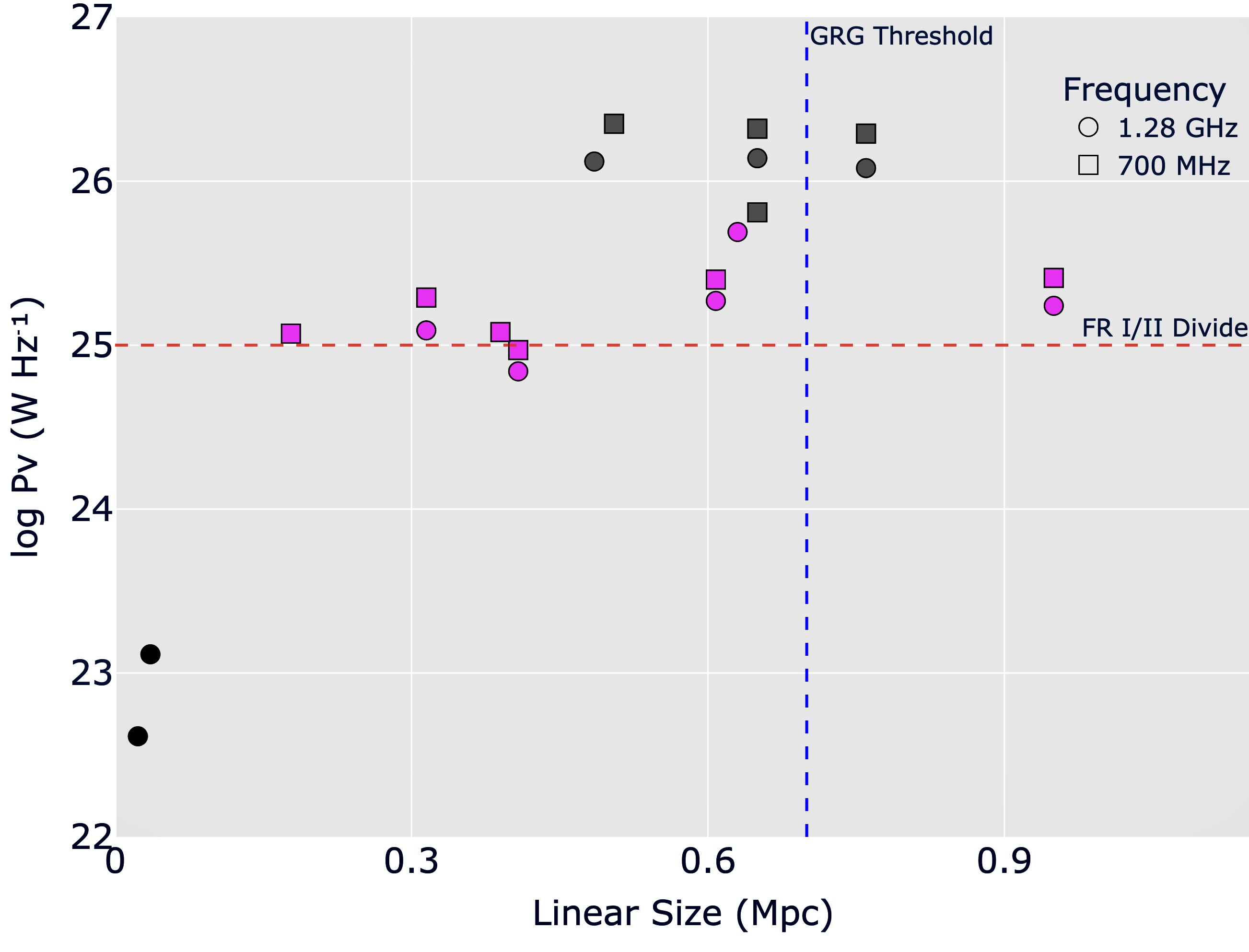}
    \caption{Radio power versus linear size of the sample of radio galaxies. The morphological types are colour-coded: FR\,I (magenta), FR\,II (grey). The MeerKAT and uGMRT observations are denoted by circles and square markers, respectively. Dashed black lines connect the frequencies of the sources observed at both bands. The red dashed horizontal line marks the FR I/II luminosity divide at $\log P_{\nu} = 25$ WHz$^{-1}$, and the blue dashed vertical line indicates the giant radio galaxy threshold at 0.7 Mpc.}
    \label{fig:power}
\end{figure}

\begin{figure}
    \centering
    \includegraphics[width=1\linewidth]{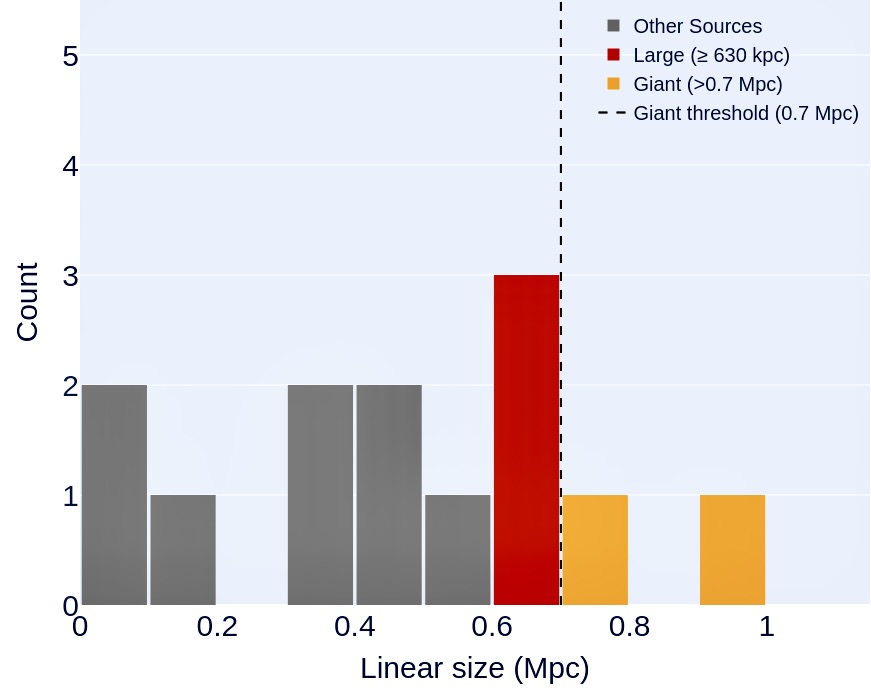}
    \caption{Distribution of the linear size in our sample. The blue dotted line marks the threshold of 0.7 Mpc.}
    \label{fig:dist}
\end{figure}

The projected linear sizes of the radio structures, measured at the 9$\sigma$ contour level and excluding spurious low-surface brightness features, span $\sim$160 kpc to 0.98 Mpc,
%for the extended source, 
as clear from Figures \ref{fig:power} and \ref{fig:dist}. 
Adopting the commonly used threshold of 0.7 Mpc for defining Giant Radio Galaxies (GRGs) under the current cosmological framework \citep[e.g.][]{dabhade2020giant, simonte2022giant, DDRG}, among the radio galaxies presented here only CGCG0\,47$-$067 qualifies as bona fide GRG, with a projected size of $\sim$0.98 Mpc, which adds to 4C\,12.02 presented in Paper I. {Moreover, 4C\,12.03 (presented in Paper I) and the two radio galaxies 3C\,198 and 3C\,445 (presented in this paper) approach the giant regime, with linear sizes $\ge$630 kpc}.
{The distribution of the largest linear sizes, colour-coded by morphology, is shown in Figure \ref{fig:dist}, where the cumulative distribution indicates that the two cluster-dominant radio galaxies in our sample fall below the giant threshold.}

\subsection{\texorpdfstring{The case of 3C\,198}{The case of 3C 198}}
\label{dead}

3C\,198 is an interesting case. From the morphological point of view, it lacks both a detectable core and hotspots, suggesting that the radio emission is no longer sustained by active acceleration processes \citep{parma2007search, murgia2011dying}. Its steep spectrum and break frequency at ~200 MHz lead to a radiative age of $\sim$2.4$\times$10$^8$ yr (Sect. \ref{ages}), confirming that it is an old, most likely dying radio galaxy. The overall properties of the radio emission, i.e. the very low and fairly uniform surface brightness, and featuresless lobes with filamentary structure, together with its very old age, suggest that 3C\,198 could be a transition phase between an active double radio galaxy and the very steep spectrum, faint, filamentary and amorphous sources which are now being detected with the current generation of radio interferometers like LOFAR and MeerKAT \citep[e.g,][]{oozeer2021discovery, morganti2024learnedlifecycleradio, brienza2025non}.

\subsection{Restarted radio galaxies}
Two sources in our sample present morphology suggestive of restarted activity: 3C\,445 and 4C\,12.03 (reported in Paper I). The first shows a north-south well-aligned orientation of the inner and outer brightness peaks, previously reported in literature \citep[e.g.,][]{schoenmakers2000radio, saikia2010recurrent}. The 1.28~GHz radio image reveals the inner structure of 3C\,445, labelled as the inner hotspot in Figure \ref{INtesity1}, with a total extent of $\sim$143 kpc. The image of this source allowed only a morphological study. %\citet{dabhade2020giant} interpret the structure as a flow interruption in a bipolar jet; however, its underlying mechanism remains unclear; the nature of this phenomenon needs to be unveiled through spectral analysis. 
4C\,12.03 falls among the restarting X-shaped galaxies with no peak brightness at the lobe end \citep{saripalli2009genesis}. In both sources, the newly propagating jet pair is aligned with the direction of the earlier activity, consistent with a behaviour seen in DDRGs.

\subsection{Morphological consideration and the role of environment}

Table \ref{tab:radio_properties} summarises the properties of the radio structures detected in our images.
The radio morphologies revealed by our MeerKAT and uGMRT observations demonstrate a broad range of structural complexity while largely conforming to the classical FR framework \citep{fanaroff&riley}. Tailed radio galaxies in the sample display morphologies consistent with strong environmental influence, supporting the view that ram pressure and interactions with the ICM play a dominant role in shaping their radio emission \citep[e.g,][]{jones1979hot, giacintucci2009tailed, missaglia2022high}. In contrast, most FR I and FR II sources retain their characteristic core- or edge-brightened structures, indicating that, despite environmental effects, the intrinsic jet power and dynamics remain key determinants of large-scale morphology.

A small subset of sources, notably 3C\,403.1, exhibits hybrid characteristics that blur the traditional FR I/II distinction \citep{2008MNRAS.390.1105L}. The coexistence of FR\,I like jet properties on one side and FR II like termination features on the other suggests asymmetric jet-environment interactions or temporal variations in jet power \citep{o1985constraints, Sebastian_2017}. Such systems provide valuable insight into transitional phases of radio-galaxy evolution and highlight the role of local conditions in modifying otherwise symmetric outflows.

The presence of hotspots with multiple emission peaks, particularly in FR II sources, further stresses the importance of jet-medium interactions. These features are consistent with scenarios involving multiple shock-like features, deflected backflows, or sites associated with changes in jet direction \citep{kraft2005chandra, cheung2007electron}. Generally, the morphological details of the two hotspots in  FR\,II galaxies often differ.  We observe a bifurcation in the backflow of one hotspot in 3C\,105 and 3C\,227, as well as a similar feature in the fading lobe of 3C\,198. The distinct morphology of 3C\,227, including aligned components perpendicular to the jet axis and X-ray–detected synchrotron emission in both hotspots \citep{hardcastle2007hot}, supports models in which hydrodynamic instabilities or stream-splitting events \citep{horton20203d} redistribute energy while maintaining efficient particle acceleration at multiple sites \citep{leahy}.

Filamentary substructures are a ubiquitous feature across the sample, appearing both in classical radio galaxies and in tailed sources. Their presence within diffuse lobes and tails suggests localised enhancements in magnetic field strength \citep{gendron2021vla} or turbulence within the surrounding medium \citep{gendron2021vla}. In cluster environments, such filaments may trace interactions with the ICM or remnants of earlier jet activity indicates that similar processes operate beyond dense environments. These ordered structures may represent well defined paths that connect active hotspots to older lobe plasma, pointing to a complex and dynamic energy transport within radio lobes \citep{hardcastle2019ngc, rudnick2022intracluster, maccagni2020flickering}.

Large-scale asymmetries in lobe brightness and extent, most clearly observed in 3C\,105 and 3C\,227, provide further evidence that environmental factors significantly influence radio-galaxy evolution. The disparity between opposing lobes in 3C\,105 cannot be readily explained by relativistic beaming alone and instead points to gas density (and hence pressure) gradients in the ambient medium on scales of hundreds of kiloparsecs. In 3C\,227, the presence of symmetric lobe truncation near the core and a central surface-brightness trough \citep[e.g., 3C\,198,][]{black1992study} suggests the existence of a dense, disk-like gaseous structure that constrains backflowing plasma \citep{black1992study, wiita2004asymmetries}. While relativistic effects may account for some degree of jet-sidedness, they are insufficient to explain the observed hotspot and lobe asymmetries without invoking environmental influences \citep{o2009physical, wiita2004asymmetries}. 
\newline
{The general structure and radio power of FR\,0 discussed in Sect.~\ref{fr0} aligns with the defining properties of this morphological class. Our observations are not conclusive, and the nature of these sources remains elusive. It is unclear whether they represent a separate population or a young /shortlived phase of FRI objects \citep{baldi, Miraghaei2017The}. Recent observations increasingly reveal faint, low surface brightness extensions in some FR\,0s, as seen in the small-scale jet structures of SDSS~J\,091754+040742 \citep[e.g.,][]{giovanni}. Such  weak extended emissions may reflect intrinsically low jet power, frustrated jets, or short duty cycles that limit the persistence of jet activity.}

{Albeit the morphological diversity and the prevalence of substructure, the radio power distribution of the sample remains broadly consistent with the FR classification scheme. Giant radio galaxies span both morphological classes, and hybrid FR\,I/II systems occupy the expected transitional region \citep{kapinska2017radio, ceglowski2013orientation}. These results support the necessity of the FR framework while emphasising that radio power alone does not uniquely determine morphology \citep{ledlow199620cm, hardcastle2007hot, saripalli2012understanding}. Instead, the observed structures likely reflect a combination of jet power, environmental conditions, and the evolutionary history of the central engine.
The role of environment in tailed radio galaxies is well assessed, but the case of CGCG\,047$-$067 is extreme: the sharp transition of the surface brightness and spectral behaviour of the northern jet does suggest a substantial change in the surrounding medium. Another clear example of interaction between the radio emission and the ICM in our sample is NGC\,7503, whose jets show wiggles and distortions which could be explained by invoking local interactions as the jets propagate through a turbulent ICM, where velocity shear at the jet-ICM interface can significantly perturb the flow (see also the case of NGC\,1265, \citealt{gendron2021vla}).  The local environment of NGC\,7503 is noteworthy, as two nearby bright galaxies NGC\,7499 (V=13.00, $z=0.039$) and NGC\,7501 (V=13.4, $z=0.043$) are also radio loud. NGC\,7499 is a typical FR\,I radio galaxy \citep{hogan}, extending over $\sim$70 kpc with a total radio power of P$_{\rm 700\,MHz}=4\times10^{23}$ W\,Hz$^{-1}$. Its two lobes are detached from the nuclear emission coincident with the optical host and are connected to the core by thin, faint filaments. The lobes are misaligned, forming an angle of $\sim120^\circ$, suggestive of motion toward the north in the plane of the sky. NGC\,7501 is characterised by compact radio emission with P$_{\rm 700~MHz}=2.6\times10^{22}$ W\,Hz$^{-1}$ and may be classified as an FR\,0 radio galaxy.
\newline
Another interesting case is 4C\,$-$03.43, a radio galaxy belonging to the galaxy cluster A\,1308 associated with a m=15.22 galaxy, which is the dominant galaxy in the cluster. Its special location, together with the morphological features we report here, suggests that we might see a WAT radio galaxy viewed almost face-on. Alternative interpretations are found in the literature. \cite{rec}  suggests that the orientation rises due to the source's rotational dynamics, as opposed to motion through the surrounding external medium when viewed in projection. The jets are aligned along the north-east and south-west directions. The radio galaxy shows bright spots resembling `knots', likely caused by clumpiness in the underlying medium, leading to turbulence within the jet, further enriching the details of the jet morphology. The 4C\,$-$03.43 morphology shares similarities with the structure observed in the radio galaxy M\,84. The jets in M\,84 are speculated to have undergone a possible precession \citep{bambic2023agn}. 
The 4C\,$-$03.43 northern jet bend point bears similarities with the morphology of MRC\,0600$-$399 as revealed by MeerKAT \citep{Chibueze}, which has been interpreted as the result of interaction with a density contact discontinuity, thus suggesting that a change in density within the surrounding medium may influence the jet's trajectory. Alternatively, the radio source can be interpreted as a WAT viewed face-on.}

\begin{table}
\centering
\caption{Synchrotron parameters. Column 1: source name. Column 2: break frequency derived from the fits. Column 3: equipartition magnetic field. Column 4: source age.}
\label{ages}
\setlength{\tabcolsep}{7pt}
\renewcommand{\arraystretch}{1.25}
\begin{tabular}{lccc}
\toprule
Source name & $\nu_{\mathrm{br}}$ & $B_{\mathrm{eq}}$ & $t_{\mathrm{rad}}$ \\
            & (GHz)               & ($\mu$G)          & (Myr)              \\
\midrule
3C\,105          & $\brfreq{6.00}{0.77}{0.66}$  & 4.1 & 39.7  \\
3C\,198          & $\brfreq{0.22}{0.05}{0.04}$ & 1.5 & 242.2 \\
3C\,227          & $\brfreq{1.56}{0.15}{0.14}$  & 5.1 & 67.3  \\
CGCG\,047$-$067  & $\brfreq{7.54}{1.80}{1.30}$  & 1.5 & 46.2  \\
3C\,403.1        & $\brfreq{3.26}{1.50}{0.96}$  & 2.6 & 69.2  \\
3C\,445          & $\brfreq{1.13}{0.12}{0.11}$  & 2.5 & 118.7 \\
NGC\,7503        & $\brfreq{3.27}{0.52}{0.43}$  & 2.6 & 71.6  \\
\bottomrule
\end{tabular}
\end{table}

\section{Conclusions}

In this paper, we presented MeerKAT L-band and uGMRT Band-4 observations of 10 radio galaxies which, together with the sources reported in Paper I, bring our investigation to a total of 14 out of 17 radio galaxies of the selected sample. Our observations cover the frequency range 550 -- 1712~MHz and the resulting images yield an angular resolution of  4$^{\prime \prime}$ – 10$^{\prime \prime}$, allowing detailed analysis of the features within the radio emission of each source. 
\newline
We discussed our results in light of all 14 radio galaxies studies so far.
Our main conclusions are reported below:

\begin{itemize}[leftmargin=*]
\item The classical FR classification remains a valuable reference; however, the richness of details which is imaged at the sensitivity and angular resolution of MeerKAT and uGMRT provides more depth to the scheme.
\item Our analysis of low surface brightness emission around FR\,0
  reveal no significant extension, i.e., the objects largely remain compact and unresolved at the current sensitivity. SDSS J0917$+$1331 give hints of marginal faint extended emission, which requires deeper low-frequency observation to fully characterise its structure. 
\item Overall, the observations highlight the interplay between intrinsic jet properties and external conditions in shaping radio galaxies. While the FR dichotomy provides a useful first-order classification, the detailed morphologies, particularly filaments, asymmetries, and hybrid structures, emphasise the need for models that incorporate environmental complexity and episodic jet activity to fully describe radio-galaxy evolution. 
\item {The broad frequency coverage available for our sample enabled spectral ageing analysis across the sample. The radiative ages in our sample span the range $\sim$40 -- 242 Myr, with one of them, 3C\,198, showing age and morphological features suggestive of a transition stage between a dying radio galaxy and an amorphous radio source in the ICM. These results indicate that the sample likely spans multiple evolutionary stages from actively fueled sources  to remnant/dying objects, as well as those with evidence of restarted activities.  Although the spectral index distributions across the sample are generally narrow, the expected variations could not be fully explored due to resolution mismatches, making detailed spectral index mapping unfeasible at present.}
\end{itemize}
 Polarization measurements will provide further insight into the environments of these sources, complementing the structural information obtained from total intensity imaging. Future observations with MeerKAT’s UHF and S-band receivers will offer improved frequency coverage and resolution, enabling more comprehensive investigations of spectral properties and environmental interactions in these radio galaxies.
\section*{Acknowledgements}

{ The authors thank the anonymous referee for the insightful comments and suggestions which improved the clarity of the manuscript.}
D.V.L. acknowledge the support of the Department of Atomic Energy, Government of India, under project no. 12-R\&D-TFR-5.02-0700.
We thank the MeerKAT staff for the observations. TV and PPL acknowledge the support from the Ministero degli Affari Esteri della Cooperazione Internazionale - Direzione Generale per la Promozione del Sistema Paese Progetto di Grande Rilevanza  ZA18GR02.
This work is based on research supported by the National Research Foundation of South Africa (Grant Number 113121). OMS's research is supported by the South African Research Chairs Initiative of the Department of Science and Technology and National Research Foundation (Grant No. 81737). Basic research in radio astronomy at the Naval Research Laboratory is supported by 6.1 Base funding. The South African Radio Astronomy Observatory, facility of the National Research Foundation, an agency of the Department of Science and Innovation, operates the MeerKAT telescope. We thank the staff of the GMRT, which is run by the National Centre for Radio Astrophysics of the Tata Institute of Fundamental Research.
%uGMRT is run by the National Centre for Radio Astrophysics of the Tata Institute of Fundamental Research. 
This research has made use of the NASA Extragalactic Database, which is operated by the Jet Propulsion Laboratory, Caltech, under contract with NASA, and NASA’s Astrophysics Data System. 
%%%%%%%%%%%%%%%%%%%%%%%%%%%%%%%%%%%%%%%%%%%%%%%%%%

\section*{DATA AVAILABILITY}
The raw MeerKAT visibilities are available from the SARAO archive (\url{https://archive.sarao.ac.za/}) under proposal ID SCI-20210212-BF-02. GMRT data can be accessed via the GMRT Online Archive (\url{https://naps.ncra.tifr.res.in/naps}) using project code 43\_058. Images may be shared upon reasonable request to the corresponding author. All software packages employed in this work are publicly available, with URLs provided in the main text.

%%%%%%%%%%%%%%%%%%%% REFERENCES %%%%%%%%%%%%%%%%%%

% The best way to enter references is to use BibTeX:

\bibliographystyle{mnras}
\bibliography{ref} % if your bibtex file is called example.bib

% Alternatively you could enter them by hand, like this:
% This method is tedious and prone to error if you have lots of references
%\begin{thebibliography}{99}
%\bibitem[\protect\citeauthoryear{Author}{2012}]{Author2012}
%Author A.~N., 2013, Journal of Improbable Astronomy, 1, 1
%\bibitem[\protect\citeauthoryear{Others}{2013}]{Others2013}
%Others S., 2012, Journal of Interesting Stuff, 17, 198
%\end{thebibliography}

%%%%%%%%%%%%%%%%%%%%%%%%%%%%%%%%%%%%%%%%%%%%%%%%%%

%%%%%%%%%%%%%%%%% APPENDICES %%%%%%%%%%%%%%%%%%%%%
%\newpage
%\clearpage
\appendix

\section{Properties of the radio galaxies in the sample}\label{ngc75}

In the following, we briefly summarise the literature on the ten galaxies presented in this paper.

\medskip\noindent

\medskip\noindent
\textbf{3C\,105} is a classical FR\,II source, associated with a m$_r$=17.6 galaxy  \citep{3c105_magnitude} located at  z=0.089 \citep{leahy}. The host is a narrow-line radio galaxy (NLRG) with
only nuclear emission lines \citep{baum, smith&hec, tad}. The separation between the hotspots is $\sim$500 kpc, and both hot spots show multiple peaks. \cite{masaro2010} detected X-ray emission from the core and south-eastern hotspot, both from where the radio jet appears to enter the hotspot region and from the brightest radio emission at the terminal hotspot itself. \cite{orienti2012} detected near-infrared
emission  also associated with the same hotspot. Recently, HI in absorption has been detected in this radio galaxy
\citep{murry2021}, as well as a nuclear and extended outflow of ionized gas \citep{speranza}.

\medskip\noindent
\textbf{3C\,198} is classified as FR\,I radio galaxy, it is associated with a m$_g$=17.3 \citep{adelman2008sixth} galaxy located at z=0.081 and is the brightest member of the group MSPM04593 \citep{smith}.
No recent deep imaging and detailed study of this source has been carried out in the radio band.
The optical counterpart shows a star-forming spectrum 
\citep{baldi_capetti, buttiglione, runge}.

\medskip\noindent
\textbf{3C\,227} is an FR\,II radio galaxy associated with a m$_g$=16.7 quasar at z=0.085 \citep{adelman2008sixth}. It is aligned in East-West, with multiple peaks of emission in each hotspot. The multiple hotspots are detected up to very high energy as a result of particle re-acceleration (NIR, optical, X-ray emission, see \cite{miglio,2020orienti} respectively). The hotspots are polarized \citep{2020orienti}.

\medskip\noindent
\textbf{4C\,--03.43} is a classical FR\,I radio galaxy at z=0.0554 \citep[m=15.22,][]{bcg}. It is characterised by two symmetric jets which open up and spread into two lobes. 
The radio galaxy is associated with the brightest cluster galaxy (BCG) in the cluster Abell\,1308 at z=0.0519 \citep{bcg}.
Very little literature information is available for this object.

\medskip\noindent
\textbf{CGCG\,047$-$067} is an FR\,I radio galaxy hosted by an optical
counterpart with magnitude m$_g$=14.9 at redshift z=0.055901 \citep{AdelmanMcCarthy2007}. Very little information is available in the literature for this radio galaxy, and no detailed radio studies exist. \cite{kuzmicz2018updated} included it in a sample of giant radio galaxies, reporting a total size ok 0.7 Mpc.
The source appears in the G4\, Jy sample as G4\,Jy\,1173, classified as a bent tail morphology, with double elongated and possibly L-shaped structure \citep{SARA1, sara2}. Very little is known on its environment. A galaxy group is reported in the literature at the redshift and position of CGCG\,047$-$067 (MSPM\,01896, \cite{smith}. \\
\newline\noindent
\textbf{3C\,403.1} is an FR\,I radio galaxy associated with a 17.5 magnitude galaxy \citep{3c403.1magni} at redshift z=0.055. \cite{missaglia2022high} recently conducted a radio/optical study and concluded that the radio galaxy is part of a low-mass poor group. The radio galaxy belongs to the MURALES sample \citep{speranza,2019Balmaverde,2021Balmaverde}, investigated in search for nuclear outflows. None was detected.\\
\newline\noindent
\textbf{3C\,445} is another FR\,II radio galaxy located at z=0.0568 \citep{vla_paper}. The associated optical counterpart,
a m$_r$=15.2 \citep{3c105_magnitude} elliptical galaxy possessing an exceptionally bright and compact central region \citep{Hubble_paper_of_3CR},
is a typical BLRG emitting very strong broad emission lines \citep{vla_paper_cf_rudy}.
With a linear size of $\sim$630 kpc, it belongs to the class of giant radio galaxies. It is aligned along  north-south
direction, and the northern hotspot has multiple peaks. Both lobes are characterised by X-ray and infrared emission \citep{orienti2012, Mack} and have been investigated in detail with the JVLA at 22 GHz to study the distribution of
the strong shocks in the hotspot region. The radio galaxy was imaged at high angular resolution in \cite{leahy},
where the two inner compact components located north and south of the core and aligned along the jet direction
(labelled N1 and S1 in their paper) were identified. These two components and their alignment with the direction of the
outer hot spots were suggestive of a double-double radio galaxy \citep{DDRG}. \\
\newline
\noindent
\textbf{NGC\,7503} is a narrow-angle tailed radio galaxy associated with a bright (V=13.5) galaxy at z=0.044 in the Pegasus II cluster (Zw2307.6$+$0713, Z\,8852). The cluster shows X--ray emission \citep[MCXC\, J2310.4$+$0734,][]{Piffaretti}. The optical counterpart is offset by approximately  $\sim5^{\prime}$  in projection from the X-ray centre of the cluster and is dominated by the brightest cluster galaxy (BCG), NGC 7499.\\
\newline
\noindent
 \textbf{SDSS\,J\,1120$+$0407} is associated with the galaxy CGCG\,039$-$127, with m$_g$=15.4 and z=0.049655 \citep{j11_mag}, and \textbf{SDSS\,J\,0917$+$1331} is associated with WISEA J\,091754.26$+$133145.5, with magnitude 17.79 and located at similar redshift  \citep[z=0.049932,][]{adelman2008sixth}.
 In both cases, the optical counterpart is an early-type galaxy, and the radio source is classified as FR\,0. They are characterized by the lack of extended radio lobes and show the same core properties as FR\,I \citep{baldi}. {\cite{capetti2020large} reported on the LOFAR detection of both sources at 144 MHz, with no emission revealed in either case.}
 \clearpage
\section{Integrated spectrum}
\begin{minipage}{2\linewidth}
\centering
\label{tab:appendix}
\begin{tabular}{@{}|lllllllll@{}}
   \toprule
 Source Name& Frequency& flux density& Reference&Source Name& Frequency& flux density& Reference\\
 &GHz& Jy &&&GHz& Jy \\
 \hline
3C\,105 & 31.4 & $0.39\pm0.07$ & 1 &  & 2.7 & 0.9 & 6 \\
 & 14.9 & $0.71\pm0.01$ & 2 &  & 2.7 & $0.98\pm0.15$ & 8 \\
 & 14.9 & $0.0198\pm5.0E-4$ & 3 &  & 1.41 & 2.3 & 6 \\
 & 10.7 & $1.49\pm0.1$ & 4 &  & 1.4 & $2.1\pm0.1$ & 8 \\
 & 10.7 & $1.5\pm0.05$ & 5 &  & 1.4 & 1.8 & 15 \\
 & 8.4 & 1.22 & 6 &  & 1.4 & $2.12\pm0.08$ & 16 \\
 & 5.01 & $2.45\pm0.1$ & 7 &  & 1.28 & -- & $\ast$ \\
 & 5.01 & $2.5\pm0.06$ & 7 &  & 0.75 & $3.78\pm0.06$ & 16 \\
 & 5 & $2.16\pm0.11$ & 8 &  & 0.75 & $3.6\pm0.54$ & 8 \\
 & 5 & 2.02 & 6 &  & 0.7 & $3.91\pm0.07$ & $\ast$ \\
 & 5 & $2.16\pm0.05$ & 9 &  & 0.408 & $4.65\pm0.31$ & 19 \\
 & 5 & $2.14\pm0.11$ & 7 &  & 0.408 & 6.6 & 6 \\
 & 4.85 & $2.06\pm0.288$ & 10 &  & 0.178 & 18 & 6 \\
 & 4.85 & $2.03\pm0.305$ & 11 &  & 0.178 & $9\pm1.12$ & 20 \\
 & 4.85 & $2.23\pm0.099$ & 12 &  & 0.178 & $9.7\pm1.46$ & 8 \\
 & 2.7 & 3.28 & 6 &  & 0.16 & 12 & 21 \\
 & 2.7 & $3.44\pm0.17$ & 7 &  & 0.145 & 16 & 22 \\
 & 2.7 & $3.55\pm0.12$ & 13 & 3C\,227 & 31.4 & $0.53\pm0.13$ & 1 \\
 & 2.7 & $3.74\pm0.06$ & 14 &  & 22.4 & $0.012\pm3.0E-4$ & 3 \\
 & 2.7 & $3.36\pm0.17$ & 8 &  & 14.9 & $0.014\pm3.0E-4$ & 3 \\
 & 2.7 & $3.4\pm0.05$ & 7 &  & 14.9 & $0.96\pm0.19$ & 2 \\
 & 2.65 & $3.79\pm0.07$ & 14 &  & 10.7 & $1.51\pm0.05$ & 4 \\
 & 1.41 & 5.1 & 6 &  & 8 & $2.05\pm0.0492$ & 24 \\
 & 1.41 & $5.84\pm0.19$ & 7 &  & 6.7 & $1.93\pm0.048$ & 25 \\
 & 1.4 & 5.34 & 15 &  & 5.01 & $2.75\pm0.11$ & 7 \\
 & 1.4 & $5.21\pm0.29$ & 16 &  & 5.01 & $2.68\pm0.1$ & 7 \\
 & 1.4 & $5.2\pm0.5$ & 7 &  & 5 & $2.58\pm0.13$ & 7 \\
 & 1.4 & $5.31\pm0.15$ & 7 &  & 5 & 2.6 & 6 \\
 & 1.4 & $5.1\pm0.77$ & 8 &  & 5 & $2.6\pm0.13$ & 8 \\
 & 1.4 & $0.114\pm0.004$ & 17 &  & 4.85 & $2.14\pm0.322$ & 11 \\
 & 1.28 & $6.60\pm0.30$ & $\ast$ &  & 4.85 & $2.58\pm0.099$ & 12 \\
 & 0.75 & $7.7\pm0.39$ & 8 &  & 4.85 & $2.19\pm0.304$ & 10 \\
 & 0.75 & $8.11\pm0.19$ & 16 &  & 4.78 & 1.71 & 23 \\
 & 0.75 & $8.1\pm0.4$ & 7 &  & 2.7 & $4.63\pm0.06$ & 14 \\
 & 0.75 & $8.59\pm0.2$ & 7 &  & 2.7 & $4.26\pm0.21$ & 7 \\
 & 0.7 & -- & * &  & 2.7 & 4.1 & 6 \\
 & 0.635 & 8.9 & 6 &  & 2.7 & $4.16\pm0.21$ & 8 \\
 & 0.635 & $10\pm0.36$ & 7 &  & 2.7 & $4.21\pm0.1$ & 7 \\
 & 0.408 & $10.3\pm2.55$ & 18 &  & 2.65 & $4.72\pm0.07$ & 14 \\
 & 0.408 & 9.35 & 6 &  & 1.41 & 7.1 & 6 \\
 & 0.408 & $9.35\pm0.29$ & 19 &  & 1.41 & $7.92\pm0.22$ & 7 \\
 & 0.318 & $14.2\pm0.56$ & 7 &  & 1.4 & 3.12 & 26 \\
 & 0.178 & $17.8\pm1.78$ & 8 &  & 1.4 & 7.62 & 26 \\
 & 0.178 & $18\pm1.8$ & 7 &  & 1.4 & $7.1\pm0.13$ & 16 \\
 & 0.178 & $19\pm0.9$ & 7 &  & 1.4 & $6.8\pm0.34$ & 8 \\
 & 0.178 & 16.2 & 6 &  & 1.4 & $7\pm0.3$ & 7 \\
 & 0.178 & $9.9\pm1.24$ & 20 &  & 1.4 & 6.94 & 15 \\
 & 0.16 & 23.5 & 21 &  & 1.4 & $7.5\pm0.22$ & 7 \\
 & 0.16 & $18\pm2.7$ & 7 &  & 1.28 & -- & * \\
 & 0.145 & 46.2 & 22 &  & 0.96 & $10.7\pm0.18$ & 7 \\
3C\,198 & 10.7 & $0.19\pm0.05$ & 4 &  & 0.75 & $11.6\pm0.58$ & 8 \\
 & 5 & 0.32 & 6 &  & 0.75 & $12.1\pm0.6$ & 7 \\
 & 5 & $0.46\pm0.07$ & 8 &  & 0.75 & $12.8\pm0.14$ & 7 \\
 & 4.85 & $0.305\pm0.042$ & 10 &  & 0.75 & $12.1\pm0.13$ & 16 \\
 & 4.85 & $0.539\pm0.081$ & 11 &  & 0.7 & $12.46\pm0.02$ & $\ast$ \\
 & 4.85 & $0.436\pm0.025$ & 12 &  & 0.468 & $18\pm1.46$ & 7 \\
 & 4.78 & 0.451 & 23 &  & 0.408 & $19.4\pm3.91$ & 18 \\																			
\end{tabular}
\end{minipage}

\clearpage
\begin{minipage}{2\linewidth}
\centering
\begin{tabular}{@{}|lllllllll@{}}
   \toprule
Source Name& Frequency& flux density& Reference&Source Name& Frequency& flux density& Reference\\
&GHz& Jy &&&GHz& Jy \\
\hline
 & 0.408 & $15.5\pm0.48$ & 19 &  & 1.4 & $1.79\pm0.12$ & 16 \\
 & 0.408 & 22.1 & 6 &  & 1.28 & -- & $\ast$ \\
 & 0.178 & $32.6\pm1.6$ & 7 &  & 0.75 & $2.83\pm0.11$ & 16 \\
 & 0.178 & $33.3\pm3.3$ & 7 &  & 0.75 & $2.7\pm0.41$ & 8 \\
 & 0.178 & 30 & 6 &  & 0.7 & $1.45\pm0.08$ & $\ast$ \\
 & 0.178 & $16.5\pm1.32$ & 20 &  & 0.408 & 4.4 & 6 \\
 & 0.178 & $30.4\pm3.04$ & 8 &  & 0.178 & 13.5 & 6 \\
 & 0.16 & $39.6\pm5.2$ & 7 &  & 0.178 & $13.5\pm2.02$ & 8 \\
 & 0.16 & 45.2 & 21 &  & 0.178 & $3.3\pm0.825$ & 20 \\
 & 0.145 & 44.2 & 22 &  & 0.16 & 9.5 & 21 \\
4C\,-03.43 & 22.5 & $0.0621\pm0.001$ & 27 & 3C\,445 & 22.4 & $0.0161\pm0.001$ & 3 \\
 & 19.9 & $0.108\pm0.006$ & 28 &  & 22 & $0.045\pm0.005$ & 35 \\
 & 8.46 & $0.107\pm0.001$ & 27 &  & 18.5 & $0.039\pm0.002$ & 35 \\
 & 8.4 & 0.048 & 29 &  & 14.9 & 0.027 & 3 \\
 & 5 & 0.3 & 6 &  & 10.7 & $0.82\pm0.03$ & 4 \\
 & 4.86 & $0.159\pm0.002$ & 27 &  & 8.4 & 0.32 & 6 \\
 & 4.85 & $0.37\pm0.022$ & 12 &  & 8 & $1.34\pm0.079$ & 24 \\
 & 2.7 & 0.56 & 6 &  & 5.01 & $2.18\pm0.09$ & 7 \\
 & 1.4 & 1.11 & 15 &  & 5 & $2.04\pm0.1$ & 8 \\
 & 1.4 & $0.32\pm0.01$ & 30 &  & 5 & $2.03\pm0.1$ & 7 \\
 & 1.28 & $1.08\pm0.07$ & $\ast$ &  & 5 & 2.12 & 6 \\
 & 0.7 & $1.45\pm0.07$ & $\ast$ &  & 4.85 & $1.26\pm0.066$ & 12 \\
 & 0.408 & 1.18 & 6 &  & 2.7 & $3.3\pm0.16$ & 7 \\
 & 0.408 & $1.18\pm0.07$ & 19 &  & 2.7 & $3.46\pm0.12$ & 13 \\
 & 0.365 & $0.434\pm0.054$ & 31 &  & 2.7 & 3.46 & 6 \\
 & 0.186 & $1.76\pm0.036$ & 32 &  & 2.7 & $3.34\pm0.33$ & 36 \\
 & 0.178 & 2 & 6 &  & 2.7 & $3.22\pm0.16$ & 8 \\
 & 0.178 & $2\pm0.3$ & 20 &  & 1.41 & 5.1 & 6 \\
 & 0.16 & 2.5 & 21 &  & 1.4 & $5.75\pm0.12$ & 7 \\
 & 0.155 & $2.28\pm0.051$ & 32 &  & 1.4 & $5.3\pm0.27$ & 8 \\
CGCG\,047-067 & 8 & $0.52\pm0.0452$ & 24 &  & 1.4 & $5.5\pm0.3$ & 7 \\
 & 5 & 0.66 & 6 &  & 1.4 & 6.16 & 15 \\
 & 4.85 & $0.557\pm0.031$ & 12 &  & 1.4 & $5.33\pm0.1$ & 16 \\
 & 4.85 & $0.481\pm0.072$ & 11 &  & 1.280 & $6.460\pm0.320$ & $\ast$ \\
 & 4.85 & $0.511\pm0.071$ & 10 &  & 0.75 & $8.7\pm0.43$ & 8 \\
 & 4.78 & 0.575 & 23 &  & 0.75 & $9.67\pm0.18$ & 7 \\
 & 2.7 & 1.06 & 6 &  & 0.75 & $9.1\pm0.5$ & 7 \\
 & 1.41 & 2.1 & 6 &  & 0.75 & $9.13\pm0.17$ & 16 \\
 & 1.4 & 2.01 & 33 &  & 0.7 & -- & $\ast$ \\
 & 1.4 & 1.79 & 34 &  & 0.635 & 12.2 & 6 \\
 & 1.28 & $2.39\pm0.01$ & $\ast$ &  & 0.408 & 17.5 & 6 \\
 & 0.7 & $3.53\pm0.16$ & $\ast$ &  & 0.408 & $14.5\pm3.18$ & 18 \\
 & 0.408 & 5.8 & 6 &  & 0.408 & $7.15\pm0.18$ & 19 \\
 & 0.408 & $2.82\pm0.13$ & 19 &  & 0.178 & $26.1\pm5.2$ & 7 \\
 & 0.178 & 3.5 & 6 &  & 0.178 & $25.2\pm3.24$ & 7 \\
 & 0.178 & $3.5\pm0.525$ & 20 &  & 0.178 & 23.5 & 6 \\
 & 0.16 & 7.9 & 21 &  & 0.178 & $23.5\pm1.88$ & 20 \\
3C\,403.1 & 8.4 & 0.1 & 6 &  & 0.178 & $24.8\pm3.72$ & 8 \\
 & 5 & $0.45\pm0.11$ & 8 &  & 0.16 & $33.5\pm4.4$ & 7 \\
 & 5 & 0.58 & 6 &  & 0.145 & 36.5 & 22 \\
 & 4.85 & $0.344\pm0.021$ & 12 & NGC\,7503 & 8 & $0.41\pm0.039$ & 24 \\
 & 4.85 & $0.58\pm0.087$ & 11 &  & 5 & 0.52 & 6 \\
 & 2.7 & 0.7 & 6 &  & 4.85 & $0.544\pm0.082$ & 11 \\
 & 2.7 & $0.96\pm0.14$ & 8 &  & 4.85 & $0.557\pm0.031$ & 12 \\
 & 1.41 & 1.5 & 6 &  & 4.85 & $0.547\pm0.076$ & 10 \\
 & 1.4 & $1.8\pm0.27$ & 8 &  & 2.7 & 0.96 & 6 \\
 & 1.4 & 1.85 & 15 &  & 1.41 & 1.7 & 6 \\

\end{tabular}

\end{minipage}
%%%%%%%%%%%%%%%%%%%%%%%%%

\clearpage
\begin{minipage}{2\linewidth}
\centering
\captionsetup{width=\linewidth} % caption matches table width
\begin{tabular}{@{}|lllllllll@{}}
   \toprule
   Source Name & Frequency & flux density & Reference & Source Name & Frequency & flux density & Reference \\
   & GHz & Jy &&& GHz & Jy \\
   \hline
 &1.4 & 1.63 & 15 \\
&1.28 & -- & $\ast$ \\
&0.7 & $2.58\pm0.02$ & $\ast$ \\
 &0.408 & $3.88\pm0.18$ & 19 \\
 &0.408 & 3.88 & 6 \\
 &0.178 & $4.5\pm0.562$ & 20 \\
 &0.178 & 4.5 & 6 \\	
   \hline
\end{tabular}
\vspace{0.5em}
\captionof{table}{The integrated flux densities of the radio sources at various frequencies presented in the literature. 
1. Geldzahler and Witzel (1981), 
2. Genzel et al. (1976), 
3. Dicken et al. (2008), 
4. Kellermann and Pauliny-Toth (1973), 
5. Pauliny-Toth et al. (1978), 
6. Wright and Otrupcek (1990), 
7. Kuehr et al. (1981), 
8. Kellermann et al. (1969), 
9. Pauliny-Toth et al. (1972), 
10. Gregory and Condon (1991), 
11. Becker et al. (1991), 
12. Griffith et al. (1995), 
13. Wall et al. (1971), 
14. Wills (1975), 
15. White and Becker (1992), 
16. Pauliny-Toth et al. (1966), 
17. Condon et al. (1998), 
18. Ekers (1969), 
19. Large et al. (1981), 
20. Gower et al. (1967), 
21. Slee (1995), 
22. Jacobs et al. (2011), 
23. Bennett et al. (1986), 
24. Stull (1971), 
25. Carter et al. (2009), 
26. Ching et al. (2017), 
27. Lin et al. (2009), 
28. Murphy et al. (2010), 
29. Healey et al. (2007), 
30. Douglas et al. (1996), 
31. Tingay et al. (2016), 
32. Yuan et al. (2017), 
33. Lin et al. (2018), 
34. Ku\'{z}micz et al. (2018), 
35. Ricci et al. (2006), 
36. Witzel et al. (1971).
}
\end{minipage}

\section{Spectral index error}\label{error}

\begin{minipage}{2\linewidth}

  \begin{tabular}{@{}|lccc@{}}
\includegraphics[width=0.35\textwidth]{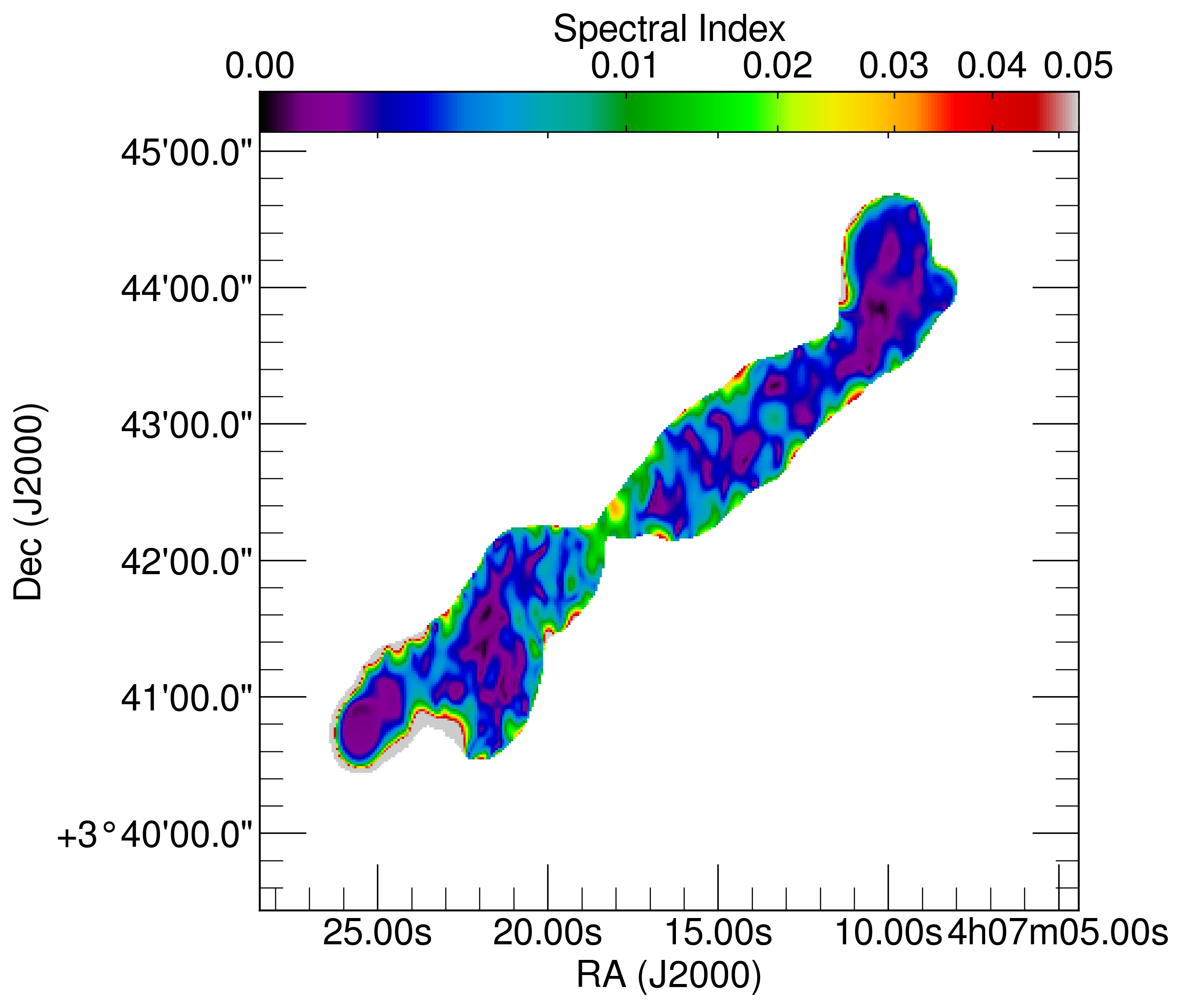} &
\hspace*{-0.9cm}
       \includegraphics[width=0.35\textwidth]{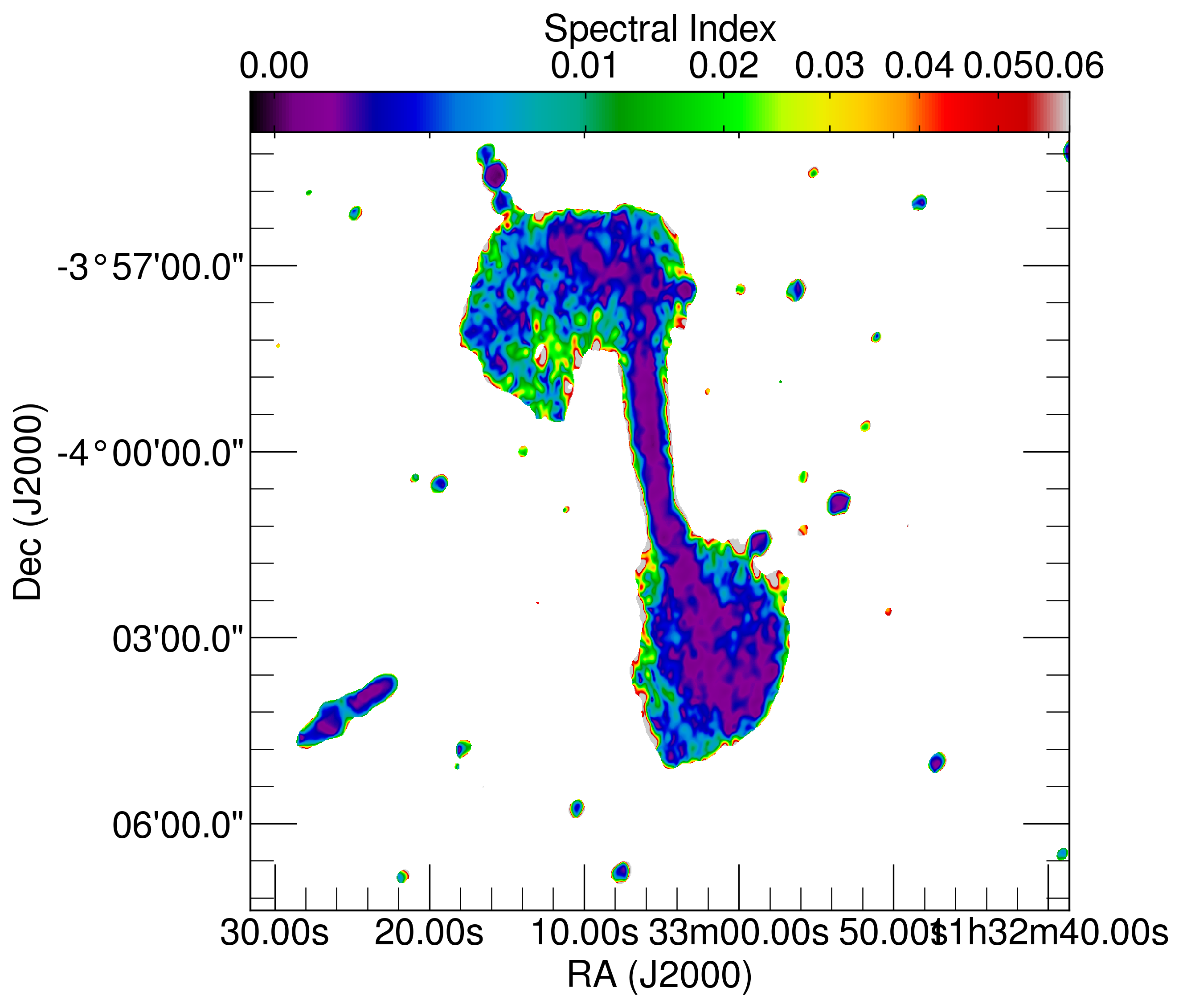} &
       \hspace*{-0.9cm} \includegraphics[width=0.325\textwidth]{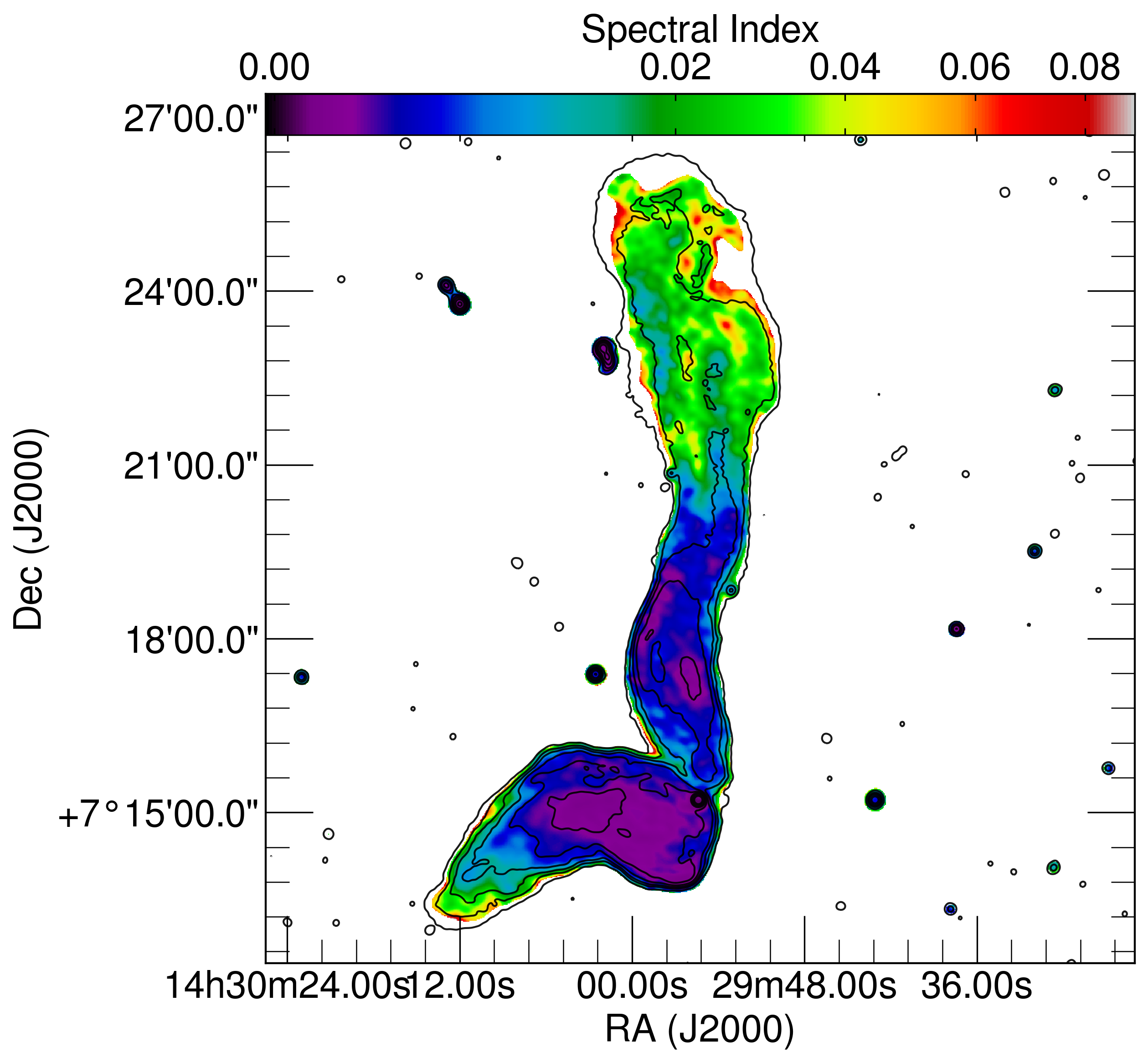} 
\end{tabular}
\label{fig:specinderror}
  \captionof{figure}{MeerKAT in-band spectral index error maps for the sources 3C\,105, 4C\,$-$03.43, and CGCG\,047$-$067.}  
%\end{figure}
\end{minipage}

%%%%%%%%%%%%%%%%%%%%%%%%%%%%%%%%%%%%%%%%%%%%%%%%%%

% Don't change these lines
%\bsp	% typesetting comment
\label{lastpage}
\end{document}